\documentclass[12pt]{article}
\pdfoutput=1

\usepackage{putex}
\usepackage{autobreak}
\usepackage{graphicx}
\usepackage{caption}
\usepackage{amsmath}
\usepackage{array}
\usepackage{subcaption}
\usepackage{epstopdf}
\usepackage{enumerate}
\usepackage{cite}
\usepackage{youngtab}
\usepackage{tensor}
\usepackage{slashed}
\usepackage[aligntableaux=center]{ytableau}
\usepackage[utf8]{inputenc}
\usepackage{rotating}
\usepackage{multirow}
\usepackage[
      colorlinks=true,
      linkcolor=blue,
      urlcolor=blue,
      filecolor=black,
      citecolor=red,
      ]{hyperref}

\newcommand {\be} {\begin {equation}}
\newcommand {\ee} {\end {equation}}

\newcommand {\bes} {\begin {equation*}}
\newcommand {\ees} {\end {equation*}}

\newcommand{\es}[2] {\begin{equation} \label{#1} \begin{split} #2 \end{split} \end{equation}}

\newcommand{\cN}{{\mathcal N}}
\newcommand{\cO}{{\mathcal O}}
\newcommand{\cP}{{\mathcal P}}

\newcommand{\cR}{{\mathcal R}}
\newcommand{\cS}{{\mathcal S}}
\newcommand{\cT}{{\mathcal T}}

\newcommand{\beq}{\begin{equation}}
\newcommand{\eeq}{\end{equation}}

\def\ie{\begin{equation}\begin{aligned}}
\def\fe{\end{aligned}\end{equation}}

\numberwithin{equation}{section}

\def\<{\langle}
\def\>{\rangle}

\newcommand{\ak}{\alpha}        
         
\newcommand{\gk}{\gamma}        \newcommand{\Gk}{\Gamma}
\newcommand{\dk}{\delta}        \newcommand{\Dk}{\Delta}
   
\newcommand{\zk}{\zeta}         
          
\newcommand{\qk}{\theta}        
         
\newcommand{\kk}{\kappa}        
\newcommand{\lk}{\lambda}

\newcommand{\fk}{\phi}          \newcommand{\Fk}{\Phi}

\newcommand{\nb}{\partial}      \newcommand{\Nb}{\nabla}

\newcommand{\fcy}[1]{\mathcal{#1}}
\newcommand{\bb}[1]{\mathbb{#1}}

\begin{document}

\preprint{PUPT-2624 \\ LCTP-21-06}

\institution{PU}{Joseph Henry Laboratories, Princeton University, Princeton, NJ 08544, USA}
\institution{Exile}{Department of Particle Physics and Astrophysics, Weizmann Institute of Science, \cr Rehovot, Israel}
\institution{UMich}{Leinweber Center for Theoretical Physics, Randall Laboratory of Physics, \cr Department of Physics, University of Michigan, Ann Arbor, MI 48109, USA}

\title{
ABJ Correlators with Weakly Broken Higher Spin Symmetry
}

\authors{Damon J.~Binder,\worksat{\PU} Shai M.~Chester,\worksat{\Exile} and Max Jerdee\worksat{\PU, \UMich}  }

\abstract{
We consider four-point functions of operators in the stress tensor multiplet of the 3d $\mathcal{N}=6$ $U(N)_k\times U(N+M)_{-k}$ or $SO(2)_{2k}\times USp(2+2M)_{-k}$ ABJ theories in the limit where $M$ and $k$ are taken to infinity while $N$ and $\lambda\sim M/k$ are held fixed.  In this limit, these theories have weakly broken higher spin symmetry and are holographically dual to $\mathcal{N}=6$ higher spin gravity on $AdS_4$, where $\lambda$ is dual to the bulk parity breaking parameter.  We use the weakly broken higher spin Ward identities, superconformal Ward identities, and the Lorentzian inversion formula to fully determine the tree level stress tensor multiplet four-point function up to two free parameters. We then use supersymmetric localization to fix both parameters for the ABJ theories in terms of $\lambda$, so that our result for the tree level correlator interpolates between the free theory at $\lambda=0$ and a parity invariant interacting theory at $\lambda=1/2$.  We compare the CFT data extracted from this correlator to a recent numerical bootstrap conjecture for the exact spectrum of $U(1)_{2M}\times U(1+M)_{-2M}$ ABJ theory (i.e. $\lambda=1/2$ and $N=1$), and find good agreement in the higher spin regime.
}
\date{}

\maketitle

\tableofcontents

\section{Introduction and Summary}
\label{intro}

There are two known theories of quantum gravity with dynamical gravitons: string theory (including M-theory) and higher spin gravity. The former has massless particles of spin two and fewer, while the latter has massless particles of all spins.\footnote{Or an infinite subset of spins, such as all even spins. See \cite{Giombi:2016ejx} for a review.} The AdS/CFT duality relates string theory on Anti-de Sitter space (AdS) to conformal field theories (CFTs) with matrix degrees of freedom such as 4d $\mathcal{N}=4$ SYM \cite{Maldacena:1997re} and 3d ABJM theory \cite{Aharony:2008ug}, while higher spin gravity is holographically dual to CFTs with vector degrees of freedom such as the singlet sector of the critical $O(N)$ model \cite{Klebanov:2002ja}. Such vector models are usually easier to study, and AdS/CFT has even been recently derived for the simplest vector models \cite{Aharony:2020omh}. A key question in quantum gravity is if string theory and higher spin gravity are in fact limits of the same universal theory, and if the derivation of AdS/CFT for higher spin gravity can be extended to the richer case of string theory.

The ABJ triality of \cite{Chang:2012kt} proposes a precise relationship between these theories of quantum gravity. String theory, M-theory and higher spin gravity are each conjectured to be related by holography to different regimes of the 3d $\mathcal{N}=6$ ABJ family of CFTs \cite{Aharony:2008gk} with gauge group $U(N)_k\times U(N+M)_{-k}$. The original ABJM paper \cite{Aharony:2008gk} proposed that the large $N$ and finite $k$ and $M$ limit is dual to weakly coupled M-theory on $AdS_4\times S^7/\mathbb{Z}_k$, while the large $N,k$ and finite $M$ and $N/k$ limit is dual to Type IIA string theory on $AdS_4\times \mathbb{CP}^3$.\footnote{These two limits can be considered as different regimes of the more universal large $N,k$ and finite $\mu\equiv N/k^5$ limit considered in \cite{Binder:2019mpb}, where small $\mu$ recovers the strongly coupled (i.e. large $N/k$) Type IIA string theory limit, and large $\mu$ recovers the M-theory limit.} In both cases ABJ has a matrix-like large $N$ limit, as both gauge groups become large. In \cite{Chang:2012kt} a third limit of ABJ was considered, where $M$ and $k$ are large, while $N$ and $\lambda\equiv M/k$ are finite. In this case only one gauge group becomes large and so the theory has a vector-like large $N$ limit. This is dual to an $\cN=6$ theory of higher spin gravity on AdS$_4$, where $\lambda$ is dual to the bulk parity breaking parameter.

Another family of $\cN=6$ theories, the $SO(2)_{2k}\times USp(2+2M)_{-k}$ family of ABJ theories, also has a vector-like limit when $M,k$ are large while $\lambda\equiv (M+1/2)/k$ is held fixed.\footnote{For simplicity, we will use the same symbol $\lambda$ for both the $U(N)_k\times U(N+M)_{-k}$ and $SO(2)_{2k}\times USp(2+2M)_{-k}$ theories, which should be clear by context. The shifted $\lambda$ is the natural variable according to Seiberg duality for the theory \cite{Honda:2017nku}.} In \cite{Honda:2017nku} it was conjectured that this was related to the same $\cN=6$ theory of higher spin gravity on AdS$_4$ as the one in the ABJ triality, but this time with an orientifold. 


Unfortunately, ABJ theory is strongly coupled for all the ranges of parameters of interest to the ABJ triality, except for the weakly coupled limit when $\lambda$ is small,\footnote{More generally, the theory is weakly coupled when $k$ is much bigger than $N$ or $M$. See \cite{Chester:2021gdw,Gorini:2020new} for recent calculations in the weak coupling limit.} which has made the triality difficult to study. Progress on probing the strongly coupled regime of ABJ(M) has been made recently using the analytic conformal bootstrap, which was originally applied to $\mathcal{N}=4$ SYM in \cite{Rastelli:2017udc}. In particular, tree level correlators of single trace operators are fixed at large $N$, or equivalently large stress tensor two-point coefficient $c_T$, in terms of single trace exchange Witten diagrams plus contact terms. For the supergravity limit, which describes both the M-theory and string theory limits at leading order at large $c_T$, the only single trace operators have spin two and less, and their exchange diagrams are completely fixed by superconformal symmetry \cite{Zhou:2017zaw,Alday:2020dtb}. The contact diagrams are restricted by the flat space limit to have two derivatives or less \cite{Penedones:2010ue}, and such contact diagrams are in fact forbidden by superconformal symmetry. For the correlator $\langle SSSS\rangle$ of the stress tensor multiplet superprimary $S$, higher derivative corrections to the supergravity limit were then fixed in terms of a finite number of contact terms \cite{Chester:2018aca}, whose coefficients were computed in either the M-theory \cite{Chester:2018aca,Binder:2018yvd} or string theory \cite{Binder:2019mpb} limits using constraints from supersymmetric localization \cite{Pestun:2007rz,Kapustin:2009kz}.

In this paper we extend these tree level calculations to the higher spin limit of $\<SSSS\>$. As in the supergravity limit, the tree level correlator is fixed in terms of single trace exchange diagrams plus contact diagrams. Unlike the supergravity limit, the higher spin limit has single trace particles of every spin, their exchange diagrams are not completely fixed by superconformal symmetry, and the contact terms can no longer be fixed using the flat space limit as it does not exist for higher spin gravity \cite{Bekaert:2010hw}. We will resolve these problems by combining slightly broken higher spin Ward identities with the Lorentzian inversion formula \cite{Caron-Huot:2017vep}, as in the recent calculation of the analogous non-supersymmetric correlator in \cite{Turiaci:2018nua,Li:2019twz}.\footnote{See also \cite{Silva:2021ece} for a similar calculation of a spinning non-supersymmetric correlator, as well as \cite{Kalloor:2019xjb} for a more direct diagrammatic approach.} In particular, we will first compute tree level three-point functions of single trace operators in terms of $c_T$ and another free parameter using weakly broken higher spin symmetry, which generalizes the non-supersymmetric analysis of \cite{Maldacena:2012sf} to $\mathcal{N}=6$ theories.\footnote{Note that \cite{Maldacena:2012sf} applies to higher spin theories with only one single trace operator of each spin. This excludes the $\mathcal{N}=6$ higher spin theories we consider, whose single trace spectrum includes one higher spin multiplet of each spin plus the stress tensor multiplet, whose component operators includes multiple operators of each spin.} We then use these three-point functions to fix the infinite single trace exchange diagrams that appear in $\<SSSS\>$. Finally, we use the Lorentzian inversion formula to argue that only contact diagrams with six derivatives or less can appear, of which only a single linear combination is allowed by $\mathcal{N}=6$ superconformal symmetry. In sum, we find that $\<SSSS\>$ is fixed at leading order in large $c_T$ in the higher spin limit in terms of two free parameters. 


We then fix these two parameters for the $U(N)_k\times U(N+M)_{-k}$ and $SO(2)_{2k}\times USp(2+2M)_{-k}$ ABJ theories using the mass deformed free energy $F(m_+,m_-) = -\log Z(m_+,m_-)$, which was computed for these theories using supersymmetric localization in \cite{Kapustin:2009kz}. In particular, \cite{Binder:2019mpb} derived two constraints that relate certain integrals of $\langle SSSS\rangle$ to $\partial_{m_\pm}^4F\big\vert_{m_\pm=0}$ and $\partial_{m_+}^2\partial_{m_-}^2F\big\vert_{m_\pm=0}$. Following \cite{Hirano:2015yha,Binder:2020ckj}, we compute these constraints and find them redundant, so that they only fix one of the two unknown parameters. We then use the slightly broken higher spin Ward identities to relate $\langle SSSS\rangle$ to $\langle SSSP\rangle$, where $P$ is a pseudoscalar that appears in the stress tensor multiplet. For parity preserving theories, such as ABJ with $\lambda=0,\frac12$,\footnote{Seiberg duality make these theories periodic in $\lambda$ with period 1 \cite{Hirano:2015yha,Chang:2012kt}.} the superprimary $S$ is parity even and $P$ is parity odd, so $\langle SSSP\rangle$ vanishes in this case, but is nonzero for a generic parity breaking $\lambda$. We derive a new integrated constraint that relates $\langle SSSP\rangle$ to $\nb_{m_\pm}^3\nb_{m_\mp} F(m_+,m_-)\big\vert_{m_\pm=0}$, and then use to this to fully fix the second unknown coefficient in $\<SSSS\>$. When written in terms of $\lambda$ and $c_T\sim M$,\footnote{The precise value of $c_T$ in the large $M$ limit can also be computed using localization, as we will discuss in the main text.} our final result for $\langle SSSS\rangle$ in the tree level higher spin limit then takes the same form for both the $U(N)_{k}\times U(N+M)_{-k}$ and $SO(2)_{2k}\times USp(2+2M)_{-k}$ theories:
\es{finalAnswer}{
\langle S(\vec x_1) S(\vec x_2) S(\vec x_3) S(\vec x_4)\rangle_\text{tree}=\frac{8}{c_T}\frac{\sum_{i=1}^6 \mathcal{B}_i}{|\vec x_{12}|^2|\vec x_{34}|^2}\Big[(2-\sin^2(\pi\lambda))\mathcal{S}^i_\text{free}+\sin^2(\pi\lambda)\mathcal{S}^i_\text{scal} \Big]\,.}
Here, $\mathcal{B}_i$ are certain R-symmetry invariants given in \eqref{BasisElems}, $\mathcal{S}^i_\text{free}$ is the connected part of the correlator for a free $\mathcal{N}=6$ CFT (e.g. $\lambda=0$ ABJ), while $\mathcal{S}^i_\text{scal}$ consists of scalar exchange diagrams. In this basis, $\langle SSSS\rangle$ is uniquely fixed by crossing symmetry in terms of $\mathcal{S}^1$ and $\mathcal{S}^4$, which for the free connected and exchange terms are
\es{finalAnswer2}{
\mathcal{S}^1_\text{free}&=0\,,\qquad\qquad\qquad\qquad \quad\;\;\mathcal{S}^4_\text{free}=\frac{U}{V}\,,\\
\mathcal{S}^1_\text{scal}&=-\frac{2U}{\pi^{\frac52}}\bar{D}_{1,1,\frac12,\frac12}(U,V)\,,\qquad \mathcal{S}^4_\text{scal}=\frac{U}{\pi^{\frac52}}\left[\bar{D}_{\frac12,1,1,\frac12}(U,V)+\bar{D}_{1,\frac12,1,\frac12}(U,V)\right]\,,}
where $U,V$ are the usual conformal cross ratios, and the $\bar D$ functions are the usual exchange diagrams for scalars. From $\langle SSSS\rangle$, we can use the superconformal Ward identities to derive the result for $\langle PPPP\rangle$:
\es{finalAnswerP}{
\langle P(\vec x_1) P(\vec x_2) P(\vec x_3) P(\vec x_4)\rangle_\text{tree}=\frac{8}{c_T}\frac{\sum_{i=1}^6 \mathcal{B}_i}{|\vec x_{12}|^4|\vec x_{34}|^4}(2-\sin^2(\pi\lambda))\mathcal{P}^i_\text{free} \,,}
which is written in the same R-symmetry basis as $\langle SSSS\rangle$. We define $\mathcal{P}^i_\text{free}$ to be the connected part of the free correlator for $\langle PPPP\rangle$, whose independent terms up to crossing are
\es{finalAnswer2P}{
\mathcal{P}^1_\text{free}&=0\,,\qquad\qquad\qquad\qquad \quad\;\;\mathcal{P}^4_\text{free}=\frac{U^2(U-V-1)}{V^{\frac32}}\,.\\}
Our results for $\langle SSSS\rangle$ and $\langle PPPP\rangle$ are analogous to those of the quasi-bosonic and quasi-fermionic non-supersymmetric correlators derived in \cite{Turiaci:2018nua}, which we discuss further in the conclusion.

We then compare our analytic tree level result for $\<SSSS\>$ to non-perturbative predictions for this quantity coming from the numerical conformal bootstrap \cite{Rattazzi:2008pe,Poland:2018epd,Chester:2019wfx}. By comparing the $\mathcal{N}=6$ numerical bounds on $\<SSSS\>$ to certain protected CFT data known exactly via supersymmetric localization, \cite{Binder:2020ckj} conjectured that the low-lying spectrum of the $U(1)_{2M}\times U(1+M)_{-2M}$ ABJ theory could be numerically computed for any $c_T\sim M$. We find that the large $c_T$ regime of this finite $c_T$ bootstrap result compares well to our our tree level analytic results at $\lambda=1/2$ for both protected and unprotected low-lying CFT data\footnote{Note that our tree level result does not depend on $N$ when written in terms of $c_T$ and $\lambda$.}, as summarized in Table \ref{comparisonTable}. This nontrivial check of the the conjectured non-perturbative solution of the $U(1)_{2M}\times U(1+M)_{-2M}$ theory generalizes the analogous check of the supergravity limit in \cite{Chester:2018lbz}, which matched the tree level supergravity correlator of \cite{Zhou:2017zaw} to the conjectured $\mathcal{N}=8$ numerical bootstrap solution in \cite{Chester:2014fya,Chester:2014mea,Agmon:2017xes,Agmon:2019imm} of the $U(N)_2\times U(N+1)_{-2}$ ABJ theory in the large $c_T\sim N^{3/2}$ supergravity regime.

The rest of this paper is organized as follows. In Section~\ref{weakHS}, we derive the general form of $\<SSSS\>$ and $\<SSSP\>$ using the constraints of weakly broken higher spin symmetry. In Section~\ref{LOCALIZATION}, we use localization constraints from the mass deformed free energy in the higher spin limit to fix the unknown coefficients in $\<SSSS\>$. In Section~\ref{numSec}, we compare our results at $\lambda=1/2$ to the numerical conformal bootstrap results of \cite{Binder:2020ckj} for the $U(1)_{2M}\times U(1+M)_{-2M}$ theory. We end with a discussion of our results in Section~\ref{disc}.  Many technical details are relegated to the Appendices.

\section{Weakly Broken Higher Spin Symmetry}
\label{weakHS}

In this section we discuss the constraints of weakly broken higher spin symmetry on any 3d $\mathcal{N}=6$ CFT whose single trace spectrum consists of the stress tensor multiplet as well long multiplets with superprimaries $B_\ell$ for each spin $\ell\geq0$, which in the strict higher spin limit become conserved current multiplets. We start in Section \ref{consec} with a discussion of conserved current multiplets for 3d $\mathcal{N}=6$ CFTs. In Section \ref{PSEUDOCURRENT} we then discuss the constraints from weakly broken higher spin symmetry at tree level. In Section \ref{SCALAR3pt} we use these constraints to fix the tree level three-point functions of certain single trace operators. In Section \ref{SSSSweak} we use these three-point functions and the Lorentzian inversion formula to fix the tree level $\langle SSSS\rangle$ in terms of two coefficients $a_1(\lambda)$ and $a_2(\lambda)$. Finally, in Section \ref{SSSPweak} we use weakly broken higher spin symmetry to relate $\langle SSSS\rangle$ to $\langle SSSP\rangle$, which is then also fixed in terms of the same $a_1(\lambda)$ and $a_2(\lambda)$.

\subsection{$\cN = 6$ Conserved Currents}
\label{consec}

The $\mathfrak{osp}(6|4)$ superalgebra allows two kinds of unitary conserved current multiplets. The stress tensor multiplet, which is a 1/3-BPS operator, contains conserved currents only up to spin two and is found in all local 3d $\cN=6$ theories. This multiplet contains two scalars: the superconformal primary $S_a{}^b(\vec x)$ with dimension $1$, and the operator $P_a{}^b(\vec x)$ with dimension $2$, both transforming in the adjoint ${\bf 15}$ of $\mathfrak{so}(6)_R$. We use indices, $a,b=1\,,\dots\,,4$ to denote $\mathfrak{su}(4)\approx \mathfrak{so}(6)$ fundamental (lower) and anti-fundamental (upper) indices. To avoiding carrying around indices, we find it convenient to contract them with an auxiliary matrix $X$, defining
\begin{equation}\label{SXDef}
S(\vec{x}, X) \equiv X_a{}^b S_b{}^a(\vec{x})\,,\qquad P(\vec{x}, X) \equiv X_a{}^b P_b{}^a(\vec{x}) \,.
\end{equation}
We normalize $S(\vec{x}, X)$ and $P(\vec{x}, X)$ such that their two-point functions are 
\begin{equation}\label{2pS}
  \langle S(\vec x_1,X_1)S(\vec x_2,X_2)\rangle = \frac{\text{tr}(X_1X_2)}{x_{12}^2}\,,\qquad  \langle P(\vec x_1,X_1)P(\vec x_2,X_2)\rangle = \frac{\text{tr}(X_1X_2)}{x_{12}^4} \,,
\end{equation}
where we define $x_{ij} = |\vec x_i-\vec x_j|$. Apart from these two scalars, the other bosonic operators in the multiplet are the $R$-symmetry current $J_1^\mu(\vec x,X)$, a $U(1)$ flavor current $j^\mu(\vec x)$, and finally the stress tensor itself, $T_2^{\mu\nu}(\vec x)$.

\begin{table}
\begin{center}
\hspace{-.4in}
{\renewcommand{\arraystretch}{1.2}
\begin{tabular}{c|c|c||c|c|c}
Type & $\Delta$           & Spin   & Multiplet & $\mathfrak{so}(6)_R$ & BPS    \\\hline
Long & $>\Delta_B+\ell+1$ & $\ell$ & $\text{Long}$   & $[a_1a_2a_3]$        & 0      \\\hline
$A$  & $\Delta_B+\ell+1$  & $\ell$ & $(A,1)$   & $[a_1a_2a_3]$        & $1/12$ \\
     &                    &        & $(A,2)$   & $[0a_2a_3]$          & $1/6$  \\
     &                    &        & $(A,+)$   & $[0a_20]$            & $1/4$  \\
     &                    &        & $(A,-)$   & $[00a_3]$            & $1/4$  \\
     &                    &        & $(A, \text{cons.})$ & $[000]$              & $1/3$  \\\hline
$B$  & $\Delta_B$         & 0      & $(B,1)$   & $[a_1a_2a_3]$        & $1/6$  \\
     &                    &        & $(B,2)$   & $[0a_2a_3]$          & $1/3$  \\
     &                    &        & $(B,+)$   & $[0a_20]$            & $1/2$  \\
     &                    &        & $(B,-)$   & $[00a_3]$            & $1/2$  \\
     &                    &        & Trivial   & $[000]$              & $1$    
\end{tabular}}
\caption{Multiplets of $\mathfrak{osp}(6|4)$ and the quantum numbers of their superconformal primary, where $\Delta_B = a_1 + \frac 12 (a_2+a_3)$.}
\label{n6mults}
\end{center}
\end{table}

Unlike the stress tensor multiplet, all other $\cN=6$ conserved current multiplets are semishort rather than short, and contain conserved currents with spin greater than two. For every $\ell > 0$, there is a conserved current multiplet\footnote{We use the notation $\fcy M^{\bf r}_{\Dk,\ell}$ to denote the $\cN=6$ supermultiplet with shortening condition $\fcy T$, whose superconformal primary has spin $\ell$, conformal dimension $\Dk$ and transforms in the representation $\bf r$ under $\mathfrak{so}(6)$. A full list of unitary supermultiplets is given in Table~\ref{n6mults}.} $(A,\text{cons})^{[000]}_{\ell+1,\ell}$ whose superconformal primary is a spin-$\ell$ conserved current $B_\ell(\vec x)$. The bosonic descendants of $B_\ell(\vec x)$ are conserved currents $H_{\ell+1}(\vec x,X)$, $J_{\ell+2}(\vec x,X)$, and $T_{\ell+3}(\vec x)$ with spins $\ell+1$, $\ell+2$ and $\ell+3$ respectively. The bottom and top components $B_\ell$ and $T_{\ell+3}$ are $R$-symmetry singlets, while the middle two components $H_{\ell+1}$ and $J_{\ell+2}$ transform in the ${\bf 15}$. There is also a scalar higher spin multiplet $(A,\text{cons})^{[000]}_{1,0}$ whose primary $B_0(\vec x)$ is a dimension 1 scalar. This multiplet has the same structure as the $\ell>0$ higher spin multiplets, except that it also contains an additional scalar $C_0(\vec x)$ with dimension~$2$. We will normalize all of these operators so that
\begin{equation}\begin{split}
\<\fcy J_\ell^{\mu_1\dots\mu_\ell}(\vec x_1) \fcy J_\ell^{\nu_1\dots\nu_\ell}(\vec x_2) \> &= \left(\frac{I^{(\mu_1\dots\mu_n)(\nu_1\dots\nu_n)}(x_{12})}{x_{12}^{2\ell-1}}-\text{traces}\right)\,, \\
\<\fcy K_\ell^{\mu_1\dots\mu_\ell}(\vec x_1,X_1) \fcy K_\ell^{\nu_1\dots\nu_\ell}(\vec x_2,X_2) \> &= \text{tr}(X_1X_2)\left(\frac{I^{(\mu_1\dots\mu_n)(\nu_1\dots\nu_n)}(x_{12})}{x_{12}^{2\ell-2}}-\text{traces}\right)  \,,\\
\text{ where } I^{\mu_1\dots\mu_n\nu_1\dots\nu_n}(x_{12}) &= \left(\dk^{\mu_1\nu_1}-\frac{x_{12}^{\mu_1}x_{12}^{\nu_1}}{x_{12}^2}\right)\dots\left(\dk^{\mu_n\nu_n}-\frac{x_{12}^{\mu_n}x_{12}^{\nu_n}}{x_{12}^2}\right)\,,
\end{split}\end{equation}
for operators $\fcy J_\ell$ and $\fcy K_\ell$ transforming in the $\bf 1$ and $\bf 15$ of the $\mathfrak{so}(6)_R$ $R$-symmetry respectively.

We assume that the single-trace operators consist of a stress tensor multiplet, along with a single higher spin multiplet $(A,\text{cons})^{[000]}_{\ell+1,\ell}$ for each $\ell = 0\,,1\,,2\,,\dots$. We list the single-trace operator content of such theories in Table~\ref{SingleTraceTable}. Observe that for each spin $\ell\geq2$ the bosonic conserved currents come in pairs, so that for each $B_\ell(\vec x)$ and $H_{\ell}(\vec x)$ there is a $T_\ell(\vec x)$ and $J_\ell(\vec x)$ respectively with the same quantum numbers but belonging to different SUSY multiplets. As we shall see, these pairs of operators are mixed by the higher spin conserved currents.

\begin{table}
\begin{center}
\begin{tabular}{ c | c c c c c c }
 & \multicolumn{4}{c}{Higher-Spin Multiplet} \\
Spin    & Stress-tensor & Spin 0 & Spin 1 & Spin 2 & Spin 3 & $\hdots$ \\ \hline
 $0$    & ${\bf15}+{\bf15}$ & ${\bf1}+{\bf1}$  & & & \\
 $1/2$  & ${\bf6}+{\bf10} + \overline{\bf 10}$  & ${\bf6}$ & & & \\
 $1$    & ${\bf1}+{\bf15}$  & ${\bf15}$ & ${\bf1}$ & & \\
 $3/2$  & ${\bf6}$     & ${\bf10} + \overline{\bf 10}$ & ${\bf6}$ & & \\
 $2$    & ${\bf1}$     & ${\bf15}$ & ${\bf15}$ & ${\bf1}$ & \\
 $5/2$  &         & ${\bf6}$  & ${\bf10} + \overline{\bf 10}$ & ${\bf6}$                       & \\
 $3$    &         & ${\bf1}$  & ${\bf15}$                     & ${\bf15}$                      & ${\bf 1}$ & \\
 $7/2$  &         &           & ${\bf6}$                      & ${\bf10} + \overline{\bf 10}$  & ${\bf6}$  &  \\
 $4$    &         &           & ${\bf1}$                      &  ${\bf15}$                     & ${\bf15}$ & $\hdots$ \\
 $\vdots$ &       &           &                               & $\vdots$                       & $\vdots$ & $\ddots$ \\
\end{tabular}
\caption{Single trace operators for higher spin $\mathcal{N}=6$ CFTs. }
\label{SingleTraceTable}
\end{center}
\end{table}

Let us now consider three-point functions between the scalars $S$, $P$ and a conserved current $\fcy J$. Conformal invariance, $R$-symmetry, and crossing symmetry together imply that
\begin{equation}\begin{split}\label{SS3pts}
\<\fk(\vec x_1,X_1)\fk(\vec x_2,X_2)\fcy J_\ell^{\mu_1\dots\mu_\ell}(\vec x_3)\> &= \begin{cases}
\lk_{\fk\fk\fcy J} \text{tr}\left(X_1X_2\right)\fcy C_{\fk\fk\ell}^{\mu_1\dots\mu_\ell}(x_i) & \text{ even } \ell  \\
0 & \text{ odd } \ell 
\end{cases}\\
\<\fk(\vec x_1,X_1)\fk(\vec x_2,X_2)\fcy K_\ell^{\mu_1\dots\mu_\ell}(\vec x_3,X_3)\> &= 
\begin{cases}
\lk_{\fk\fk\fcy K} \text{tr}\left(\{X_1,X_2\}X_3\right)\fcy C_{\Dk_\fk\Dk_\fk\ell}^{\mu_1\dots\mu_\ell}(x_i )& \text{ even } \ell  \\
\lk_{\fk\fk\fcy K} \text{tr}\left([X_1,X_2]X_3\right)\fcy C_{\Dk_\fk\Dk_\fk\ell}^{\mu_1\dots\mu_\ell}(x_i) & \text{ odd } \ell  \\ 
\end{cases} \\
\end{split}\end{equation}
where we define the conformally covariant structure\footnote{Our choice of prefactors multiplying $\fcy C_{\fk_1\fk_2\ell}$ is such that the three-point coefficients $\lk_{\fk_1\fk_2\fcy O}$ match the OPE coefficients multiplying the conformal blocks in \eqref{SDecomp}.}
\begin{equation}\label{cfk1fk2}
\fcy C_{\Dk_1\Dk_2\ell}^{\mu_1\dots\mu_\ell}(x_i) = \sqrt{\frac{(1/2)_\ell}{2^{\ell+2}\ell!}}\left(\frac{x_{13}^{\mu_1}}{x_{13}^2}-\frac{x_{23}^{\mu_1}}{x_{23}^2}\right)\dots\left(\frac{x_{13}^{\mu_\ell}}{x_{13}^2}-\frac{x_{23}^{\mu_\ell}}{x_{23}^2}\right)\frac1{x_{12}^{\Dk_1+\Dk_2-1}x_{23}^{\Dk_2-\Dk_1+1}x_{31}^{\Dk_1-\Dk_2+1}}\,.
\end{equation}
Note that $\<SP\fcy J\>$ automatically vanishes when $\fcy J$ is a conserved current, as $\fcy C_{\Dk_1\Dk_2\ell}$ is not conserved unless $\Dk_1 = \Dk_2$.

Supersymmetry relates the OPE coefficients of operators in the same supermultiplet. By using the superconformal blocks for $\<SSSS\>$ computed in \cite{Binder:2020ckj}, we find for every integer $\ell$ there is a unique superconformal structure between two $S$ operators and the $(A,\text{cons})^{[000]}_{\ell+1,\ell}$ supermultiplet.  For even $\ell$ the OPE coefficients are all related to $\lk_{SSB_\ell}$ via the equations\footnote{The superconformal blocks themselves relate $\lk_{SSH_{\ell+1}}^2$ or $\lk_{SST_{\ell+3}}^2$ to $\lk_{SS\cO}^2$ and $\lk_{SS\cO}\lk_{PP\cO}$ for all superdescendants $\cO$ of $B_\ell$. Although the superconformal blocks do not fix the sign of $\lk_{SS\cO}$, we can always redefine $\cO\rightarrow-\cO$ so that $\lk_{SS\cO}/\lk_{SSB_\ell}$ or $\lk_{SS\cO}/\lk_{SST_{\ell+3}}$ is positive.}
\begin{equation}\label{evenblock}\begin{aligned}
\lk_{SSB_\ell} &= \lk_{SSH_{\ell+1}} = \lk_{SSJ_{\ell+2}}\,,\qquad &\lk_{SST_{\ell+3}} &= 0\,, \\
\lk_{PPB_\ell}  &=  -\ell \lk_{SSB_\ell}\,,\qquad     &\lk_{PPH_{\ell+1}}  &=  -(\ell+1)\lk_{SSB_\ell}\,, \\
\lk_{PPJ_{\ell+2}}  &=  -(\ell+2)\lk_{SSB_\ell}\,,\qquad     &\lk_{PPT_{\ell+3}}  &=  0\,,  \\
\end{aligned}\end{equation}
while for odd $\ell$ the OPE coefficients are related to $\lk_{SST_{\ell+3}}$:
\begin{equation}\label{oddblock}\begin{aligned}
\lk_{SSH_{\ell+1}} &= \lk_{SSJ_{\ell+2}} = \lk_{SST_{\ell+3}}\,,\qquad  & \lk_{SSB_\ell} &= 0\,, \\
\lk_{PPB_\ell}  &=  0\,,\qquad     &\lk_{PPH_{\ell+1}}  &=  (\ell+1) \lk_{SST_{\ell+3}}\,,  \\
\lk_{PPJ_{\ell+2}}  &=  (\ell+2)\lk_{SST_{\ell+3}}\,,\qquad     &\lk_{PPT_{\ell+3}}  &=  (\ell+3)\lk_{SST_{\ell+3}}\,.  \\
\end{aligned}\end{equation}
Note that $\lk_{SST_{\ell+3}}$ vanishes for even $\ell$, and $\lk_{SSB_{\ell}}$ for odd $\ell$, simply as a consequence of $1\leftrightarrow2$ crossing symmetry. The superconformal blocks for the stress tensor and the scalar conserved current have the same structure (where we treat the stress tensor block as having spin $-1$), with the additional equations
\begin{equation}\label{scalarBlock}\begin{aligned}
\lk_{SSS} &= \lk_{SST_2} \,,  \qquad &\lk_{SSP} = \lk_{PPS} = \lk_{PPP} = 0\,,         \\
\lk_{SSC_0} &= \lk_{PPC_0} = 0\,, \\
\end{aligned}\end{equation}
for the scalars $S$ and $P$, and dimension $2$ scalar $C_0$, in the stress tensor and the scalar conserved current multiplet respectively.

Finally, we note that due to the superconformal Ward identities, $\lk_{SST_2}$ can be expressed in terms of the coefficient of the canonically normalized stress tensor,\footnote{We define $c_T$ so that the stress tensor satisfies the Ward identity \cite{Osborn:1993cr}
\begin{equation}
\label{ctdef}
4\pi\sqrt{\frac{c_T}{3}}\int d^3x\,\<\Nb^\mu T_{\mu\nu}(x)\cO_1(y_1)\dots\cO_n(y_n)\> = -\sum_i \<\cO_1(y_1)\dots\nb_\nu\cO_i(y_i)\dots\cO_n(y_n)\>
\end{equation}
for any arbitrary string of operators $\cO_i(y_i)$.} so that
\begin{equation}\label{SSSCT}
\lk_{SSS} = \lk_{SST_2} =  \frac {8}{\sqrt{c_T}}\,.
\end{equation}
For a single scalar or Majorana fermion $c_T = 1$ in our normalization. The $\cN=6$ free-field theory (which in fact has $\mathcal{N}=8$ supersymmetry) consists of $8$ scalars and $8$ Majorana fermions and so has $c_T = 16$.

\subsection{The $\mathfrak{so}(6)$ Pseudocharge}
\label{PSEUDOCURRENT}

Having reviewed the properties of conserved current multiplets in $\cN=6$ theories, we now consider what happens when the higher spin symmetries are broken to leading order in $1/c_T$. We will follow the strategy employed in \cite{Maldacena:2012sf} and use the weakly broken higher spin symmetries to constrain three-point functions. Unlike that paper however, which studies the non-supersymmetric case and so considers the symmetries generated by a spin 4 operator, we will instead focus on the spin 1 operator $H_1^\mu(\vec x,X)$. While itself not a higher spin conserved current, it is related to the spin 3 current $T_3(\vec x)$ by supersymmetry.

We begin by using $H_1(x)$ to define a pseudocharge:
\begin{equation}
\tilde\dk(X) \fcy O(0) = \frac1{4\pi}\int_{|x| = r}dS\cdot H_1(\vec x,X)\,\fcy O(0)\bigg|_{\text{finite as }r\rightarrow0}\,.
\end{equation}
The action of $\tilde\dk(X)$ is fixed by the 3-points functions $\<H_1\fcy O\fcy O'\>$. Because $H_1^\mu$ has spin $1$, it must act in the same way on conformal primaries as would any other spin $1$ conserved current. In particular, it relates conformal primaries to other conformal primaries with the same spin and conformal dimension.

Now consider the action of $\tilde\dk(X)$ on an arbitrary three-point function. We can use the divergence theorem to write:
\begin{equation}\label{HWard1}\begin{split}
 \tilde\dk(X)\big\<\fcy O_1(\vec y_1)&\fcy O_2(\vec y_2)\fcy O_3(\vec y_3)\big\> \\
 &= -\frac1{4\pi}\int_{\fcy R_r} d^3x \left\<\Nb\cdot H_1(\vec x,X)\fcy O_1(\vec y_1)\fcy O_2(\vec y_2)\fcy O_3(\vec y_3)\right\>\bigg|_{\text{finite }r\rightarrow0}\,,
\end{split}\end{equation}
where $\fcy R_r$ is the set of $\vec x\in\bb R^3$ for which $|\vec x-\vec y_i|>r$ for each $y_i$. If the operator $H_1^\mu(x,X)$ were conserved, the right-hand side of this expression would vanish and we would find that correlators were invariant under $\tilde\dk(X)$. When the higher spin symmetries are broken, however, $\Nb\cdot H_1$ will no longer vanish and so \eqref{HWard1} will gives us a non-trivial identity.

In the infinite $c_T$ limit, $\Nb\cdot H_1$ will become a conformal primary distinct from $H_1^\mu$. In order to work out what this primary is, we can use the $\cN=6$ multiplet recombination rules \cite{Binder:2020ckj,Cordova:2016emh}:
\begin{equation}\begin{split}
\text{Long}_{\Dk,0}^{[000]}    &\underset{\Dk\rightarrow1}\longrightarrow (A,\text{cons})^{[100]}_{1,0} \oplus (B,1)_{2,0}^{[200]}\,, \\
\text{Long}_{\Dk,\ell}^{[000]} &\underset{\Dk\rightarrow\ell+1}\longrightarrow (A,\text{cons})^{[100]}_{\ell+1,\ell} \oplus (A,1)_{\ell+3/2,\ell-1/2}^{[100]}\quad \text{ for } \ell>0\,.
\end{split}\end{equation}
From this we see that, unlike the other conserved current multiplets, the scalar conserved current multiplet recombines with a B-type multiplet, the $(B,1)_{2,0}^{[200]}$. The only such multiplet available in higher spin $\mathcal{N}=6$ CFT at infinite $c_T$ is the double-trace operator $S^{[a}{}_{[b}S^{c]}{}_{d]}$, whose descendants are also double-traces of stress tensor operators. From this we deduce that
\begin{equation}\label{H1Recom}\begin{split}
\Nb\cdot H_1(\vec x,X) &= -\frac{\ak}{\sqrt {c_T}}\Fk(\vec x,X) + \text{fermion bilinears} + O(c_T^{-1})\\
\text{ with } \Fk(\vec x,X) &= X^a{}_b\left(S^b{}_c(\vec x)P^c{}_a(\vec x)-P^b{}_c(\vec x)S^c{}_a(\vec x)\right)\,,
\end{split}\end{equation}
and where $\ak$ is some as yet undetermined coefficient.  We then conclude that 
\begin{equation}
\tilde \delta(X)\<\fcy O_1\fcy O_2\fcy O_3\> = \frac{\ak}{4\pi\sqrt{c_T}}\int d^3x\, \<\Fk(\vec x,X)\fcy O_1\fcy O_2\fcy O_3\> + \text{fermion bilinears} + O({c_T}^{-{3/2}})\,,
\end{equation}
where we have left the regularization of the right-hand integral implicit. 

We will begin by considering the case where $\fcy O_1$, $\fcy O_2$ and $\fcy O_3$ are any three bosonic conserved currents. In this case, $\<\Fk\fcy O_1\fcy O_2\fcy O_3\> \sim c_T^{-3/2}$ and so
\begin{equation}\label{conservered3pt}
\tilde \delta(X)\<\fcy O_1\fcy O_2\fcy O_3\> = O(c_T^{-3/2})\,.
\end{equation}
We thus find that, at leading order in the $1/c_T$ expansion, these three-point functions are invariant under $\tilde\delta(X)$.  This is a strong statement, allowing us to import statements about conserved currents and apply them to $H_1$. 

Consider now the $R$-symmetry current $J_1$, which has the same quantum numbers as $H_1$, and let us define
\begin{equation}
\dk(X) \fcy O(0) = \frac1{4\pi}\int_{|x| = r}dS\cdot J_1(x)\fcy O(0)\bigg|_{\text{finite as }r\rightarrow0}\,,
\end{equation}
which generates the $\mathfrak{so}(6)_R$ symmetry. Because any correlator of both $J_1$ and $H_1$ is conserved under $\dk(X)$ and $\tilde\dk(X)$ at leading order, the (pseudo)charges $\dk(X)$ and $\tilde\dk(X)$ form a semisimple Lie algebra.\footnote{Note that this Lie algebra structure only holds when $\dk(X)$ and $\tilde\dk(X)$ act on spinning single trace operators, so that \eqref{conservered3pt} holds. In particular, equation \eqref{DkComs} and \eqref{tildeDkComs} are true when $\dk(X)$ and $\tilde\dk(X)$ act on such operators.} The $\mathfrak{so}(6)_R$ symmetry implies the commutator relations
\begin{equation}\label{DkComs}
[\dk(X),\dk(Y)] = \zeta\dk([X,Y])\,,\qquad [\dk(X),\tilde\dk(Y)] = \zeta\tilde\dk([X,Y])\,,
\end{equation}
for some non-zero constant $\zeta$, while
\begin{equation}\label{tildeDkComs}
[\tilde\dk(X),\tilde\dk(Y)] = \zeta\dk([X,Y]) + 2\gk\tilde\dk([X,Y])\,,
\end{equation}
for some additional $\gk$. Note that both the second equation in \eqref{DkComs} and the first term in \eqref{tildeDkComs} are fixed by the same conformal structure in the three-point function $\<H_1H_1J_1\>$, which is why they are both proportional to $\zeta$. We can now define charges $\dk_L(X)$ and $\dk_R(X)$ by the equations 
\begin{equation}\label{zetaDef}\begin{aligned}
\dk(X) &= \zeta\left(\dk_L(X) + \dk_R(X)\right)\,, \qquad &\tilde\dk(X) &= \zeta_L\dk_L(X) + \zeta_R\dk_R(X) \,, \\
\text{ with }\quad \zk_L &= \gk  + \sqrt{\zeta^2+\gk^2}\,,\qquad &\zk_R &= \gk - \sqrt{\zeta^2+\gk^2}\,,
\end{aligned}\end{equation}
which satisfy the commutator relations
\begin{equation}\begin{split}
[\dk_L(X),\dk_L(Y)] &= \dk_L([X,Y])\,,\qquad [\dk_L(X),\dk_R(Y)] = 0\\
[\dk_R(X),\dk_R(Y)] &= \dk_R([X,Y])\,.
\end{split}\end{equation}
These are precisely the commutation relations of an $\mathfrak{so}(6)\times\mathfrak{so}(6)$ Lie algebra, where the $\delta_L(X)$ generates the left-hand and $\delta_R(X)$ the right-hand $\mathfrak{so}(6)$ respectively.

As we have showed previously, three-point functions of bosonic conserved currents are $\tilde\dk(X)$ invariant at leading order in the large $c_T$ expansion. As a consequence, the higher spin operators $H_\ell(\vec x,X)$ and $J_\ell(\vec x,X)$ will together form representations of $\mathfrak{so}(6)\times\mathfrak{so}(6)$. There are two possibilities. Either both operators transform in the adjoint of the same $\mathfrak{so}(6)$, or instead the operators split into left and right-handed operators
\begin{equation}\label{qkRot}\begin{split}
\fcy J_\ell^L(\vec x,X) &= \color{white}-\color{black}\cos(\qk_\ell)H_\ell(\vec x,X)  + \sin(\qk_\ell)J_\ell(\vec x,X)\,,  \\
\fcy J_\ell^R(\vec x,X) &= -\sin(\qk_\ell)H_\ell(\vec x,X) + \cos(\qk_\ell) J_\ell(\vec x,X)\,,
\end{split}\end{equation}
with some mixing angle $\qk_\ell$, such that 
\begin{equation}
\tilde\dk(X)\fcy J_\ell^L(\vec y,Y) = \zeta_L \fcy J_\ell^L(\vec y,[X,Y])\,,\qquad \tilde\dk(X)\fcy J_\ell^R(\vec y,Y) = \zeta_R \fcy J_\ell^R(\vec y,[X,Y])\,.
\end{equation}
As we shall see in the next section, it is this latter possibility which is actually realized in all theories for which $\lk_{SSB_0}\neq0$.

\subsection{Three-Point Functions}
\label{SCALAR3pt}

So far we have been avoiding the scalars $S$ and $P$. Because the $H_1^\mu(\vec x,X)$ eats a bilinear of $S$ and $P$, correlators involving these scalars are not automatically conserved at leading order, and so we can not assign these operators well defined $\mathfrak{so}(6)\times\mathfrak{so}(6)$ transformation properties. The action of $\tilde\delta(X)$ is, however, still fixed by the delta function appearing in the three-point functions
$$ \<S\cO\Nb\cdot H_1\>\qquad \text{ and } \qquad \<S\hat\cO\Nb\cdot H_1\>$$
when $\cO$ and $\hat\cO$ are scalars of dimension 1 and 2 respectively. In Appendix~\ref{DELTASP} we systematically work through the possibilities, finding that\footnote{Throughout this section, we will abuse notation slightly and use $\lk_{\fcy O_1\fcy O_2\fcy O_3}$ to refer to the leading large $c_T$ behavior of the OPE coefficient, which for three single trace operators scales as $c_T^{-1/2}$.}
\begin{equation}\label{chiEq}\begin{split}
\tilde\delta(X) S(\vec y,Y) &= -\frac14 \lk_{SSB_0} S(\vec y,[X,Y])   \\
\tilde\delta(X) P(\vec y,Y) &= \frac14 \lk_{SSB_0} P(\vec y,[X,Y]) + \kk_1 S^2(\vec y,[X,Y]) + \kk_2 SB_0(\vec y,[X,Y])\,,  \\
\end{split}\end{equation}
where we define the double trace operators
\begin{equation}\begin{split}
S^2(\vec y,Y) &\equiv Y^a{}_b \left(S^b{}_c(\vec y)S^c{}_a(\vec y) - \frac14\delta^a{}_bS^c{}_d(\vec y)S^d{}_c(\vec y)\right)\,,\\ 
SB_0(\vec y,Y) &\equiv Y^a{}_b S^b{}_a(\vec y)B_0(\vec y)\,,
\end{split}\end{equation}
and where $\lk_{SSB_0} = \lk_{SSH_1} \sim c_T^{-1/2}$ and $\kk_i\sim c_T^{-1}$. 
By suitably redefining the sign of the conserved current multiplet operators, we can always fix $\lk_{SSB_0}\geq0$.

Let us now consider the three-point function of two scalars with a spin $\ell$ conserved current $\fcy O_\ell^L(\vec y,Y)$ transforming in the left-handed $\bf 15$, so that
\begin{equation}
\tilde\dk(X)\fcy O_\ell^L(\vec y,Y) = \zeta_L \fcy O^L_\ell(\vec y,[X,Y])\,.
\end{equation}
We can then consider the weakly broken $\tilde \dk(X)$ Ward identity:
\begin{equation}\label{SSVar}\begin{split}
\tilde \dk(X)\<&S(\vec y_1,Y_1)S(\vec y_2,Y_2)\fcy O_\ell^L(\vec y_3,Y_3)\> \\
&= \frac{\ak}{4\pi\sqrt{{c_T}}} \int d^3x\, X^a{}_b\<S(\vec y_1,Y_1)S^b{}_c(\vec x)\>\<P^c{}_a(\vec x)S(\vec y_2,Y_2)\fcy O^L_\ell(\vec y_3,Y_3)\> \\
&- \frac{\ak}{4\pi\sqrt{{c_T}}} \int d^3x\, X^a{}_b\<S(\vec y_1,Y_1)S^c{}_a(\vec x)\>\<P^b{}_c(\vec x)S(\vec y_2,Y_2)\fcy O^L_\ell(\vec y_3,Y_3)\> + 1\leftrightarrow2 \,.
\end{split}\end{equation}
Defining the operators $\tilde S(\vec x,X)$ and $\tilde P(\vec x,X)$ to be the ``shadow transforms'' \cite{Ferrara:1972uq} of $S(\vec x,X)$ and $P(\vec x,X)$ respectively:
\begin{equation}\label{ShadowSP}
\tilde S(\vec x,X) = \frac1{4\pi}\int \frac{d^3z}{|\vec x-\vec z|^4} S(\vec z,X)\,,\qquad \tilde P(\vec x,X) = \frac1{4\pi}\int \frac{d^3z}{|\vec x-\vec z|^2} P(\vec z,X)\,,
\end{equation}
we can then rewrite the Ward identity as:
\begin{equation}\label{SSVar2}\begin{split}
\tilde \dk&(X)\<S(\vec y_1,Y_1)S(\vec y_2,Y_2)\fcy O_\ell^L(\vec y_3,Y_3)\> \\
&= \frac{\ak}{\sqrt{{c_T}}}\left(\<\tilde P(\vec y_1,[X,Y_1])S(\vec y_2,Y_2)\fcy O^L_\ell(\vec y_3,Y_3)\> + \<S(\vec y_1,Y_1)\tilde P(\vec y_2,[X,Y_2])\fcy O^L_\ell(\vec y_3,Y_3)\>\right)\,.
\end{split}\end{equation}

Our task now is to expand correlators in this Ward identity in terms of conformal and $R$-symmetry covariant structures. Using \eqref{SS3pts}, the LHS of \eqref{SSVar2} becomes
\begin{equation}\begin{split}\label{SSLHS}
\tilde \dk(X)\<S(\vec y_1,Y_1)&S(\vec y_2,Y_2)\fcy O_\ell^L(\vec y_3,Y_3)\>\\
&= \left(\zeta_L+\frac{\lk_{SSH_1}}4\right)\lk_{SS\fcy O_\ell^L}\text{tr}\left([Y_1,Y_2]_\pm [X,Y_3]\right)\fcy C_{11\ell}(\vec y_i)\,,
\end{split}\end{equation}
where $[Y_i,Y_j]_\pm$ is a commutator when $\ell$ is even, and anticommutator when $\ell$ is odd. To evaluate the RHS we first note that, using both conformal and $R$-symmetry invariance,
\begin{equation}\begin{split}
\<S(\vec y_1,Y_1)&P(\vec y_2,Y_2)\fcy O_\ell^L(\vec y_3,Y_3)\>  \\
&=\left(\lk^+_{SP\fcy O_\ell^L} \text{tr}\left([Y_1,Y_2]_\pm Y_3\right) + \lk^-_{SP\fcy O_\ell^L} \text{tr}\left([Y_1,Y_2]_\mp Y_3\right)\right)  \fcy C_{12\ell}(\vec y_i)\,,
\end{split}\end{equation}
for some OPE coefficients $\lk^\pm_{SP\fcy O_\ell^L}$. Because the operator $\fcy O^L_\ell(\vec y,Y)$ is not conserved at finite $c_T$, this three-point function does not necessarily vanish at $O(c_T^{-1/2})$. We can then compute the shadow transform using the identity \cite{Dobrev:1977qv}
\begin{equation}\begin{split}\label{shad3pt}
\int &\frac{d^3z}{|\vec z-\vec x_1|^{2\Dk_1-6}}\fcy C_{\Dk_1,\Dk_2,\ell}^{\mu_1\dots\mu_\ell}(\vec z_1,\vec x_2,\vec x_3) \\
&\qquad\qquad = \frac{\pi^{3/2}\Gk\left(\Dk_1-\frac32\right)\Gk\left(\frac{\Dk_2-\Dk_1+2}2\right)\Gk\left(\frac{4+2\ell-\Dk_1-\Dk_2}2\right)}{\Gk\left(3-\Dk_1\right)\Gk\left(\frac{\Dk_1+\Dk_2-1}2\right)\Gk\left(\frac{2\ell+\Dk_1-\Dk_2+1}2\right)}\fcy C_{3-\Dk_1,\Dk_2\ell}^{\mu_1\dots\mu_\ell}(\vec x_1,\vec x_2,\vec x_3)\,.
\end{split}\end{equation}
Putting everything together, we conclude that
\begin{equation}
\lk_{SP\fcy O_\ell^L}^+ = -\frac{(\lk_{SSB_0}+4\zeta_L)\ell!}{\pi^{3/2}\Gamma(\ell+1/2)} \frac{\lk_{SS\fcy O_\ell^L}\sqrt{c_T}}{\ak}\,,\qquad \lk_{SP\fcy O_\ell^L}^- = 0\,.
\end{equation}

So far we have consider the weakly broken Ward identity for the three-point function $\<SS\fcy O_\ell^L\>$, but it is straightforward to repeat this exercise with the variations
\begin{equation}\begin{split}
\tilde\dk\<SP\fcy O_\ell^L\> &= \frac{\ak}{\sqrt{c_T}}\left(\<\tilde PP\fcy O_\ell^L\>-\<S\tilde S\fcy O_\ell^L\>\right)\,, \\
\tilde\dk\<PP\fcy O_\ell^L\> &= -\frac{\ak}{\sqrt{c_T}}\left(\<\tilde SP\fcy O_\ell^L\>+\<P\tilde S\fcy O_\ell^L\>\right)\,.
\end{split}\end{equation}
Expanding each of these correlators and using \eqref{shad3pt}, we find that 
\begin{equation}\label{cOEqsL}
\lk_{PP\fcy O_\ell^L} = \left(\frac{4\zeta_L+\lk_{SSB_0}}{4\zeta_L-\lk_{SSB_0}}\right)\ell \lk_{SS\fcy O_\ell^L}\,,\qquad 16\zeta_L^2 - \lk_{SSB_0}^2 = \frac{2\ak^2\pi^2}{c_T}\,.
\end{equation}
Applying the same logic to a right-handed operator $\fcy O_\ell^R(\vec x,X)$, we immediately see that
\begin{equation}\label{cOEqsR}
\lk_{PP\fcy O_\ell^R} = \left(\frac{4\zeta_R+\lk_{SSB_0}}{4\zeta_R-\lk_{SSB_0}}\right)\ell \lk_{SS\fcy O_\ell^R}\,,\qquad 16\zeta_R^2 - \lk_{SSB_0}^2 = \frac{2\ak^2\pi^2}{c_T}\,.
\end{equation}
In particular, taking the last equations of \eqref{cOEqsL} and \eqref{cOEqsR} and combining them with \eqref{zetaDef}, we find that
\begin{equation}
\zeta_L = -\zeta_R = \zeta\,.
\end{equation}
As we saw in the previous section, $\zeta$ fixes the action of the $R$-symmetry charge $\delta(X)$, which, unlike $\tilde\dk(X)$, is exactly conserved in any $\cN=6$ theory. We can therefore relate it to the three-point function $\<SSJ_1\>$, and thus to the OPE coefficient $\lk_{SSS}$:
\begin{equation}
\zeta = -\frac14\lk_{SSJ_1} = -\frac14\lk_{SSS}\,. 
\end{equation}

We now apply \eqref{cOEqsL} and \eqref{cOEqsR} to the operators $H_\ell$ and $J_\ell$. Recall that these operators either transform identically under $\mathfrak{so}(6)$, or they split into left-handed and right-handed operators. Let us begin with the possibility that they transform identically under $\mathfrak{so}(6)\times\mathfrak{so}(6)$, and assume without loss of generality that both are left-handed.  Combining \eqref{cOEqsL} with the superblocks \eqref{evenblock} and \eqref{oddblock}, we find that
\begin{equation}
(-1)^\ell\lk_{SSH_\ell} = \left(\frac{\lk_{SSS}-\lk_{SSB_0}}{\lk_{SSS}+\lk_{SSB_0}}\right)\lk_{SSH_\ell}\,,\quad -(-1)^\ell\lk_{SSJ_\ell} = \left(\frac{\lk_{SSS}-\lk_{SSB_0}}{\lk_{SSS}+\lk_{SSB_0}}\right)\lk_{SSJ_\ell}\,.
\end{equation}
Because $\lk_{SSS}\neq0$, the only way to satisfy these equations is if $\lk_{SSB_0} = 0$. We know however that $\lk_{SSB_0}$ is nonzero for generic higher spin $\mathcal{N}=6$ CFTs, such as the ABJ theory, and in particular does not vanish in free field theory. We therefore conclude that is not possible for $H_\ell$ and $J_\ell$ to transform identically under $\mathfrak{so}(6)\times\mathfrak{so}(6)$. 

We now turn to the second possibility, that $H_\ell$ and $J_\ell$ recombine into left and right-handed multiplets $\fcy J^L_\ell$ and $\fcy J^R_\ell$ under $\mathfrak{so}(6)\times\mathfrak{so}(6)$, satisfying
\begin{equation}\begin{split}\label{shadCon}
\lk_{PP\fcy J^L_\ell} &= \left(\frac{\lk_{SSS}-\lk_{SSB_0}}{\lk_{SSS}+\lk_{SSB_0}}\right) \ell \lk_{SS\fcy J^L_\ell}\,,\qquad 
\lk_{PP\fcy J^R_\ell} = \left(\frac{\lk_{SSS}+\lk_{SSB_0}}{\lk_{SSS}-\lk_{SSB_0}}\right)\ell \lk_{SS\fcy J^R_\ell}\,.
\end{split}\end{equation}
We can then use the superconformal blocks \eqref{evenblock} and \eqref{oddblock} to find that
\begin{equation}\label{supcon}
\lk_{PPH_\ell} = (-1)^{\ell}\ell \lk_{SSH_\ell}\,,\qquad \lk_{PPJ_\ell} = (-1)^{\ell+1}\ell \lk_{SSJ_\ell}\,,
\end{equation}
and, from \eqref{qkRot}, we see that
\begin{equation}\label{rotcon}\begin{aligned}
&\lk_{SS\fcy J^L_\ell}  = \lk_{SSH_\ell} \cos\qk_\ell + \lk_{SSJ_\ell} \sin\qk_\ell\,,\qquad &\lk_{SS\fcy J^R_\ell}  &= -\lk_{SSH_\ell} \sin\qk_\ell + \lk_{SSJ_\ell} \cos\qk_\ell\\
&\lk_{PP\fcy J^L_\ell}  = \lk_{PPH_\ell} \cos\qk_\ell + \lk_{PPJ_\ell} \sin\qk_\ell\,,\qquad &\lk_{PP\fcy J^R_\ell}  &= -\lk_{PPH_\ell} \sin\qk_\ell + \lk_{PPJ_\ell} \cos\qk_\ell\,.\\
\end{aligned}\end{equation}
Combining \eqref{shadCon}, \eqref{supcon}, and \eqref{rotcon} together, we have 8 equations which are linear in the 8 OPE coefficients. Generically, the only solution to these equations will be the trivial one where all of OPE coefficients vanish. However, if we set
\begin{equation}
\qk_\ell = \frac\pi 4 + \frac{n \pi}2 \text{ for } n\in\bb Z\,,
\end{equation}
then we find that the equations become degenerate, allowing non-trivial solutions.  By suitably redefining the conserved currents $H_\ell\rightarrow -H_\ell$ we can always fix $n = 0$ so that $\lk_{SSH_\ell}\geq0$, and can then solve the equations to find that
\begin{equation}
\lk_{SSH_\ell} = \begin{cases} 
\frac{\lk_{SSS}}{\lk_{SSB_0}} \lk_{SSJ_\ell} & \ell \text{ is even} \\
\frac{\lk_{SSB_0}}{\lk_{SSS}} \lk_{SSJ_\ell} & \ell \text{ is odd} \,. \\
\end{cases}
\end{equation}

To complete our derivation, we simply note that from the superblocks \eqref{evenblock}, \eqref{oddblock} and \eqref{scalarBlock} that
\begin{equation}
\lk_{SSJ_{\ell+2}} = \lk_{SSH_{\ell+1}}\,,\qquad \lk_{SSJ_1} = \lk_{SSS}\,,\qquad \lk_{SSH_1} = \lk_{SSB_0}\,,
\end{equation}
so that 
\begin{equation}\label{sbSol}
\lk_{SSH_{\ell+1}} = \begin{cases} 
\lk_{SSS} & \ell \text{ is even} \\
\lk_{SSB_0} & \ell \text{ is odd}\,. \\
\end{cases}
\end{equation}

Let us now apply \eqref{sbSol} to two special cases: free field theory and parity preserving theories. In free field theory the higher spin currents remain conserved, so that $\ak = 0$ and hence, using \eqref{cOEqsL}, we find that $\lk_{SSB_0} = \lk_{SSS}$. We thus find that each conserved current supermultiplet contributes equally to $\<SSSS\>$. 

For parity preserving theories, supersymmetry requires that $S$ is a scalar but that $P$ is a pseudoscalar. As we see from \eqref{H1Recom}, the operator $H_1^\mu$ eats a pseudoscalar, and so is also a pseudovector rather than a vector. Parity preservation then requires that $\lk_{SSB_0} = 0$, and so we conclude that for parity preserving theories only conserved current supermultiplet with odd spin contribute to $\<SSSS\>$. Note that this does not apply to free field theory (which is parity preserving), because $H_1^\mu$ remains short.

We will conclude by noting that, unlike all previous calculations in this section, the variations $\tilde\dk\<SSP\>$ and $\tilde\dk\<SPB_0\>$ involve the double trace operators in \eqref{chiEq}. In Appendix~\ref{OPMIX} we use these variations to compute $\kk_1$ and $\kk_2$, and find that
\begin{equation}
\kk_1 = \frac{\ak \lk_{SSS}}{4\sqrt {c_T}}\,, \qquad \kk_2 = -\frac{\ak\lk_{SSB_0}}{4\sqrt{c_T}}\,.
\end{equation}

\subsection{The $\<SSSS\>$ Four-Point Function}
\label{SSSSweak}

In the previous section we showed that the OPE coefficient between two $S$ operators and a conserved current is completely fixed by $H_1^\mu$ pseudo-conservation in terms of $\lk_{SSS}$ and $\lk_{SSB_0}$. Our task now is to work out the implications of this for the $\<SSSS\>$ four-point function.

Conformal and R-symmetry invariance imply that \cite{Binder:2019mpb}
\begin{equation}\label{SSSSSt}
   \langle S(\vec x_1,X_1)  S(\vec x_2,X_2) S(\vec x_3,X_3) S(\vec x_4,X_4)\rangle = \frac{1}{x_{12}^2 x_{34}^2} \sum_{i=1}^6 {\cal S}^i(U, V) {\cal B}_i \,,\\
\end{equation}
where we define the R-symmetry structures
\begin{equation}\begin{split}\label{BasisElems}
  {\cal B}_1 &= \tr (X_1 X_2) \tr (X_3 X_4) \,, \quad {\cal B}_2 = \tr (X_1 X_3) \tr (X_2 X_4) \,, \quad
  {\cal B}_3 = \tr (X_1 X_4) \tr (X_2 X_3) \,, \\
  {\cal B}_4 &= \tr (X_1 X_4 X_2 X_3) + \tr (X_3 X_2 X_4 X_1) \,, \quad {\cal B}_5 = \tr (X_1 X_2 X_3 X_4) + \tr (X_4 X_3 X_2 X_1) \,, \\
  {\cal B}_6 &= \tr (X_1 X_3 X_4 X_2) + \tr (X_2 X_4 X_3 X_1) \,,
\end{split}\end{equation}
and where $\cS^i(U,V)$ are functions of the conformally-invariant cross-ratios
\begin{equation}\label{crossRatio}
U \equiv \frac{x_{12}^2x_{34}^2}{x_{13}^2x_{24}^2} \,, \qquad  V \equiv \frac{x_{14}^2x_{23}^2}{x_{13}^2x_{24}^2} \,.
\end{equation}
Crossing under $1\leftrightarrow3$ and $2\leftrightarrow3$ relates the different $\cS^i(U,V)$:
\begin{equation}\begin{split}\label{CrossSSSS}
 \cS^2(U,V) &= U\cS^1\left(\frac{1}{U},\frac{V}{U}\right)\,, \quad \cS^3(U,V) = \frac{U}{V}\cS^1(V,U)  \,, \\
 \cS^5(U,V) &= U\cS^4\left(\frac{1}{U},\frac{V}{U}\right)\,, \quad \cS^6(U,V) = \frac{U}{V}\cS^4(V,U) \,,
 \end{split}\end{equation}
so that $\langle SSSS\rangle$ is uniquely specified by $\cS^1(U,V) $ and $\cS^4(U,V)$. The $\cS^i(U,V)$ furthermore satisfy certain differential equation imposed by the supersymmetric Ward identities, computed in \cite{Binder:2019mpb}.

Another useful basis for the R-symmetry structures corresponds to the $\mathfrak{so}(6)$ irreps that appear in the $S\times S$ tensor product
\begin{equation}\begin{split}\label{irreps}
{\bf15}\otimes{\bf15}&={\bf1}_s\oplus{\bf15}_a\oplus{\bf15}_s\oplus{\bf20'}_s\oplus({\bf45}_a\oplus{\bf\overline{45}}_a)\oplus{\bf84}_s\,,
\end{split}\end{equation}
where $s/a$ denotes if the representation is in the symmetric/antisymmetric product. We define $\cS_{\bf{r}}$ to receive contributions only from operators in the $s$-channel OPE that belong to $\mathfrak{so}(6)_R$ irrep $\bf{r}$. This is related to the basis $\cS^i(U,V)$ by the equation \cite{Binder:2019mpb} 
\begin{equation}\label{toSi}
\begin{pmatrix} \cS^1\\\cS^2\\\cS^3\\\cS^4\\\cS^5\\\cS^6 \end{pmatrix}
= 
\begin{pmatrix}
1&0&-1&2&0&\frac4{15}\\
0&0&0&6&-4&4  \\
0&0&0&6&4&4 \\
0&0&0&-6&0&4 \\
0&1&1&-3&-1&-\frac23\\
0&-1&1&-3&1&-\frac23\\
\end{pmatrix}\begin{pmatrix}\cS_{{\bf 1}_s} \\  \cS_{{\bf 15}_a} \\ \cS_{{\bf 15}_s} \\ \cS_{{\bf 20}'_s} \\ \cS_{{\bf 45}_a \oplus {\overline{\bf 45}}_a} \\ \cS_{{\bf 84}_s} \end{pmatrix}\,.
\end{equation}
Each $\cS_{\bf r}(U,V)$ can be expanded as a sum of conformal blocks
\begin{equation}\begin{split}\label{SDecomp}
\cS_{\bf r}(U,V) = \sum_{\text{conformal primaries ${\cal O}_{\Delta, \ell, {\bf r}}$}} \lk^2_{SS\fcy O}g_{\Delta,\ell}(U,V) \,,
\end{split}\end{equation}
where ${\cal O}_{\Delta, \ell, {\bf r}}$ is a conformal primary in $S\times S$ with scaling dimension $\Delta$, spin $\ell$, and $\mathfrak{so}(6)$ irrep $\bf r$, $\lk_{SS\fcy O}$ is its OPE coefficient, and $g_{\Delta,\ell}(U,V)$ are conformal blocks normalized as in \cite{Kos:2013tga}. Note that due to $1\leftrightarrow2$ crossing symmetry, even (odd) spin operators contribute to $\cS_{\bf r}$ only if ${\bf r}$ appears symmetrically (anti-symmetrically) in \eqref{irreps}.

Our task is to write down the most general ansatz for $\cS^1(U,V)$ and $\cS^4(U,V)$ compatible with both supersymmetry and with the constraints from weakly broken higher spin symmetry computed in the previous section. As shown in \cite{Turiaci:2018nua} using the Lorentzian inversion formula \cite{Caron-Huot:2017vep}, $\cS^i(U,V)$ is fully fixed by its double discontinuity up to a finite number contact interactions in AdS. More precisely, we can write:
\begin{equation}
\fcy S^i(U,V) = \fcy S^i_{\text{GFFT}}(U,V) + \frac 1{c_T}\Big(\fcy S^i_{\text{exchange}}(U,V) + \fcy S^i_{\text{contact}}(U,V)\Big) + O(c_T^{-2})\,,
\end{equation}
where the generalized free field theory correlator is
\begin{equation}
\fcy S^i_{\text{GFFT}}(U,V) = \begin{pmatrix} 1 & U & \frac UV &0 &0 &0\end{pmatrix}\,.
\end{equation}
The $\fcy S^i_{\text{exchange}}(U,V)$ term is any CFT correlator with the same single trace exchanges as $\fcy S^i(U,V)$, and with good Regge limit behavior so that the Lorentzian inversion formula holds. Finally, $\fcy S^i_{\text{contact}}(U,V)$ is a sum of contact interactions in AdS with at most six derivatives, which contribute to CFT data with spin two or less.  We will focus on each of these two contributions in turn.

Let us begin with the exchange term. In higher spin $\mathcal{N}=6$ theories the only single trace operators are conserved currents, and their contributions to $\<SSSS\>$ are fixed by the OPE coefficients computed in the previous section.  Let us define the $s$-channel superconformal blocks\footnote{A more general discussion of superconformal blocks is given in Section~\ref{ANALYTICCFT}\,.}
\begin{equation}\begin{split}
\mathfrak G_{\text{stress tensor}}^i(U,V) &= \mathfrak g_{S}^i(U,V)+\mathfrak g_{J_1}^i(U,V)+\mathfrak g_{T_2}^i(U,V)\,, \\
\mathfrak G_{\text{cons},\ell}^i(U,V) &= \mathfrak g_{B_\ell}^i(U,V)+\mathfrak g_{H_{\ell+1}}^i(U,V)+\mathfrak g_{J_{\ell+2}}^i(U,V)+\mathfrak g_{T_{\ell+3}}^i(U,V)\,, \\
\end{split}\end{equation} 
corresponding to the exchange of conserved current multiplets, and where 
\begin{equation}\begin{split}
\mathfrak g_{B_\ell}^i(U,V) &= \mathfrak g_{T_\ell}^i(U,V) = 
g_{\ell+1,\ell}(U,V)\times\begin{cases}
\begin{pmatrix} 1&0&0&0&0&0\end{pmatrix} & \text{ even }\ell  \\
\ 0 & \text{ odd }\ell 
\end{cases} \\
\mathfrak g_{B_\ell}^i(U,V) &= \mathfrak g_{T_\ell}^i(U,V) = 
g_{\ell+1,\ell}(U,V)\times\begin{cases}
\begin{pmatrix} -1&0&0&0&1&1\end{pmatrix} & \text{ even }\ell  \\
\begin{pmatrix} 0&0&0&0&1&-1\end{pmatrix} & \text{ odd }\ell 
\end{cases}
\end{split}\end{equation}
These superconformal blocks can be derived by expanding each $\cS_{\bf r}(U,V)$ as a sum of conformal blocks using the OPE coefficients \eqref{evenblock}, \eqref{oddblock} and \eqref{scalarBlock}, and then using \eqref{toSi} to convert back to the basis $\cS^i(U,V)$ of $R$-symmetry structures.  We can now write
\begin{equation}\begin{split}
\frac1{c_T}\fcy S^i_{\text{exchange}}(U,V) &= \lk_{SSS}^2\left(\mathfrak G_{\text{stress tensor}}^i(U,V)+\sum_{\text{odd }\ell}\mathfrak{G}_{\text{cons},\ell}^i(U,V)\right)\\
& + \lk_{SSB_0}^2\sum_{\text{even }\ell}\mathfrak{G}_{\text{cons},\ell}^i(U,V) + \text{crossing} + \text{double trace terms}\,,
\end{split}\end{equation}
where the double trace terms are some combination of contact terms required so that $\fcy S^i_{\text{exchange}}(U,V)$ has good Regge behavior. 

To make further progress, we note that
\begin{equation}\label{FConn}\begin{split}
\frac14\left(\mathfrak G_{\text{stress tensor}}^i(U,V) + \sum_{\ell}\mathfrak{G}_{\text{cons},\ell}^i(U,V) + \text{crossing} \right) = \fcy S_{\text{free}}^i(U,V) \,,
\end{split}\end{equation}
where we define
\begin{equation}\begin{split}
\fcy S_{\text{free}}^i(U,V) = \begin{pmatrix}0&0&0&\frac U{\sqrt V}&\sqrt{\frac UV}&\sqrt U\end{pmatrix}\,,
\end{split}\end{equation}
to be the connected correlator in the $\cN = 6$ free field theory. This equality can be verified by performing a conformal block expansion of $\<SSSS\>$ in free field theory. Observe that, as proved at the end of the previous section, each conserved supermultiplet contributes equally. Because $\fcy S_{\text{free}}^i(U,V)$ is a correlator in a unitary CFT, it is guaranteed to have the necessary Regge behavior required for the Lorentzian inversion formula. 

Having derived an expression for the sum of odd and even conserved current superblocks, let us turn to the difference. Note that for $\ell>0$, each contribution from $B_\ell$, $J_{\ell+1}$, $H_\ell$ and $T_{\ell+1}$ appearing in an even superblock comes matched with contributions from $T_\ell$, $J_{\ell+1}$, $H_\ell$, and $B_{\ell+1}$ from an odd superblock. We thus find that if we take the difference between the odd and even blocks, the contributions from spinning operators will cancel, leaving us only with the scalar conformal blocks
\begin{equation}
\mathfrak G_{\text{stress tensor}}^i(U,V) +\sum_{\text{odd }\ell}\mathfrak{G}_{\text{cons},\ell}^i(U,V) - \sum_{\text{even }\ell}\mathfrak{G}_{\text{cons},\ell}^i(U,V) = \mathfrak g_{S}^i(U,V) - \mathfrak g_{B_0}^i(U,V)\,.
\end{equation}
On their own, the difference of two conformal blocks does not have good Regge behavior. We can however replace these conformal blocks with scalar exchange diagrams in AdS. Such exchange diagrams do have good Regge behavior, and the only single trace operators that appears in their OPE have the same quantum numbers as the exchanged particle. Using the general scalar exchange diagram computed in \cite{Penedones:2010ue}, and then using \eqref{toSi} to convert from the $s$-channel $R$
-symmetry basis to $\cS^i(U,V)$, we find that\footnote{Our conventions for $\bar D$-functions can be found in Appendix~\ref{SHADOW4PT}\,.}
\begin{equation}\begin{split}
\fcy S^1_{\text{scal}}(U,V) &= -\frac{2U}{\pi^{5/2}}\bar D_{1,1,\frac12,\frac12}(U,V)\,,\\
\fcy S^4_{\text{scal}}(U,V) &= \frac{U}{\pi^{5/2}}\left[\bar D_{\frac12,1,1,\frac12}(U,V)+\bar D_{1,\frac12,1,\frac12}(U,V)\right]\,,
\end{split}\end{equation}
which has been normalized so that the exchange of $S$ itself contributes equally to $\fcy S_{\text{free}}^i(U,V)$ and $\fcy S^1_{\text{scal}}(U,V)$. Using \eqref{SSSCT} to eliminate $\lk_{SSS}^2$ in favor of $c_T^{-1}$, we arrive at our ansatz for the exchange contribution:
\begin{equation}
\frac1{c_T}\fcy S^i_{\text{exchange}}(U,V)= \frac1{c_T}\left((16-a_1(\lambda))\fcy S_{\text{free}}^i(U,V) + a_1(\lambda) \fcy S_{\text{scal}}^i(U,V)\right)\,,
\end{equation}
where $a_1(\lambda)$ is related to $\lk_{SSB_0}^2$ by the equation
\begin{equation}\label{a1a2CT}
a_1(\lk) = 8 - \frac{c_T \lk_{SSB_0}^2}8 \,.
\end{equation}
Because $\lk_{SSB_0}^2$ is always positive in unitary theories, $a_1(\lk)\leq8$.

Now that we have an expression for the exchange terms, let us now turn to the contact terms. As already noted, $\fcy S^i_{\text{contact}}(U,V)$ must be a sum of contact Witten diagrams that contribute to CFT data of spin two or less. Furthermore, because our theory is supersymmetric these contact Witten diagrams must also preserve $\cN=6$ supersymmetry. The problem of finding such contact Witten diagrams was solved in \cite{Binder:2019mpb}, where it was shown that there is a unique such contact Witten diagram that contributes only to CFT data of spin two or less:
\begin{equation}\begin{split}
\fcy S^1_{\text{cont}}(U,V) &= 4UV\bar D_{2,2,3,1}(U,V)\,,\\ 
\fcy S^4_{\text{cont}}(U,V) &= 4U\left(\bar D_{1,1,1,3}(U,V)-\frac43 D_{1,1,2,2}(U,V)\right)\,.
\end{split}\end{equation}

Putting everything together, we arrive at our ansatz for $\<SSSS\>$ in higher spin $\mathcal{N}=6$ theories:
\begin{equation}\label{SSSSansatz}\begin{split}
\fcy S^i&(U,V) = \fcy S^i_{\text{GFFT}}(U,V)\\
&+ \frac 1{c_T}\left((16-a_1(\lk)) \fcy S_{\text{free}}^i(U,V) + a_1(\lk)\fcy S_{\text{scal}}^i(U,V) + a_2(\lk) \fcy S^i_{\text{cont}}(U,V) \right) + O(c_T^{-2})\,.
\end{split}\end{equation}
In Section~\ref{CONSOLVE} we will use localization to fix these two coefficients. These localization constraints however require us to compute not only $\<SSSS\>$ itself, but also $\<SSSP\>$ and $\<SSPP\>$. We can compute $\<SSPP\>$ in terms of $\<SSSS\>$ via superconformal Ward identities derived in \cite{Binder:2019mpb}, and so we relegate this computation to Appendix~\ref{SSPPApp}.  The correlator $\<SSSP\>$ however is not related to $\<SSSS\>$ by these superconformal Ward identities, and so to fix it we must turn once more to the weakly broken higher spin Ward identities.

\subsection{The $\<SSSP\>$ Four-Point Function}
\label{SSSPweak}

Our task in this section is to use the weakly broken higher spin Ward identities to compute $\<SSSP\>$. We begin by noting that conformal and R-symmetry invariance together imply that
\begin{equation}\label{SSSPSt}
   \langle S(\vec x_1,X_1)  S(\vec x_2,X_2) S(\vec x_3,X_3) P(\vec x_4,X_4)\rangle = \frac{x_{13}}{x_{12}^2 x_{34}^3 x_{14}} \sum_{i=1}^6 {\cal T}^i(U, V) {\cal B}_i \,,\\
\end{equation}
where the $\fcy B_i$ are defined as in \eqref{BasisElems}, and where $\cT^i(U,V)$ are functions of the cross-ratios \eqref{crossRatio}. Crossing under $1\leftrightarrow3$ and $2\leftrightarrow3$ relates the different $\cT^i(U,V)$:
\begin{equation}\begin{split}\label{CrossSSSP}
 \cT^2(U,V) &= U^{3/2}\cT^1\left(\frac{1}{U},\frac{V}{U}\right)\,, \quad \cT^3(U,V) = \frac{U}{V}\cT^1(V,U)  \,, \\
 \cT^5(U,V) &= U^{3/2}\cT^4\left(\frac{1}{U},\frac{V}{U}\right)\,, \quad \cT^6(U,V) = \frac{U}{V}\cT^4(V,U) \,,
 \end{split}\end{equation}
so that $\langle SSSP\rangle$ is uniquely specified by $\cT^1(U,V) $ and $\cT^4(U,V)$. By demanding the $Q$ supersymmetry charge \cite{Binder:2019mpb} annihilates $\<SSSF\>$, where $F$ is a fermionic descendant of $S$ defined in \cite{Binder:2019mpb}, we can derive the superconformal Ward identities
\begin{equation}\begin{split}\label{SSSPWard}
\cT^5(U,V) &= \frac1{2U}\left(-U\cT^1(U,V)+\cT^2(U,V)+(1-U)\cT^3(U,V)+2\cT^4(U,V)\right)\,, \\
\cT^6(U,V) &= \frac1{2U}\left(-U\cT^1(U,V)+(V-U)\cT^2(U,V)+V\cT^3(U,V)+2V\cT^4(U,V)\right)\,. \\
\end{split}\end{equation}

Acting with the pseudocharge $\tilde\dk(X)$ on $\<SSSP\>$, we find that
\begin{equation}\label{SSSPps}\begin{split}
\tilde\dk(X)&\<S(\vec y_1,Y_1)S(\vec y_2,Y_2)S(\vec y_3,Y_3)P(\vec y_4,Y_4)\> \\
&= \frac{\alpha}{\sqrt N}\bigg(\left\<\tilde P(\vec y_1,[X,Y_1])S(\vec y_2,Y_2)S(\vec y_3,Y_3)P(\vec y_4,Y_4)\right\> + 1\leftrightarrow2 + 1\leftrightarrow3 \\
&\qquad\qquad -\left\<S(\vec y_1,Y_1)S(\vec y_2,Y_2)S(\vec y_3,Y_3)\tilde S(\vec y_4,[X,Y_4])\right\>\bigg) \,.
\end{split}\end{equation}
To expand the right-hand side of this identity, we define
\begin{equation}\begin{split}\label{shadowSSSS}
   \langle S(\vec x_1,X_1)  S(\vec x_2,X_2) S(\vec x_3,X_3) \tilde S(\vec x_4,X_4)\rangle &= \frac{x_{13}}{x_{12}^2 x_{34}^3 x_{14}} \sum_{i=1}^6 \tilde{\cal S}^i(U, V) {\cal B}_i \,,\\
   \langle S(\vec x_1,X_1)  S(\vec x_2,X_2) \tilde P(\vec x_3,X_3) P(\vec x_4,X_4)\rangle &= \frac{x_{13}}{x_{12}^2 x_{34}^3 x_{14}} \sum_{i=1}^6 \tilde{\cal R}^i(U, V) {\cal B}_i \,,\\
\end{split}\end{equation}
where $\tilde \cS^i(U,V)$ and $\tilde \cR^i(U,V)$ can be computed by taking the shadow transform of $\<SSSS\>$ and $\<SSPP\>$. To expand the left-hand side, we use \eqref{chiEq} and $SO(6)_R$ invariance to write
\begin{equation}\label{ssspLHS}\begin{split}
\tilde\dk(X)\<S(\vec y_1,Y_1)S(\vec y_2,Y_2)&S(\vec y_3,Y_3)P(\vec y_4,Y_4)\> \\
&= \frac12\lk_{SSB_0} \left\<S(\vec y_1,Y_1)S(\vec y_2,Y_2)S(\vec y_3,Y_3)P(\vec y_4,[X,Y_4])   \right\> \\
&+ \kk_1  \left\<S(\vec y_1,Y_1)S(\vec y_2,Y_2)S(\vec y_3,Y_3)S^2(\vec y_4,[X,Y_4]) \right\> \\
&+ \kk_2  \left\<S(\vec y_1,Y_1)S(\vec y_2,Y_2)S(\vec y_3,Y_3)SB_0(\vec y_4,[X,Y_4])\right\>\,.
\end{split}\end{equation}
The two double-trace terms can each be expanded at $O(c_T^{-3/2})$ as a product of a two-point and a three-point function, so that for instance
\begin{equation}\begin{split}
\big\<S&(\vec y_1,Y_1)S(\vec y_2,Y_2)S(\vec y_3,Y_3)SB_0(\vec y_4,Y_4)\big\> \\
&= \left\<S(\vec y_1,Y_1)S(\vec y_2,Y_2)B_0(\vec y_4)\right\> \left\< S(\vec y_3,Y_3)S(\vec y_4,Y_4)\right\> + \text{ permutations } + O(c_T^{-3/2})\,.
\end{split}\end{equation}
We can then solve \eqref{SSSPps} to find that it fully fixes $\<SSSP\>$ in terms of $\<SSSS\>$ and $\<SSPP\>$:
\begin{equation}\label{TExpr}\begin{split}
\fcy T^1(U,V) &= -\frac{2\ak}{\lk_{SSB_0}\sqrt{c_T}}\left(\tilde{\fcy R}^1(U,V) + \tilde{\fcy S}^1(U,V) - \frac{\lk_{SSB_0}^2}8\sqrt U\right)\,, \\
\fcy T^4(U,V) &= -\frac{2\ak}{\lk_{SSB_0}\sqrt{c_T}}\left(\tilde{\fcy R}^4(U,V) + \tilde{\fcy S}^4(U,V) + \frac{\lk_{SSS}^2}8 U\left(1+\frac1{\sqrt V}\right)\right)\,. \\
\end{split}\end{equation}

To compute $\fcy T^i(U,V)$ for the various contributions to our $\<SSSS\>$ ansatz, we must first calculate $\<SSPP\>$, and then find the shadow transforms of both $\<SSSS\>$ and $\<SSPP\>$. Since both tasks are straightforward albeit tedious, we relegate them to Appendix~\ref{SSPPApp} and Appendix~\ref{SHADOW4PT} respectively. The only subtlety occurs for the scalar exchange diagram contribution, where we have to include the effects of operator mixing between $P$ and the double traces $S^2$ and $SB_0$. We can then furthermore use \eqref{cOEqsL} and \eqref{a1a2CT} to simplify the prefactor\footnote{Because equation \eqref{cOEqsL} gives an expression for $\alpha^2$, it only fixes $\alpha$ up to an overall sign. Note that this sign is determined by the sign convention for $P$, such that by redefining $P\rightarrow-P$ we can always fix $\alpha\geq0$. This choice turns out to also be consistent with our conventions in Section~\ref{LOCALIZATION}.}
\begin{equation}\label{rkDef}
\frac{2\ak}{\lk_{SSB_0}\sqrt{c_T}} = \frac{1}{\pi}\sqrt{\frac{2a_1(\lk)}{8-a_1(\lk)}}\,.
\end{equation}
Once the dust settles, we find that
\begin{equation}\label{SSSPAnz}\begin{split}
\fcy T^i&(U,V) = -\frac{1}{\pi}\sqrt{\frac{2a_1(\lk)}{8-a_1(\lk)}}\\
&\times\Bigg((16-a_1(\lk))\fcy T^i_{\text{free}}(U,V) + a_1(\lk)\fcy T^i_{\text{scal}}(U,V) + a_2(\lk)\fcy T^i_{\text{cont}}(U,V)\Bigg) + O(c_T^{-2})\,,
\end{split}\end{equation}
where we define
\begin{equation}\label{freeSSSP}
\begin{split}
\fcy T^1_{\text{free}}(U,V) &= -\sqrt U\,,\qquad \fcy T^4_{\text{free}}(U,V) =  -\frac12\left(\sqrt{\frac{U^3}V} - U - \frac{U}{\sqrt V}\right)\,, \\
\fcy T^1_{\text{scal}}(U,V) &= +\sqrt U\,,\qquad \fcy T^4_{\text{scal}}(U,V) =  \frac12\left(\sqrt{\frac{U^3}V} - U - \frac{U}{\sqrt V}\right)\,, \\
\fcy T^1_{\text{cont}} (U,V) &= \frac{8\pi^{1/2}}{3}UV\Big(2U\bar D_{2,3,2,2}+2 V\bar D_{1,3,3,2}-2 \bar D_{2,2,3,2}-3 \bar D_{1,2,2,2}\Big)  \,,\\ 
\fcy T^4_{\text{cont}}(U,V) &= -\frac{32\pi^{1/2}}{3}  U^2\left(U\bar D_{3,3,1,2}-\bar D_{2,2,1,2}\right) \,. \\
\end{split}
\end{equation}
It is not hard to check that each of these contributions individually satisfies the $\<SSSP\>$ superconformal Ward identity \eqref{SSSPWard}.

We conclude by applying \eqref{freeSSSP} to parity preserving theories, where $\<SSSP\>$ must vanish. Using \eqref{freeSSSP}, we see that this is possible only if $a_1 = 8$ and $a_2 = 0$, and so conclude that
\begin{equation}\label{SSSS1/2}\begin{split}
\fcy S^i(U,V) &= \fcy S^i_{\text{GFFT}}(U,V) + \frac {8}{c_T}\left(\fcy S_{\text{free}}^i(U,V) +\fcy S_{\text{scal}}^i(U,V)\right) + O(c_T^{-2})\,,
\end{split}\end{equation}
in such theories. In particular, we see that 
\begin{equation}
\lk_{SSB_0}^2 = \frac{8(8-a_1)}{c_T} = 0\,,
\end{equation}
just as argued at the end of Section~\ref{SCALAR3pt}.

\section{Constraints from Localization}
\label{LOCALIZATION}

In the previous section, we fixed the tree level $\<SSSS\>$ and $\<SSSP\>$ in terms of two coefficients $a_1(\lambda)$ and $a_2(\lambda)$ for 3d $\mathcal{N}=6$ CFTs with weakly broken higher spin symmetry. We now determine these coefficients using supersymmetric localization applied to the $U(N)_{k}\times U(N+M)_{-k}$ or $SO(2)_{2k}\times USp(2+2M)_{-k}$ ABJ theories. We will start by reviewing the relation of certain integrals of $\langle SSSS\rangle$ to $\partial_{m_\pm}^4\log Z\big\vert_{m_\pm=0}$ and $\partial_{m_+}^2\partial_{m_-}^2\log Z\big\vert_{m_\pm=0}$ derived in \cite{Binder:2019mpb}, as well as deriving a new relation between a certain integral of $\langle SSSP\rangle$ and $\partial_{m_\pm}^3\partial_{m_\mp}\log Z\big\vert_{m_\pm=0}$. We then compute these derivatives of the mass deformed sphere free energy in the higher spin limit of the $U(N)_{k}\times U(N+M)_{-k}$ or $SO(2)_{2k}\times USp(2+2M)_{-k}$ ABJ theories following \cite{Hirano:2015yha}. We find that $a_2(\lambda)=0$, while $a_1(\lambda)$ is the same for each theory, which completes our derivation of the tree level correlator.

\subsection{Integrated Correlators}
\label{INTCOR}

In this section, we will review the integrated constraints derived in \cite{Binder:2019mpb}. We will then extend the results of that paper to include constraints on the parity odd correlator $\<SSSP\>$. 

Any $\cN = 6$ theory on $S^3$ admits three real mass deformations, preserving a $\cN=2$ subgroup of the full $\cN=6$ supersymmetry. We will focus on two of these real mass deformations, $m_\pm$, under which the $S^3$ partition function for the $U(N)_k\times U(N+M)_{-k}$ ABJ theory is 
 \cite{Kapustin:2009kz,Hama:2011ea}:
\begin{equation}\begin{split}\label{ABJv1}
Z_{M,N,k}&(m_+,m_-)  \\
&=\int d^{M+N}\mu\, d^N\nu\  \frac{e^{-i\pi k(\sum_i \mu_i^2 - \sum_a \nu_a^2)}\prod_{i<j}4\sinh^2\left[\pi(\mu_i-\mu_j)\right]\prod_{a<b}4\sinh^2\left[\pi(\nu_a-\nu_b)\right]}{\prod_{i,a}4\cosh\left[\pi(\mu_i-\nu_a)+\frac{\pi m_+}2\right]\cosh[\pi(\mu_i-\nu_a)+\frac{\pi m_-}2]}\,,
\end{split}\end{equation}
up to an overall $m_\pm$-independent normalization factor. The ABJ mass deformations on a unit-sphere take the form
\begin{equation}
m_+\int_{S^3}(iJ_+ + K_+) + m_-\int_{S^3}(iJ_- + K_-) + O(m_\pm^2)\,,
\end{equation}
where we define
\begin{equation}\label{JKDef}\begin{split}
J_\pm(\vec x) = \frac{\sqrt{c_T}}{2^5\pi}S(\vec x,X_\pm)\,,\qquad K_\pm(\vec x) = \frac{\sqrt{2c_T}}{2^5\pi}P(\vec x,X_\mp)\,, \\
\text{ with } X_+ = \text{diag}\{-1,-1,0,0\}\,,\qquad X_- = \text{diag}\{0,0,1,-1\}\,.
\end{split}\end{equation}
By taking derivatives of the $S^3$ partition function $Z(m_+,m_-)$ with respect to $m_+$ and $m_-$, we can thus compute certain integrated correlators of $S$ and $P$. Taking two derivatives of $Z$ and then integrating the resultant two-point function over the sphere, we find that \cite{Closset:2012vg}
\begin{equation}
\left.\frac{\nb^2\log Z}{\nb m_+^2}\right|_{m_\pm=0} = \left.\frac{\nb^2\log Z}{\nb m_-^2}\right|_{m_\pm=0} = -\frac{\pi^2c_T}{64}\,,
\end{equation}
allowing us to compute $c_T$ as a function of the parameters in the ABJ Lagrangian.

Evaluating fourth derivatives is more involved. In \cite{Binder:2019mpb}, derivatives with an even number of $m_+$'s were evaluated, where it was shown that
\begin{equation}\begin{split}\label{mixDiv4}
\left.\frac{\nb^4\log Z}{\nb m_\pm^4}\right|_{m_\pm=0} & = \frac{\pi^4c_T^2}{2^{13}}I_{++}[\cS^i]\,, \\
\left.\frac{\nb^4\log Z}{\nb m_+^2\nb m_-^2}\right|_{m_\pm=0} &= \frac{\pi^2c_T^2}{2^{11}}I_{+-}\left[\cS^i\right]\,, \\
\end{split}\end{equation}
where $I_{++}[\cS^i]$ and $I_{+-}[\cS^i]$ are the linear functionals:
\begin{equation}\begin{split}\label{IDefs}
I_{++}[\cS^i]  &= 2\lk^2_{(B,2)_{2,0}^{[022]}} - 4\,,\\
I_{+-}[\cS^i] &= \int_0^\infty dr\int_0^\pi d\theta\,\sin\theta\frac{\cS^1\left(1+r^2-2r\cos\qk,r^2\right)}{1+r^2-2r\cos\qk}\,,
\end{split}\end{equation}
and where $\lk^2_{(B,2)_{2,0}^{[022]}}$ is the OPE coefficient of the 1/3-BPS multiplet whose superprimary transforms in the $\bf84$ under $\mathfrak{so}(6)_R$.

We now list the contributions of each term in \eqref{SSSSansatz} to both $I_{++}[\cS^i]$ and $I_{+-}[\cS^i]$. For the free term $\cS_{\text{free}}$ and contact term $\cS_{\text{cont}}$ these were computed in \cite{Binder:2018yvd} and \cite{Binder:2019mpb} respectively, leaving just\footnote{The generalized free term $\cS_{\text{GFFT}}(U,V)$ is fully disconnected and so does not contribute to the integrated constraints.} the exchange term $\cS_{\text{scal}}$, which we compute in Appendix~\ref{SDevenLoc}. In sum we find that:
\begin{equation}\begin{aligned}
I_{++}\left[\fcy S^i_{\text{free}}\right] &= 4 \,,\qquad I_{+-}\left[\fcy S^i_{\text{free}}\right] = 0\,, \\
I_{++}\left[\fcy S^i_{\text{scal}}\right] &= 0 \,,\qquad I_{+-}\left[\fcy S^i_{\text{scal}}\right] = -\pi^2\,, \\
I_{++}\left[\fcy S^i_{\text{cont}}\right] &= \frac83 \,,\qquad I_{+-}\left[\fcy S^i_{\text{cont}}\right] = \frac23\pi^2\,. \\
\end{aligned}\end{equation}
We now have two constraints on the two coefficients $a_1(\lambda)$ and $a_2(\lambda)$:
\begin{equation}\label{LocCons1}\begin{split}
\frac1{c_T}\left(32-2a_1(\lk) + \frac43a_2(\lk)\right) &= \lk^2_{(B,2)_{2,0}^{[022]}}\Big|_{1/c_T} \,, \\
-a_1(\lk) + \frac23a_2(\lk) &= \lim_{c_T\rightarrow0}\frac{2^{11}}{\pi^4c_T}\left.\frac{\nb^4\log Z}{\nb m_+^2\nb m_-^2}\right|_{m_\pm=0}\,. \\
\end{split}\end{equation}
Note however that these equations are redundant, which implies that
\begin{equation}
\left.\frac{\nb^4\log Z}{\nb m_+^2\nb m_-^2}\right|_{m_\pm=0} = \frac{c_T\pi^4}{2^{11}}\left(16+c_T\left(2-\lk^2_{(B,2)_{2,0}^{[022]}}\right)\right) + O(c_T^0)\,,
\end{equation}
regardless of the values of $a_i(\lk)$, and so do not suffice to fully fix $\<SSSS\>$ from localization.

To find an additional constraint, we turn to the mixed mass derivative\footnote{The constraint from $\frac{\nb^4\log Z}{\nb m_-^3\nb m_+} $ is equivalent.}
\begin{equation}\label{parOddDer}
\frac{\nb^4\log Z}{\nb m_+^3\nb m_-} = \left\<\left(\int d^3x\sqrt{g}(iJ_+(\vec x)+K_+(\vec x))\right)^3\left(\int d^3x\sqrt{g}(iJ_-(\vec x)+K_-(\vec x))\right)\right\>\,.
\end{equation}
Unlike the previously considered derivatives, when we expand \eqref{parOddDer} using \eqref{JKDef} we find that it gets contributions only from the parity violating correlators $\<SSSP\>$ and $\<PPPS\>$, rather than from the parity preserving $\<SSSS\>$. Following the methods used in \cite{Binder:2019mpb} to derive \eqref{mixDiv4}, we simplify \eqref{parOddDer} in Appendix~\ref{MIXEDODD} using the 1d topological sector \cite{Chester:2014mea,Beem:2016cbd,Dedushenko:2016jxl} and the superconformal Ward identity for $\<SSSP\>$. We find that
\begin{equation}\label{oddIRes}
\frac{\nb^4 \log Z}{\nb^3 m_+\nb m_-} = -\frac{ic_T^2\pi^2}{2^{13}\sqrt 2}I_{\text{odd}}\left[\fcy T^i\right]\,,
\end{equation}
where we define
\begin{equation}\label{oddIDef}\begin{split}
I_{\text{odd}}\left[\fcy T^i\right] &= \tilde I_{\text{odd}}[\fcy T^2] + \tilde I_{\text{odd}}[\fcy T^3] + 2\tilde I_{\text{odd}}[\fcy T^4]\,, \\
\tilde I_{\text{odd}}[\fcy T^i] &= \int_0^\infty dr\int_0^\pi d\qk\,4\pi  r\sin\qk\, \fcy T^i\left(\frac1{1+r^2-2r\cos\qk},\frac{r^2}{1+r^2-2r\cos\qk}\right)\,.
\end{split}\end{equation}
In Appendix~\ref{IODD} we evaluate \eqref{oddIDef} on $\cT_{\text{free}}^i$, $\cT_{\text{scal}}^i$ and $\cT_{\text{cont}}$, and find that
\begin{equation}
I_{\text{odd}}\left[\fcy T^i_{\text{free}}\right] = -I_{\text{odd}}\left[\fcy T^i_{\text{scal}}\right]  = -2\pi^3\,,\qquad I_{\text{odd}}\left[\fcy T^i_{\text{free}}\right] = -4\pi^3 \,.
\end{equation}

Using the ansatz \eqref{SSSPAnz} for $\<SSSP\>$ derived using the weakly broken higher spin symmetry in Section \ref{SSSPweak}, we arrive at the constraint
\begin{equation}\label{LocCons2}
\sqrt{\frac{a_1(\lk)}{8-a_1(\lk)}}\left(8-a_1(\lk)+ a_2(\lk)\right) = \frac{2^{11}i}{\pi^4c_T}\frac{\nb^4 \log Z}{\nb^3 m_+\nb m_-}  + O(c_T^{-1})\,.
\end{equation}

\subsection{Localization Results}
\label{HSLOC}

In the previous subsection we saw that localization could be used to derive equations \eqref{LocCons1} and \eqref{LocCons2} relating the coefficients $a_i(\lk)$ to derivatives of the $S^3$ partition function. Our task now is to compute the relevant localization quantities for specific $\cN=6$ higher spin theories, namely the $U(N)_{k}\times U(N+M)_{-k}$ and $SO(2)_{2k}\times USp(2+2M)_{k}$ ABJ theories.

\subsubsection{$U(N)_k\times U(N+M)_{-k}$ theory}
The $U(N)_k\times U(N+M)_{-k}$ $S^3$ partition function can be written as the integral \cite{Kapustin:2009kz,Hama:2011ea,Binder:2020ckj}
\begin{equation}\label{mixdU1d}\begin{split}
Z_{M,N,k}(m_+,m_-)\qquad\qquad& \\
= \frac{e^{-\frac \pi2MN m_-}Z_0}{\cosh^N \frac{\pi m_+}2}&\int d^Ny\prod_{a<b} \frac{\sinh^2\frac{\pi(y_a-y_b)}{k}}{\cosh\left[\frac{\pi (y_a-y_b)}k +\frac{\pi m_+}{2}\right]\cosh\left[\frac{\pi (y_a-y_b)}k -\frac{\pi m_+}{2}\right]}\\
&\times\prod_{a=1}^N\left(\frac{e^{i\pi y_am_-}}{2\cosh\left(\pi y_a\right)}\prod_{l=0}^{M-1}\frac{\sinh\left[\frac{\pi\big(y_a+i(l+1/2)\big)}{k}\right]}{\cosh\left[\frac{\pi\big(y_a+i(l+1/2)\big)}k-\frac{\pi m_+}2\right]}\right)\,,
\end{split}\end{equation}
where $Z_0$ is an overall constant which is independent of $m_\pm$. In Appendix~\ref{locHS} we expand this integral at large $M$ expansion holding $M/k=\lk$ fixed, generalizing previous work in \cite{Hirano:2015yha,Binder:2020ckj}. We find that $c_T$ is given in terms of the Lagrangian parameters $M$, $N$ and $k$ by the series
\begin{equation}\begin{split}\label{ctUN}
c_T &= \frac{16N k\sin(\pi \lambda)}{\pi} + 4N^2(3+\cos(2\pi \lambda)) \\
&- \frac{\pi N\left(16-18N^2+(1-14N^2)\cos(2\lk\pi)\right)}{3k} +O(k^{-2})\,.
\end{split}\end{equation}
We invert this series to eliminate $k$ in favor of $c_T$, and so find that
\begin{equation}\begin{split}\label{UNs}
&\lambda^2_{(B,2)_{2,0}^{[022]}} =  2 + \frac{8(3+\cos(2\pi \lambda))}{c_T}-\frac{64N^2\sin^2(\pi\lk)(3+5\cos(2\pi\lk))}{c_T^2} \\
&-\frac{512N^3\left(29-\frac{121}{4N^2}+\left(44-\frac{19}{N^2}\right)\cos(2\pi\lk)+\left(23+\frac5{4N^2}\right)\cos(4\pi\lk)\right)\sin^2(\pi\lk)}{3c_T^3} + O\left(c_T^{-4}\right)\,, 
\end{split}\end{equation}
and
\begin{equation}\begin{split}\label{fourDivUN}
\frac1{c_T^2}\frac{\nb^4 \log Z}{\nb^2 m_+\nb^2 m_-} &= -\frac{\pi^4\sin^2(\pi\lk)}{256c_T} + \frac{\pi^4N^2(1-5\cos(2\pi\lk))\sin^2(\pi\lk)}{64c_T^2} + O(c_T^{-3})\,, \\
\frac1{c_T^2}\frac{\nb^4 \log Z}{\nb^3 m_+\nb m_-}   &= -\frac{i\pi^4\sin(2\pi\lk)}{512c_T} + \frac{5i\pi^4N^2\cos(\pi\lk)\sin^3(\pi\lk)}{32c_T^2} + O(c_T^{-3})\,. \\
\end{split}\end{equation}
Note that for each of these quantities, the $O(c^{-1}_T)$ term is independent of $N$, the $O(c_T^{-2})$ term is proportional to $N^2$,\footnote{This overall factor is expected for the following reason. The $U(N)_k$ gauge factor is very weakly coupled in the higher spin limit at finite $N$, so we can construct $N^2$ different ``single-trace'' operators of the $U(M+N)$ factor (which are an adjoint+singlet of $SU(N)$, with the $SU(N)$-adjoint not being a gauge-invariant operator in the full theory), and because of the weak $U(N)$ coupling the ``double-trace'' operators constructed from pairs of each of these $N^2$ ``single-trace'' operators contribute the same, so we get a factor of $N^2$. Note that it is important to distinguish the single trace operators in scare quotes from single-trace operators in the usual sense, which are gauge-invariant. We thank Ofer Aharony for discussion on this point.} and further subleading terms have more complicated $N$ dependence. Also, the $O(c_T^{-2})$ correction to $\lambda^2_{(B,2)_{2,0}^{[022]}}$ is smallest for $N=1$ and $\lambda=1/2$, as expected from the conjecture in \cite{Binder:2020ckj} that the $U(1)_{2M}\times U(1+M)_{-2M}$ theory has the minimal value of $\lambda^2_{(B,2)_{2,0}^{[022]}}$ for fixed $c_T$.

\subsubsection{$SO(2)_{2k}\times USp(2+2M)_{-k}$ theory}
Just as for the $U(N)_k\times U(N+M)_{-k}$ theory, the $SO(2)_{2k}\times USp(2+2M)_{-k}$ $S^3$ partition function can be written as a single integral \cite{Kapustin:2009kz,Gulotta:2012yd,Moriyama:2016kqi,Binder:2020ckj}
\begin{equation}\label{ZINTo2}\begin{split}
&Z_{M,k}(m_+,m_-)\propto \frac 1 {\cosh \frac{\pi m_+}{2} }\int d y\  \frac{e^{i \pi m_- y}\cosh\left[\frac{\pi y}{2k}\right]\cosh\left[\frac{\pi y}{2k}+\frac{\pi m_+}2\right]}{\sinh\left[\pi y\right]\cosh\left[\frac{\pi y}{k}+\frac{\pi m_+}{2}\right]}\\
&\times\prod_{l = -M}^M\frac{\sinh\left[\frac{\pi( y+i l)}{2k}\right]}{\cosh\left[\frac{\pi( y+ i l)}{2k}+\frac{\pi m_+}2\right]}\,,
\end{split}\end{equation}
where the overall constant of proportionality is independent of $m_\pm$. The higher spin limit for this theory is given by the large $M$ limit, where $\lk = \frac{2M+1}{2k}$ is held fixed. Just as for the ${U(N)_k\times U(N+M)_{-k}}$ theory, when $\lk = \frac12$ the ${SO(2)_{2k}\times USp(2+2M)_{-k}}$ theory preserves parity as a consequence of Seiberg duality.

The large $M$ expansion of \eqref{ZINTo2} was previously considered in \cite{Binder:2020ckj} for $m_+ = 0$, where it was found that
\begin{equation}\label{cTLamSO}\begin{split}
c_T &= \frac{32k \sin(\pi \lk)}{\pi} + 16\cos^2(\pi\lk) - \frac{\pi(15+29\cos(2\pi\lk))\sin(\pi\lk)}{3k} + O(k^{-2})\,, \\
\lambda^2_{(B,2)_{2,0}^{[022]}}  &= 2 + \frac{8(3+\cos(2\lk\pi))}{c_T}-\frac{32\sin^2(\pi\lk)(17+23\cos(2\pi\lk))}{c_T^2} + O(c_T^{-3}) \,.
\end{split}\end{equation}
We generalize to the mixed mass case in Appendix~\ref{locHS}, finding that
\begin{equation}\label{mixedSO}\begin{split}
\frac1{c_T^2}\frac{\nb^4 \log Z}{\nb^2 m_+\nb^2 m_-} &= -\frac{\pi^4\sin^2(\pi\lk)}{256c_T} - \frac{\pi^4(5+23\cos(2\pi\lk))\sin^2(\pi\lk)}{128c_T^2}+O(c_T^{-3})\,, \\
\frac1{c_T^2}\frac{\nb^4 \log Z}{\nb^3 m_+\nb m_-}   &= \frac{i\pi^4\sin(2\pi\lk)}{512c_T} - \frac{i\pi^4(30\sin(2\pi\lk)-23\sin(4\pi\lk))}{512c_T^2} + O(c_T^{-3}) \,. \\
\end{split}\end{equation}

\subsection{Solving the Constraints}
\label{CONSOLVE}

We are now finally in a position to fully fix the coefficients $a_i(\lk)$ in higher spin ABJ theory. Solving the parity even constraints \eqref{LocCons1} with either the localization results \eqref{UNs} and \eqref{fourDivUN} for the ${U(N)_k\times U(N+M)_{-k}}$ theory or \eqref{cTLamSO} and \eqref{mixedSO} for the ${SO(2)_{2k}\times USp(2+2M)_{-k}}$ theory (the localization results are identical at leading order in $c_T^{-1}$), we find that 
\begin{equation}
a_2(\lk) = \frac32a_1(\lk)+6\cos(2\lk\pi)-6\,.
\end{equation}
Substituting this equation into the parity odd constraint \eqref{LocCons2} and squaring both sides, we find that
\begin{equation}
\frac{a_1(\lk)(a_1(\lk) + 12\cos(2\pi\lk)+4)^2}{2(8-a_1(\lk))} = 32\sin^2(2\pi\lk)\,,
\end{equation}
which upon further rearrangement becomes the cubic equation
\begin{equation}
\left(a_1(\lk) - 8\sin^2(\pi\lk)\right)\left(a_1(\lk)^2 + 4(5\cos(2\pi\lk) + 3)a_1(\lk) + 256\cos^2(\pi\lk)\right)\,.
\end{equation}
This has three solutions for $a_1(\lk)$. However, two of these solutions are not real for all $\lk\in[0,\frac12]$ and so we discard them as non-physically. We therefore conclude that
\begin{equation}
a_1(\lk) = 8\sin^2(\pi\lk)\,,
\end{equation}
which in turn implies that
\begin{equation}
a_2(\lk) = 0\,.
\end{equation}
Substituting these values into our ansatz for $\<SSSS\>$, we arrive at the expression \eqref{finalAnswer} for $\<SSSS\>$ given in the introduction.

As discussed in Appendix~\ref{SSPPApp}, we can compute $\<PPPP\>$ from $\<SSSS\>$ using the superconformal Ward identities given in \cite{Binder:2019mpb}, and so can derive the expression \eqref{finalAnswerP} for $\<PPPP\>$.

\section{Tree Level CFT Data}
\label{numSec}

In the previous sections, we derived $\langle SSSS\rangle$ for $U(N)_{k}\times U(N+M)_{-k}$ or $SO(2)_{2k}\times USp(2+2M)_{-k}$ ABJ theories to leading order in the large $M,k$ limit at fixed $\lambda$ and $N$, where recall that $\lambda=M/k$ or $\lambda=(M+1/2)/k$ for each theory. When written in terms of $c_T$ and $\lambda$, the answer is the same for all theories, and is a periodic function of $\lambda$. We now extract tree level CFT data using the $\langle SSSS\rangle$ superblock expansion from \cite{Binder:2020ckj}. We will then plug in $\lambda=1/2$ and compare to the numerical bootstrap prediction from \cite{Binder:2020ckj} for the $U(1)_{2M}\times U(1+M)_{-2M}$ theory, and find a good match. 

\subsection{Analytic Results for General $\lambda$}
\label{ANALYTICCFT}

Let us start by briefly reviewing the superblock expansion for $\langle SSSS\rangle$, for more details see \cite{Binder:2020ckj}. We can expand $\langle SSSS\rangle$ as written in the R-symmetry basis \eqref{SDecomp} in superblocks as:
\es{blockExp}{
\cS_{\bf{r}}(U,V)=\sum_{I\in S\times S} \lambda^2_{I} \mathfrak{G}^{\bf{r}}_I(U,V)\,,
}
where $ \lambda^2_{I}$ is the OPE coefficient squared for each superblock $\mathfrak{G}^{\bf{r}}_I(U,V)$, and the index $I = {{\cal M}_{\Delta_0, \ell_0}^{{\bf r}_0, n} }$ encodes both the supermultiplet $ {{\cal M}_{\Delta_0, \ell_0}^{{\bf r}_0} }$ labeled by the scaling dimension $\Delta_0$, spin $\ell_0$, and $\mathfrak{so}(6)$ irrep $\bf r_0$ of its superprimary, as well as an integer $n = 1, 2, \ldots$ when there is more than one superblock for a given multiplet (this index is omitted when the superblock is unique). Each superblock is written as a linear combination of the conformal blocks for the operators in its supermultiplet:
 \es{SuperconfBlock}{
  \mathfrak{G}_I^{\bf r}(U, V)= \sum_{\substack{\text{conf primaries } \\ \text{${\cal O}_{\Delta, \ell, {\bf r}} \in {\cal M}_{\Delta_0, \ell_0}^{{\bf r}_0}$ }}} 
    a^I_{\Delta, \ell, {\bf r}}  \,  g_{\Delta, \ell}(U, V)  \,,
 }
 where the coefficients $a^I_{\Delta, \ell, {\bf r}} $ are related to the coefficients $a_{\Delta, \ell, {\bf r}} $ in the conformal block expansion \eqref{SDecomp} as
  \es{aCoeffExpansion}{
  a_{\Delta, \ell, {\bf r}} = \sum_{I}  \lambda_{I}^2 a^{I}_{\Delta, \ell, {\bf r}}\,,
 }
such that each coefficient $a^I_{\Delta, \ell, {\bf r}} $ is fixed by superconformal symmetry in terms of a certain coefficient that we normalize to one. The list of superblocks that appear in $S\times S$ along with their normalization is summarized in Table \ref{SupermultipletTable}, and the explicit values of the $a^{I}_{\Delta, \ell, {\bf r}}$ for each superblock are given in the \texttt{Mathematica} notebook attached to \cite{Binder:2020ckj}. Note that the multiple superblocks that can appear for a given supermultiplet are distinguished by their $\mathcal{P}$ and $\mathcal{Z}$ charges relative to the superprimary,\footnote{While not all 3d $\mathcal{N}=6$ SCFTs are invariant under these symmetries, they can still be used to organize superblocks for any theory.} where $\mathcal{P}$ is parity, and $\mathcal{Z}$ is another discrete symmetry that is defined in \cite{Binder:2019mpb}.

\begin{table}
\begin{center}
{\renewcommand{\arraystretch}{1.4}
\begin{tabular}{ l | c |   l|l }
Superconformal block    & {normalization} & $\mathcal{P}$ & $\mathcal{Z}$   \\  \hline
\multirow{2}{*}{$\text{Long}_{\Delta, 0}^{[000],n}$}  & $n=1:\quad(a_{\Delta, 0, {\bf 1}}, a_{\Delta+1, 0, {\bf 20'}})  = (1, 0)$ &  $+$ &  $+$    \\ 
 &    $n=2:\quad (a_{\Delta, 0, {\bf 1}}, a_{\Delta+1, 0, {\bf 20'}})  = (0, 1)$   &  $-$ &  $+$  \\ 
 \hline
$\text{Long}_{\Delta, \ell}^{[000]}$, $\ell \geq 1$ odd    &   $a_{\Delta+1, \ell+1, {\bf 15}_s} = 1$  & $+$ & $+$ \\ 
\hline
\multirow{3}{*}{$\text{Long}_{\Delta, \ell}^{[000],n}$, $\ell \geq 2$ even}    & $n=1:\quad(a_{\Delta, \ell, {\bf 1}}, a_{\Delta+1, \ell, {\bf 1}}, a_{\Delta+1, \ell, {\bf 15}_s})  = (1, 0, 0)$ &  $+$ &  $+$  \\ 
  & $n=2:\quad(a_{\Delta, \ell, {\bf 1}}, a_{\Delta+1, \ell, {\bf 1}}, a_{\Delta+1, \ell, {\bf 15}_s})  = (0, 1, 0)$&  $-$ &  $+$ \\ 
&   $n=3:\quad(a_{\Delta, \ell, {\bf 1}}, a_{\Delta+1, \ell, {\bf 1}}, a_{\Delta+1, \ell, {\bf 15}_s})  = (0, 0, 1)$&  $-$ &  $-$  \\ 
 \hline
 \multirow{2}{*}{$(A, 1)_{\ell + 2, \ell }^{[100],n}$, $\ell - \frac 12 \geq 1$ odd}
 & $n=1:\quad(a_{\ell + \frac 52, \ell + \frac 12, {\bf 1}}, a_{\ell + \frac 52, \ell + \frac 12, {\bf 15}_s} ) = (1, 0)$  &$+$  & $+$  \\
  &  $n=2:\quad(a_{\ell + \frac 52, \ell + \frac 12, {\bf 1}}, a_{\ell + \frac 52, \ell + \frac 12, {\bf 15}_s} ) = (0, 1)$ &$+$&$-$ \\
 \hline
$(A, 2)_{\ell + 2, \ell}^{[011]}$, $\ell \geq 0$ even & $a_{\ell+2, \ell, {\bf 15}_s} = 1$ & $+$ &  $-$ \\
\hline
$(A, 2)_{\ell + 2, \ell}^{[011]}$, $\ell \geq 0$ odd  & $a_{\ell+2, \ell, {\bf 15}_a} = 1$ & $+$ &  $+$ \\
\hline 
$(A, +)_{\ell + 2, \ell }^{[020]}$, $\ell - \frac 12 \geq 0$ even  & $a_{\ell+\frac 52, \ell + \frac 12, {\bf 15}_a} = 1$ & $+$ &  \\
\hline 
$(A, -)_{\ell + 2, \ell }^{[002]}$, $\ell - \frac 12 \geq 0$ even  & $a_{\ell+\frac 52, \ell + \frac 12, {\bf 15}_a} = 1$ & $+$ &  \\
\hline
$(A, \text{cons})_{\ell + 1, \ell}^{[000]}$, $\ell \geq 0$ even   & $a_{\ell + 1, \ell, {\bf 1}} = 1$ & $+$ &  $+$ \\
\hline
$(A, \text{cons})_{\ell + 1, \ell}^{[000]}$, $\ell \geq 1$ odd   & $a_{\ell + 2, \ell+1, {\bf 15}_s} = 1$ & $+$ &  $-$ \\
\hline
$(B, 1)_{2, 0}^{[200]}$  & $a_{2, 0, {\bf 20}'} = 1$ & $+$ &  $+$\\
\hline
$(B, 2)_{2, 0}^{[022]}$   & $a_{2, 0, {\bf 84}} = 1$& $+$ &  $+$ \\
\hline 
$(B, 2)_{1, 0}^{[011]}$  & $a_{1, 0, {\bf 15}_s} = 1$ & $+$ &  $-$
\end{tabular}}
\caption{A summary of the superconformal blocks and their normalizations in terms of a few OPE coefficients.  The values $a_{\Delta, \ell, {\bf r}}$ in this table correspond to $a^I_{\Delta, \ell, {\bf r}}$ in Eq.~\eqref{aCoeffExpansion}---we omitted the index $I$ for clarity. Note that the $(A,\pm)$ are complex conjugates and do not by themselves have well defined $\mathcal Z$ parity, but together they can be combined into a $\mathcal Z$-even and a $\mathcal Z$-odd structure.}
\label{SupermultipletTable}
\end{center}
\end{table}

The various long blocks are related to certain short and semishort blocks at the unitarity limit $\Delta \to \ell + 1$. These relations take the form 
 \es{LimitSpin0}{
  \mathfrak{G}_{\text{Long}_{\Delta, 0}^{[000],1} }&=  \mathfrak{G}_{(A, \text{cons})_{1,  0}^{[000]}} \,, \qquad
    \mathfrak{G}_{\text{Long}_{\Delta, 0}^{[000],2}} = \mathfrak{G}_{ (B,1)_{2,0}^{[200]}} \,,\\
    \ell \geq 1 \text{ odd:} \;\;\qquad  \mathfrak{G}_{\text{Long}_{\Delta, \ell}^{[000]}} &= \mathfrak{G}_{ (A, \text{cons})_{\ell+1,  \ell}^{[000]} }\,,\\
     \ell \geq 2 \text{ even:} \qquad \mathfrak{G}_{ \text{Long}_{\Delta, \ell}^{[000],1} }&= \mathfrak{G}_{ (A, \text{cons})_{\ell+1,  \ell}^{[000]}} \,,\\
   \mathfrak{G}_{ \text{Long}_{\Delta, \ell}^{[000],2} }&= \mathfrak{G}_{ (A,1)_{\ell+3/2,\ell-1/2}^{[100], 1}} \,, \quad    \mathfrak{G}_{ \text{Long}_{\Delta, \ell}^{[000],3}} = \mathfrak{G}_{ (A,1)_{\ell+3/2,\ell-1/2}^{[100], 2}} \,,
 }
 which respect $\mathcal{P}$ and $\mathcal{Z}$. Even though the blocks on the RHS of \eqref{LimitSpin0} involve short or semishort superconformal multiplets, they sit at the bottom of the continuum of long superconformal blocks. All other short and semishort superconformal blocks are isolated, as they cannot recombine into a long superconformal block. The distinction between isolated and non-isolated superblocks will be important when we consider the numerical bootstrap in the next section.

We will now expand the tree level correlator \eqref{finalAnswer} in superblocks. At large $c_T$ the CFT data takes the form 
\es{CFTexp}{
\lambda^2_I=\lambda^2_{I,\text{GFFT}}+\frac{1}{c_T}\lambda^2_{I,\text{tree}}+O(c_T^{-2})\,,\qquad \Delta_I=\Delta_{I,\text{GFFT}}+\frac{1}{c_T}\Delta_{I,\text{tree}}+O(c_T^{-2})\,,
}
and so using \eqref{blockExp} we find that
\es{treeExp}{
&\mathcal{S}_{\bf r}(U,V)\\
&\quad=\sum_{I\in S\times S}\left[\lambda_{I,\text{GFFT}}^2+ \frac{1}{c_T}\left(\lambda_{I,\text{tree}}^2+\lambda_{I,\text{GFFT}}^2\Delta_{I,\text{tree}}\partial_\Delta\right)+O(c_T^{-2})\right]\mathfrak{G}_I^{\bf r}(U,V)\big\vert_{\Delta=\Delta_\text{GFFT}}\,.
}
Comparing this general superblock expansion to the explicit correlator in \eqref{SSSS1/2} and \eqref{finalAnswer}, we can extract the CFT data at GFFT and tree level by expanding both sides around $U\sim 0$ and $V\sim1$. Detailed expressions are given in Appendix~\ref{Dolan}. Note that there are two cases where we cannot extract tree level CFT data from the tree level correlator. The first case is if a certain operator is degenerate at GFFT. As will be explained, this degeneracy can be lifted either by computing other correlators at tree level, or by computing $\langle SSSS\rangle$ at higher order in $1/c_T$. The second case is if an operator first appears at tree level. In this case its tree level anomalous dimension cannot be extracted from tree level $\langle SSSS\rangle$ because $\lambda_{I,\text{GFFT}}^2=0$, and so we would need to compute $\langle SSSS\rangle$ at 1-loop in order to extract the tree level anomalous dimension.

We will now show the results of the CFT data extraction. For the semishort multiplets,\footnote{We already computed the short multiplet $\lambda^{2}_{(B,2)^{[022]}_{2,0}}$ in Section \ref{LOCALIZATION} using supersymmetric localization.} we find the squared OPE coefficients:
\es{shortRes}{
&\text{$\ell\geq0$ even}:\quad\lambda^{2}_{(A,+)^{[002]}_{\ell + 5/2,\ell + 1/2}}=\frac{\pi  (\ell+1) (\ell+2) \Gamma (\ell+2)^2}{2 \Gamma
   \left(\ell+\frac{5}{2}\right)^2}\\
   &\qquad\qquad\qquad\qquad+\frac{8}{c_T}\Bigg[(2-\sin^2(\pi\lambda))\frac{4^\ell (\ell+1)^5 \Gamma \left(\frac{\ell+1}{2}\right)^4}{\pi  (\ell+2) \Gamma
   \left(\ell+\frac{5}{2}\right)^2}+\sin^2(\pi\lambda)\mathbb{S}_{(A,+)^{[002]}_{\ell + 5/2,\ell + 1/2}}\Bigg]+O(c_T^{-2})\,,\\
   &\text{$\ell\geq0$ even}:\quad\lambda^{2}_{(A,2)^{[011]}_{\ell + 2,\ell }}= \frac{\pi  (\ell+2) \Gamma (\ell+1) \Gamma (\ell+3)}{(2 \ell+3) \Gamma
   \left(\ell+\frac{3}{2}\right)^2} \\
   &\qquad\qquad\qquad\qquad+\frac{8}{c_T}\Bigg[-(2-\sin^2(\pi\lambda)) \frac{2^{2 \ell+1} (2 \ell+3) \Gamma \left(\frac{\ell+1}{2}\right)^2 \Gamma
   \left(\frac{\ell+3}{2}\right)^2}{\pi  \Gamma
   \left(\ell+\frac{5}{2}\right)^2}+\sin^2(\pi\lambda)\mathbb{S}_{(A,2)^{[011]}_{\ell + 2,\ell }}\Bigg]+O(c_T^{-2})\,,\\
   &\text{$\ell\geq0$ odd}:\quad\lambda^{2}_{(A,2)^{[011]}_{\ell + 2,\ell }}=\frac{\pi  \Gamma (\ell+2) \Gamma (\ell+4)}{\left(2 \ell^2+7 \ell+6\right) \Gamma
   \left(\ell+\frac{3}{2}\right)^2}\\
   &\qquad\qquad\qquad\qquad+\frac{8}{c_T}\Bigg[(2-\sin^2(\pi\lambda))  \frac{2^{2 \ell+1} (2 \ell+3) \Gamma \left(\frac{\ell}{2}+1\right)^4}{\pi  \Gamma
   \left(\ell+\frac{5}{2}\right)^2}+\sin^2(\pi\lambda)\mathbb{S}_{(A,2)^{[011]}_{\ell + 2,\ell }}\Bigg]+O(c_T^{-2})\,,\\
   &\text{$\ell\geq0$ even}:\quad\lambda^{2}_{(A,1)^{[100],2}_{\ell + 7/2,\ell + 3/2}}=\frac{\pi  \Gamma (\ell+3) \Gamma (\ell+5)}{\Gamma \left(\ell+\frac{5}{2}\right)
   \Gamma \left(\ell+\frac{9}{2}\right)}\\
   &\qquad\qquad\qquad\qquad+\frac{8}{c_T}\Bigg[-(2-\sin^2(\pi\lambda)) \frac{2^{2 \ell+7} \Gamma \left(\frac{\ell+3}{2}\right)^2 \Gamma
   \left(\frac{\ell+5}{2}\right)^2}{\pi  \Gamma \left(\ell+\frac{5}{2}\right)
   \Gamma \left(\ell+\frac{9}{2}\right)}+\sin^2(\pi\lambda)\mathbb{S}_{(A,1)^{[100],2}_{\ell + 7/2,\ell + 3/2}}\Bigg]+O(c_T^{-2})\,,\\
}
where the contributions $\mathbb{S}_I$  from the scalar exchange term $\mathcal{S}_\text{scal}^i$ are given in Table \ref{exchange}. Note that we did not include the result for $\lambda^{2}_{(A,1)^{[100],1}_{\ell + 7/2,\ell + 3/2}}$, since it cannot be  unambiguously extracted from $\langle SSSS\rangle$ at $O(c_T^{-1})$ due to mixing with the single trace operators, as we will discuss next.
 
\begin{sidewaystable}
\begin{center}
{\renewcommand{\arraystretch}{1.2}
\begin{tabular}{l||c|c|c|c}
 $\qquad\qquad\ell=$& $0$ & $2$ & $4$ & $6$  \\ 
\hline\hline
$\mathbb{S}_{(A,+)^{[002]}_{\ell + 5/2,\ell + 1/2}}$ & $\frac{160}{27 \pi ^2}$ & $\frac{11264}{1225 \pi ^2}$ & $\frac{104071168}{10085229 \pi ^2}$ & $\frac{60842573824}{5590245375 \pi ^2}$  \\
\hline
$\mathbb{S}_{(A,2)^{[011]}_{\ell + 2,\ell }}$ & $-\frac{32}{\pi ^2}$ & $-\frac{2048}{175 \pi ^2}$ & $-\frac{262144}{31185 \pi ^2}$ & $-\frac{2046820352}{289864575 \pi ^2}$  \\
\hline
$\mathbb{S}_{(A,2)^{[011]}_{\ell + 3,\ell+1 }}$ & $-\frac{2048}{135 \pi ^2}$ & $-\frac{65536}{3675 \pi ^2}$ & $-\frac{1375731712}{72837765 \pi ^2}$ & $-\frac{517543559168}{26609567985 \pi ^2}$ \\
\hline
$\mathbb{S}_{(A,1)^{[100],2}_{\ell + 7/2,\ell + 3/2}}$ & $-\frac{63488}{1575 \pi ^2}$ & $-\frac{29229056}{694575 \pi ^2}$ & $-\frac{318465114112}{7439857425 \pi ^2}$ & $-\frac{49162474150166528}{1138027162647375 \pi ^2}$  \\
\hline
$\mathbb{S}_{\text{Long}_{\ell+4,\ell+2}^{[000],1}}$ & $-\frac{2048 (560 \log2-613)}{7875 \pi ^2}$ & $-\frac{4194304 (45045 \log2-46819)}{1158782625 \pi ^2}$ & $-\frac{33554432 (24504480 \log2-24789439)}{4780852381305 \pi ^2}$ & $-\frac{17179869184 (1629547920 \log2-1620985787)}{157730564742926175
   \pi ^2}$  \\
\hline
$\mathbb{S}_{\text{Long}_{\ell+3,\ell+1}^{[000]}}$ & $\frac{2048}{315 \pi ^2}$ & $\frac{2490368}{848925 \pi ^2}$ & $\frac{6677331968}{4463914455 \pi ^2}$ & $\frac{820340901019648}{1138027162647375 \pi ^2}$  \\
\hline
$\mathbb{S}_{\text{Long}_{\ell+5,\ell+2}^{[000],2}}$ & $\frac{16384 (290521 \log2-166093)}{196101675 \pi ^2}$ & $\frac{262144 (8305440 \log2-4857781)}{60091156125 \pi ^2}$ & $\frac{268435456 (32580565875 \log2-19860644894)}{203408194351737375
   \pi ^2}$ & $\frac{34359738368 (16345498091104 \log2-10295061309185)}{11866103843608917978075 \pi ^2}$  \\
\hline
$\mathbb{S}_{\text{Long}_{\ell+5,\ell+2}^{[000],3}}$ & $\frac{827392}{56595 \pi ^2}$ & $\frac{82168512512}{5150670525 \pi ^2}$ & $\frac{597016157618176}{36046109224125 \pi ^2}$ & $\frac{101715902478914945024}{6013025156384371125 \pi ^2}$  \\
\end{tabular}}
\caption{Contribution $\mathbb{S}_I$ of the scalar exchange diagram $\mathcal{S}_\text{scal}^i$ defined in \eqref{finalAnswer2} to various OPE coefficients squared. Note that the $\log2$ terms for CFT data with nonzero tree level anomalous dimensions come from the $4^\Delta$ factor in the definition of our conformal block \cite{Kos:2013tga}.}
\label{exchange}
\end{center}
\end{sidewaystable}
 
For the long multiplets, we first consider the single trace approximately conserved current multiplets with superprimary $B_\ell$, starting with $\ell=0$. For generic $\lambda$ when parity is not a symmetry, we expect this multiplet at $c_T\to\infty$ to contribute to both $n=1,2$ structures of the $\mathfrak{G}_{\text{Long}_{\Delta, 0}^{[000],n}}$ superblock at unitarity $\Delta=1$, where recall from \eqref{LimitSpin0} that we can formally identify $\mathfrak{G}_{\text{Long}_{1, 0}^{[000],2}}=\mathfrak{G}_{(B,1)_{2,0}^{[200]}} $ and $\mathfrak{G}_{\text{Long}_{1, 0}^{[000],1}}=\mathfrak{G}_{{(A,\text{cons})}_{1, 0}^{[000]}}$. For each structure, we find the OPE coefficients 
\es{longp0OPE}{
\lambda^2_{\text{Long}_{1, 0}^{[000],1}}&=\frac{64}{c_T}(1-\sin^2(\pi\lambda))+O(c_T^{-2})\,,\\
\lambda^2_{\text{Long}_{1, 0}^{[000],2}}&=\frac43+O(c_T^{-1})\,,\\
}
where the $n=2$ structure starts at $O(c_T^{0})$ since $\mathfrak{G}_{\text{Long}_{1, 0}^{[000],2}}=\mathfrak{G}_{(B,1)_{2,0}^{[200]}} $ appears in the GFFT, while the $n=1$ starts at $O(c_T^{-1})$ since $\mathfrak{G}_{\text{Long}_{1, 0}^{[000],1}}=\mathfrak{G}_{{(A,\text{cons})}_{1, 0}^{[000]}}$ does not appear at GFFT. Note that $\lambda^2_{\text{Long}_{1, 0}^{[000],1}}$ is what we called $\lambda_{SSB_0}$ from Section \ref{weakHS}, and vanishes for the parity preserving $\lambda=1/2$ theory as discussed before. For $N>1$ we cannot unambiguously determine the $O(c_T^{-1})$ correction to $\lambda^2_{\text{Long}_{1, 0}^{[000],2}}$ using just tree level $\langle SSSS\rangle$,\footnote{For $N=1$, the unambiguous tree level correction will then be
\es{longp0OPEnew}{
\lambda^2_{\text{Long}_{1, 0}^{[000],2}}&=\frac43+\frac{8}{c_T}\Bigg[-\frac43(2-\sin^2(\pi\lambda)) +\frac{32}{3\pi^2}\sin^2(\pi\lambda)\Bigg]+O(c_T^{-2})\,.\\
}
}  since we cannot distinguish it from the correction to $\lambda^{2}_{(B,1)_{2,0}^{[200]}}$, which can be explicitly constructed in any $U(N)_{-k}\times U(N+M)_{k}$ theory with $N>1$.\footnote{In particular, at GFFT one can construct two $(B,1)_{2,0}^{[200]}$ operators, one using adjoints of the $SU(N)$ gauge group factor and one using singlets. The latter $(B,1)_{2,0}^{[200]}$ is what is eaten by the conserved current at tree level, while the former remains. For $N=1$, there is of course no adjoint, which is why the extra $(B,1)_{2,0}^{[200]}$ does not exist. This nonexistence was in fact observed using the numerical bootstrap in \cite{Binder:2020ckj}} To unmix these degenerate operators, we would need to compute $\langle SSSS\rangle$ at $O(c_T^{-2})$, in which case the $O(c_T^{-1})$ correction to $\lambda^2_{\text{Long}_{1, 0}^{[000],2}}$ will multiply the anomalous dimension, so that it can be unambiguously read off. The $O(c_T^{-1})$ anomalous dimension should be the same for either structure, but in practice we can only extract it from tree level $\langle SSSS\rangle$ using the $\mathfrak{G}_{{\text{Long}_{1, 0}^{[000],2}}}$ structure, because that is the only structure whose OPE coefficient is $O(c_T^0)$. From this structure we find
\es{longp0}{
\Delta_{(0,2)}&=1+\frac{128}{\pi^2c_T}\sin^2(\pi\lambda)+O(c_T^{-2})\,.\\
}
For the $\mathfrak{G}_{{\text{Long}_{1, 0}^{[000],1}}}$ structure, the tree level anomalous dimension would first appear in $\langle SSSS\rangle$ at $O(c_T^{-2})$, since the leading order OPE coefficient starts at $O(c_T^{-1})$.

Next, we consider the single trace approximately conserved currents with superprimary $B_\ell$ and even $\ell>0$. For generic $\lambda$, parity is not a symmetry and so we expect this multiplet at $c_T\to\infty$ to contribute to both $n=1,2$ structures of the $\mathfrak{G}_{\text{Long}_{\Delta, \ell}^{[000],n}}$ superblock at unitarity $\Delta=\ell+1$, where recall from \eqref{LimitSpin0} that we can formally identify $\mathfrak{G}_{\text{Long}_{\ell+1,\ell}^{[000],2}} =\mathfrak{G}_{(A,1)_{\ell+3/2,\ell-1/2}^{[100],1} }$ and $\mathfrak{G}_{\text{Long}_{\ell+1,\ell}^{[000],1}}=\mathfrak{G}_{{(A,\text{cons})}_{\ell+1, \ell}^{[000]}}$. For each structure, we find the OPE coefficients 
\es{longp0OPE}{
\text{$\ell>0$ even}:\quad\lambda^2_{\text{Long}_{\ell+1, \ell}^{[000],1}}&=\frac{64}{c_T}(1-\sin^2(\pi\lambda))+O(c_T^{-2})\,,\\
\text{$\ell>0$ even}:\quad\lambda^2_{\text{Long}_{\ell+1,\ell}^{[000],2}}&=\frac{\pi  (\ell+2) \Gamma (\ell+3) \Gamma (\ell+4)}{3 \Gamma
   \left(\ell+\frac{5}{2}\right) \Gamma \left(\ell+\frac{9}{2}\right)}+O(c_T^{-1})\,,\\
}
where the $n=2$ structure start at $O(c_T^{0})$ since $\mathfrak{G}_{\text{Long}_{\ell+1,\ell}^{[000],2}} =\mathfrak{G}_{(A,1)_{\ell+3/2,\ell-1/2}^{[100],1} }$ appears in the GFFT, while the $n=1$ starts at $O(c_T^{-1})$ since $\mathfrak{G}_{\text{Long}_{\ell+1,\ell}^{[000],1}}=\mathfrak{G}_{{(A,\text{cons})}_{\ell+1, \ell}^{[000]}}$ does not appear at GFFT. Note that $\lambda^2_{\text{Long}_{\ell+1,\ell}^{[000],1}}$ is what we called $\lambda_{SSB_\ell}$ from Section \ref{weakHS}, and vanishes for the parity preserving $\lambda=1/2$ theory as discussed before. We did not write the $O(c_T^{-1})$ correction to $\lambda^2_{\text{Long}_{\ell+1,\ell}^{[000],2}}$, since we cannot distinguish it from the correction to $\lambda^{2}_{(A,1)^{[100],1}_{\ell + 7/2,\ell + 3/2}}$, which we know exists for all $\mathcal{N}=6$ theories \cite{Binder:2020ckj}, using just tree level $\langle SSSS\rangle$. To unmix these degenerate operators, we would need to compute $\langle SSSS\rangle$ at $O(c_T^{-2})$, in which case the $O(c_T^{-1})$ correction to $\lambda^2_{\text{Long}_{\ell+1,\ell}^{[000],2}}$ will multiply the anomalous dimension, so that it can be unambiguously read off. The $O(c_T^{-1})$ anomalous dimension should be the same for either structure, but in practice we can only extract it from tree level $\langle SSSS\rangle$ using the $\mathfrak{G}_{\text{Long}_{\ell+1, \ell}^{[000],2}}$ structure, because that is the only structure which contributes at $O(c_T^0)$. From this structure we find
\es{longpbigger}{
\text{$\ell>0$ even}:\quad\Delta_{(\ell,2)}&=\ell+1+\frac{8 \ell (2 \ell+1) (2 \ell+3)^2 (2 \ell+5) (2 \ell+7)}{\pi ^2 (\ell+1)^2 (\ell+2)^3
   (\ell+3)^2c_T}\sin^2(\pi\lambda)+O(c_T^{-2})\,.\\
}
For the ${\text{Long}_{\ell+1, \ell}^{[000],1}}$ structure, the tree level anomalous dimension first contributes to $\langle SSSS\rangle$ at $O(c_T^{-2})$, as discussed above.

Finally, we consider the single trace approximately conserved current multiplets with superprimary $B_\ell$ for odd $\ell>0$. In this case there is just a single structure, which from \eqref{LimitSpin0} is identified at unitarity with $\mathfrak{G}_{\text{Long}_{\ell+1,\ell}^{[000]}}=\mathfrak{G}_{{(A,\text{cons})}_{\ell+1, \ell}^{[000]}}$. We find the tree level OPE coefficient
\es{longp0OPEodd}{
\text{$\ell>0$ odd}:\quad\lambda^2_{\text{Long}_{\ell+1, \ell}^{[000]}}&=\frac{64}{c_T}+O(c_T^{-2})\,,\\
}
which is what we called $\lambda_{SST_{\ell+3}}$ from Section \ref{weakHS}, and does not depend on $\lambda$ as discussed before. We would need to compute $\langle SSSS\rangle$ at $O(c_T^{-2})$ in order to extract the tree level anomalous dimension.

We now move on to the double trace long multiplets. We will only consider the lowest twist in each sector, since higher twist double trace long multiplets are expected to be degenerate, so we cannot extract them from just $\langle SSSS\rangle$. For twist two, we find that only $\mathfrak{G}_{\text{Long}_{\ell+2, \ell}^{[000],1}}$ receives contributions for all even $\ell$:
\es{longRes}{
\text{$\ell\geq0$ even}:\quad\Delta_{(\ell,1)}&=\ell+2-\frac{128 (2 \ell+3) (2 \ell+5)}{\pi ^2 (\ell+1) (\ell+3) (\ell+4)c_T}\sin^2(\pi\lambda)+O(c_T^{-2})\,,\\
\lambda^2_{\text{Long}_{\ell+2,\ell}^{[000],1}}&=\frac{\pi  (\ell+4) \Gamma (\ell+1) \Gamma (\ell+3)}{2 \Gamma
   \left(\ell+\frac{1}{2}\right) \Gamma \left(\ell+\frac{7}{2}\right)}\\
   &+\frac{8}{c_T}\Bigg[-(2-\sin^2(\pi\lambda))\frac{4^{\ell+1} \ell\Gamma \left(\frac{\ell+1}{2}\right)^2 \Gamma
   \left(\frac{\ell+3}{2}\right)^2}{\pi  \Gamma \left(\ell+\frac{1}{2}\right)
   \Gamma \left(\ell+\frac{7}{2}\right)} +\sin^2(\pi\lambda)\mathbb{S}_{\text{Long}_{\ell+2,\ell}^{[000],1}}\Bigg]+O(c_T^{-2})\,.\\
}
For odd $\ell$ at twist two, only $\mathfrak{G}_{\text{Long}_{\ell+2, \ell}^{[000]}}$ receives contributions:
\es{longResOdd}{
\text{$\ell>0$ odd}:\quad\Delta_{(\ell,1)}&=\ell+2+O(c_T^{-2})\,,\\
\lambda^2_{\text{Long}_{\ell+2,\ell}^{[000]}}&=\frac{\pi  \ell \Gamma (\ell+2) \Gamma (\ell+4)}{2 \Gamma
   \left(\ell+\frac{5}{2}\right) \Gamma \left(\ell+\frac{7}{2}\right)}\\
   &+\frac{8}{c_T}\Bigg[(2-\sin^2(\pi\lambda))\frac{4^{\ell+2} \Gamma \left(\frac{\ell}{2}+1\right)^2 \Gamma
   \left(\frac{\ell}{2}+2\right)^2}{\pi  \Gamma \left(\ell+\frac{5}{2}\right)
   \Gamma \left(\ell+\frac{7}{2}\right)} +\sin^2(\pi\lambda)\mathbb{S}_{\text{Long}_{\ell+2,\ell}^{[000]}}\Bigg]+O(c_T^{-2})\,,\\
}
where note that the tree level correction to the anomalous dimension vanished. For twist three, we find that both $\mathfrak{G}_{\text{Long}_{\ell+3, \ell}^{[000],2}}$ and $\mathfrak{G}_{\text{Long}_{\ell+3, \ell}^{[000],3}}$ receives contributions for all even $\ell$, though only the former receives an anomalous dimension: 
\es{others}{
\text{$\ell\geq0$ even}:&\quad\;\;\Delta'_{(\ell,2)}=\ell+3+\frac{128 (2 \ell+5) (2 \ell (\ell+4)+5)}{\pi ^2 (\ell+1) (\ell+3) (\ell+4) (\ell+5)c_T}\sin^2(\pi\lambda)+O(c_T^{-2})\,,\\
&\lambda^2_{\text{Long}_{\ell+3,\ell}^{[000],2}}=\frac{\pi  \Gamma (\ell+3) \Gamma (\ell+4)}{3 (2 \ell+3) \Gamma
   \left(\ell+\frac{1}{2}\right) \Gamma \left(\ell+\frac{9}{2}\right)}\\
   &+\frac{8}{c_T}\Bigg[(2-\sin^2(\pi\lambda)) \frac{4^{\ell+3} (\ell+2) \Gamma \left(\frac{\ell+1}{2}\right) \Gamma
   \left(\frac{\ell+3}{2}\right) \Gamma \left(\frac{\ell+5}{2}\right)^2}{3 \pi 
   (\ell+4) (2 \ell+3) \Gamma \left(\ell+\frac{1}{2}\right) \Gamma
   \left(\ell+\frac{9}{2}\right)}+\sin^2(\pi\lambda)\mathbb{S}_{\text{Long}_{\ell+3,\ell}^{[000],2}}\Bigg]+O(c_T^{-2})\,,\\
   &\lambda^2_{\text{Long}_{\ell+3,\ell}^{[000],3}}=\frac{\pi  \Gamma (\ell+2) \Gamma (\ell+5)}{(2 \ell+3) \Gamma
   \left(\ell+\frac{1}{2}\right) \Gamma \left(\ell+\frac{9}{2}\right)}\\
   &+\frac{8}{c_T}\Bigg[(2-\sin^2(\pi\lambda)) \frac{4^{\ell+3} \Gamma \left(\frac{\ell+1}{2}\right) \Gamma
   \left(\frac{\ell+3}{2}\right) \Gamma \left(\frac{\ell+5}{2}\right)^2}{\pi 
   (2 \ell+3) \Gamma \left(\ell+\frac{1}{2}\right) \Gamma
   \left(\ell+\frac{9}{2}\right)}+\sin^2(\pi\lambda)\mathbb{S}_{\text{Long}_{\ell+3,\ell}^{[000],3}}\Bigg]+O(c_T^{-2})\,,\\
}
where $\Delta'_{(\ell,2)}$ denotes that these are the second lowest dimension operators in their sector that we consider, after the single trace operators with twist one.

\subsection{Comparing $U(1)_{2M}\times U(1+M)_{-2M}$ to Numerical Bootstrap}


In \cite{Binder:2020ckj} it was shown that the exact formula for $\lambda^2_{(B,2)_{2,0}^{[022]}}$ in the $U(1)_{2M}\times U(1+M)_{-2M}$ theory, as computed from localization for finite $M$, was close to saturating the lower bound on this quantity derived from the numerical bootstrap for all values of $M$.\footnote{Note furthermore that at large $c_T$, $\lambda^2_{(B,2)_{2,0}^{[022]}}$ is smaller for ${U(1)_{2M}\times U(1+M)_{-2M}}$ than all other $U(N)_k\times U(N+M)_{-k}$ or $SO(2)_{2k}\times USp(2+2M)_{-k}$ theories, as can be seen by comparing the $O(c_T^{-2})$ corrections to \eqref{UNs} and \eqref{cTLamSO}. Hence if the bootstrap bound is saturated by a known theory then it must be saturated by the ${U(1)_{2M}\times U(1+M)_{-2M}}$.} This motivated the conjecture that in the limit of infinite numerical precision, the ${U(1)_{2M}\times U(1+M)_{-2M}}$ theory exactly saturates the bound. If this conjecture is true, the low-lying spectrum of this theory for any $M \sim c_T$ can be read off from the extremal functional. In this section, we will test this conjecture by comparing the numerical bootstrap results to the tree level CFT data for the $U(1)_{2M}\times U(1+M)_{-2M}$ theory computed in the previous section, which corresponds to the case $\lk = 1/2$. Note that the $U(1)_{2M}\times U(1+M)_{-2M}$ is special not only because it has the minimal value of $\lambda^2_{(B,2)_{2,0}^{[022]}}$ at fixed $c_T$ for all known theories, but also because it is parity invariant due to Seiberg duality.

\begin{figure}[]
\begin{center}
        \includegraphics[width=.49\textwidth]{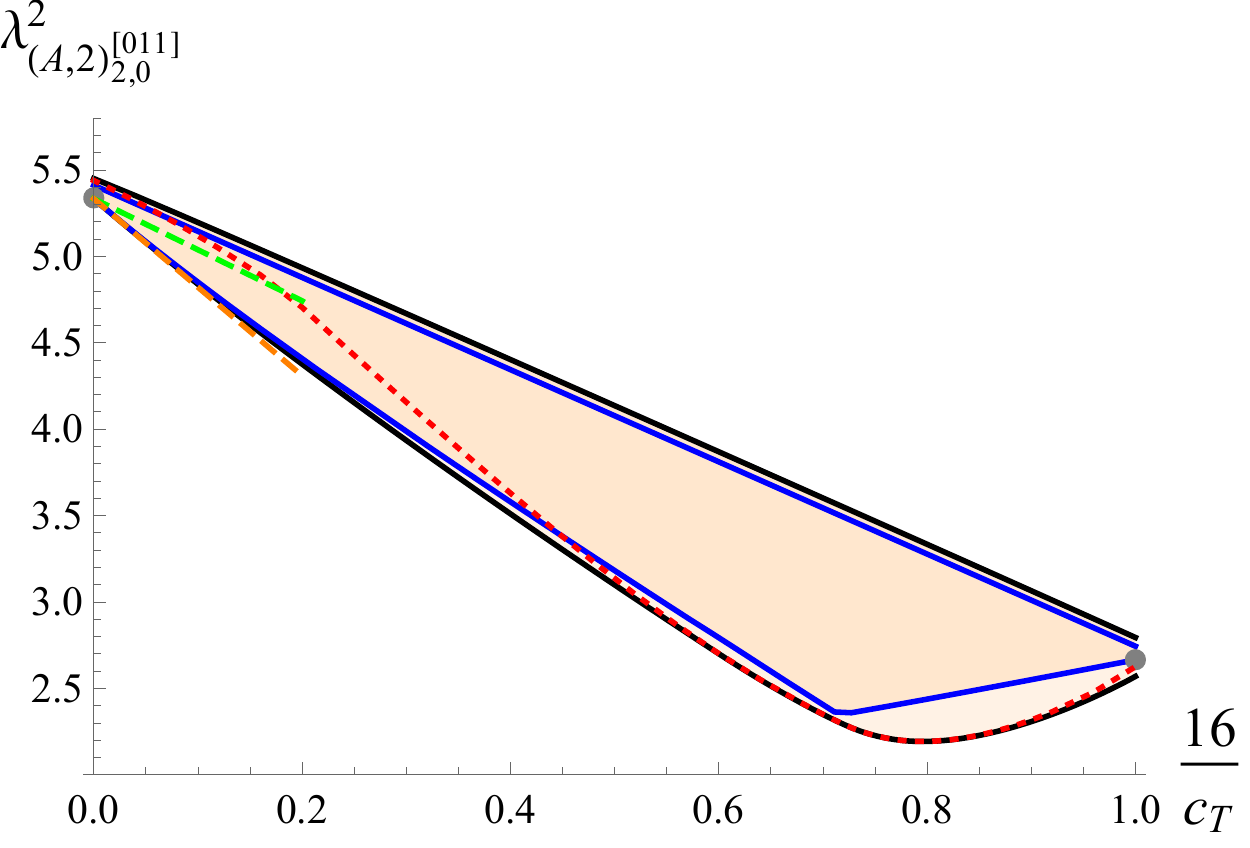} \includegraphics[width=.49\textwidth]{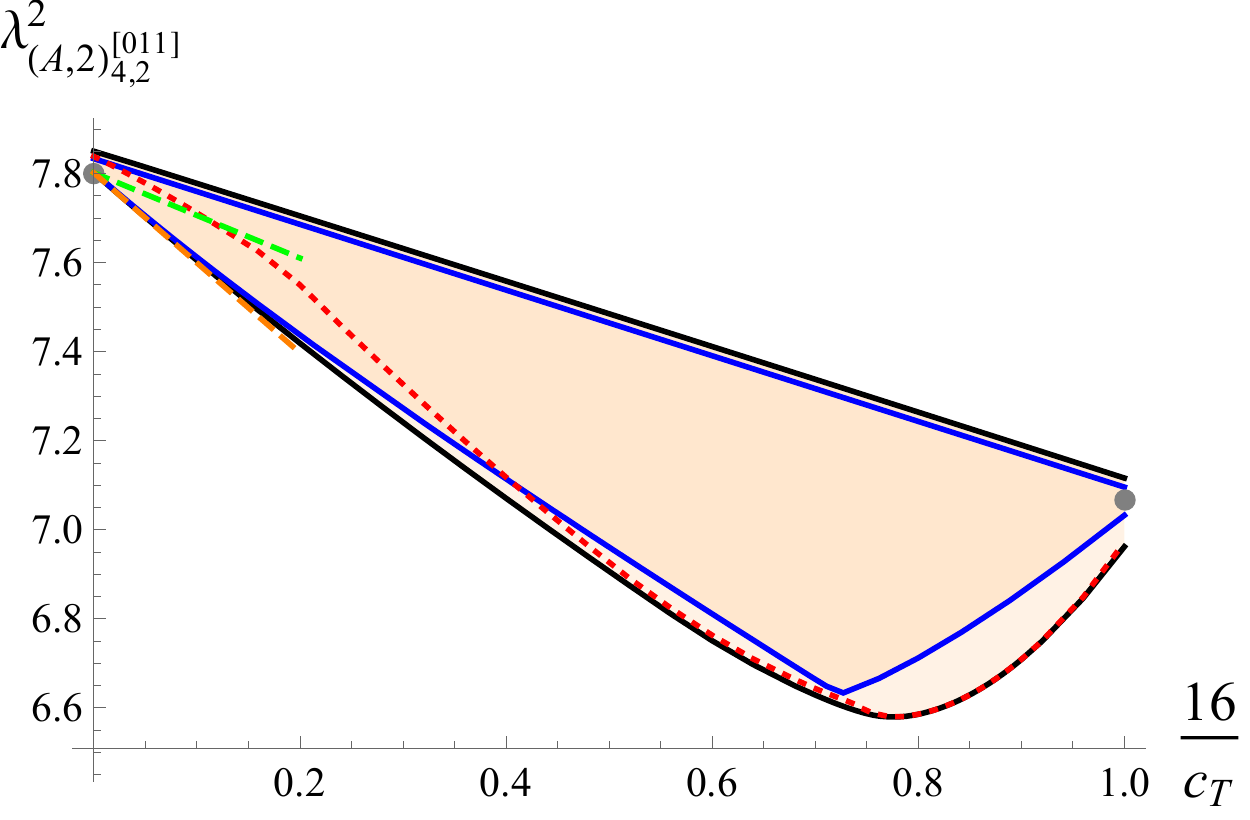}
           \includegraphics[width=.49\textwidth]{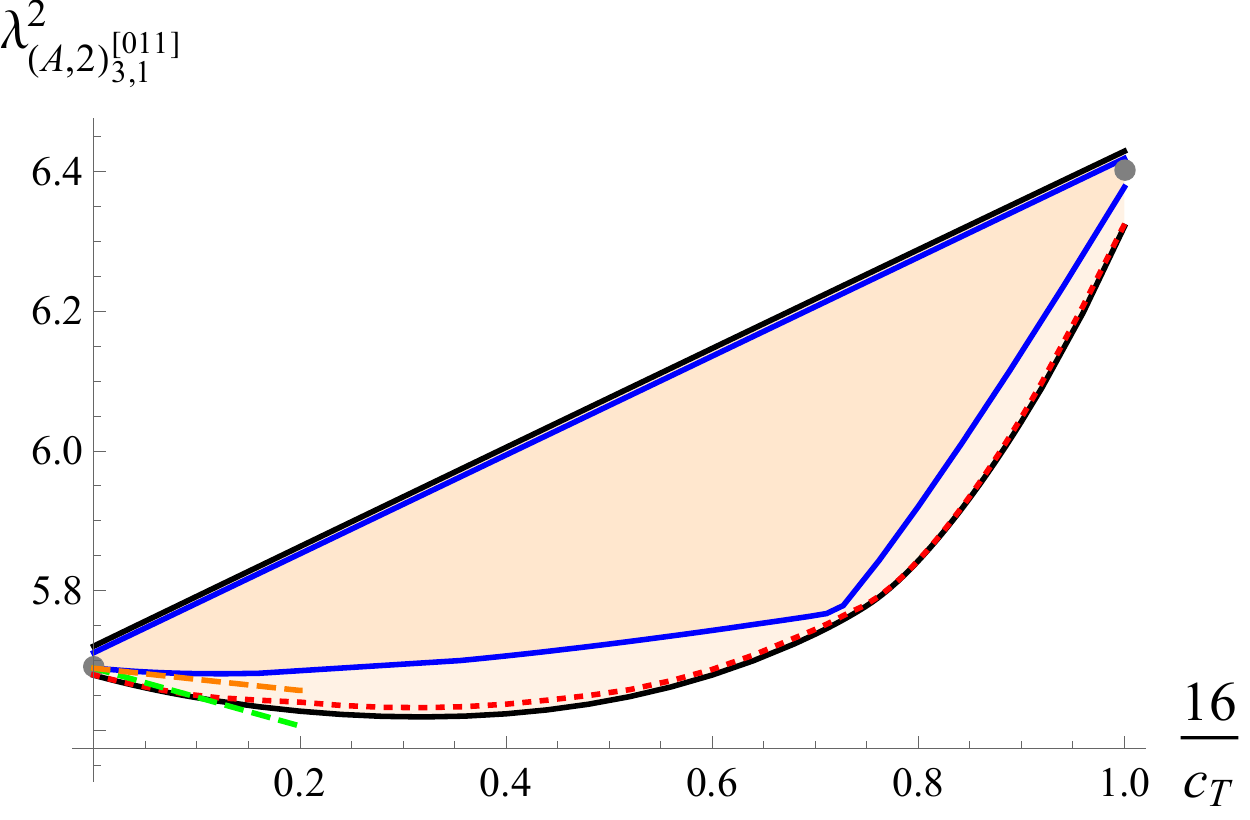} \includegraphics[width=.49\textwidth]{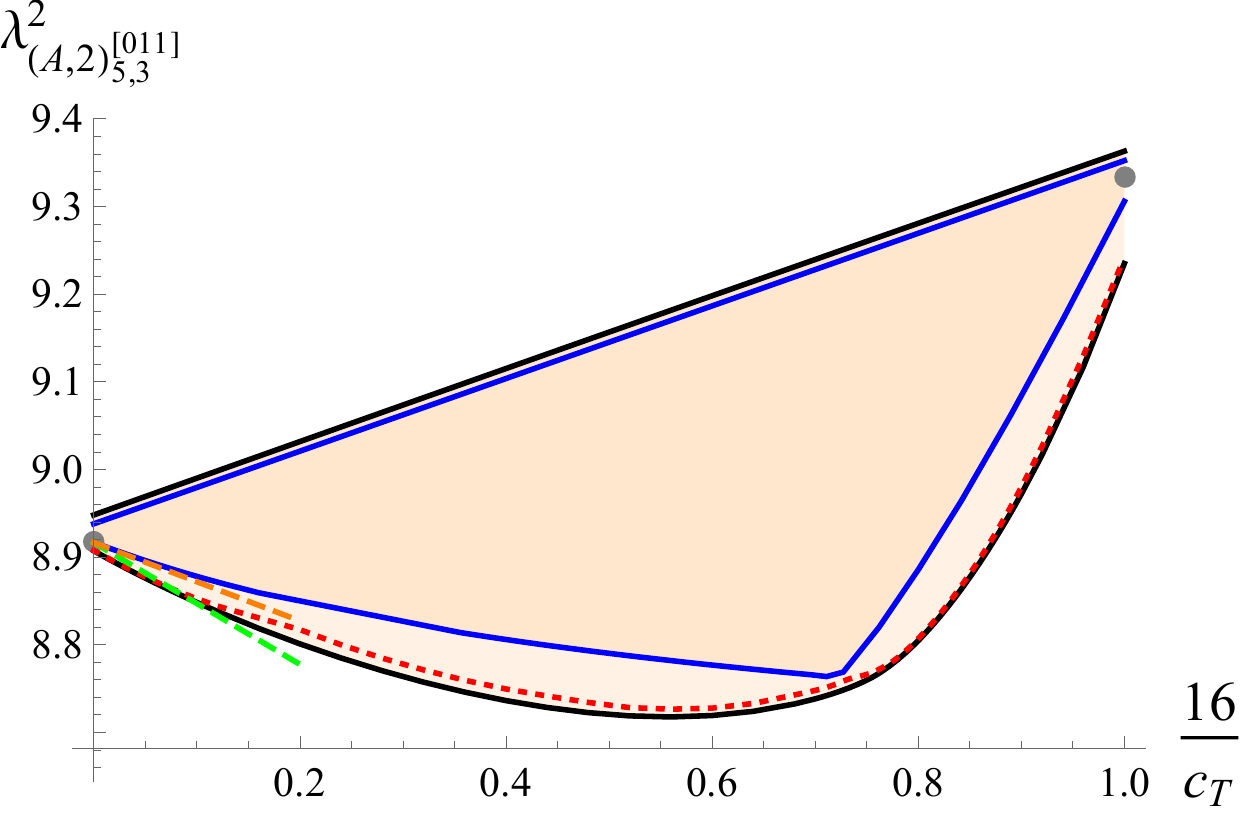}
              \includegraphics[width=.49\textwidth]{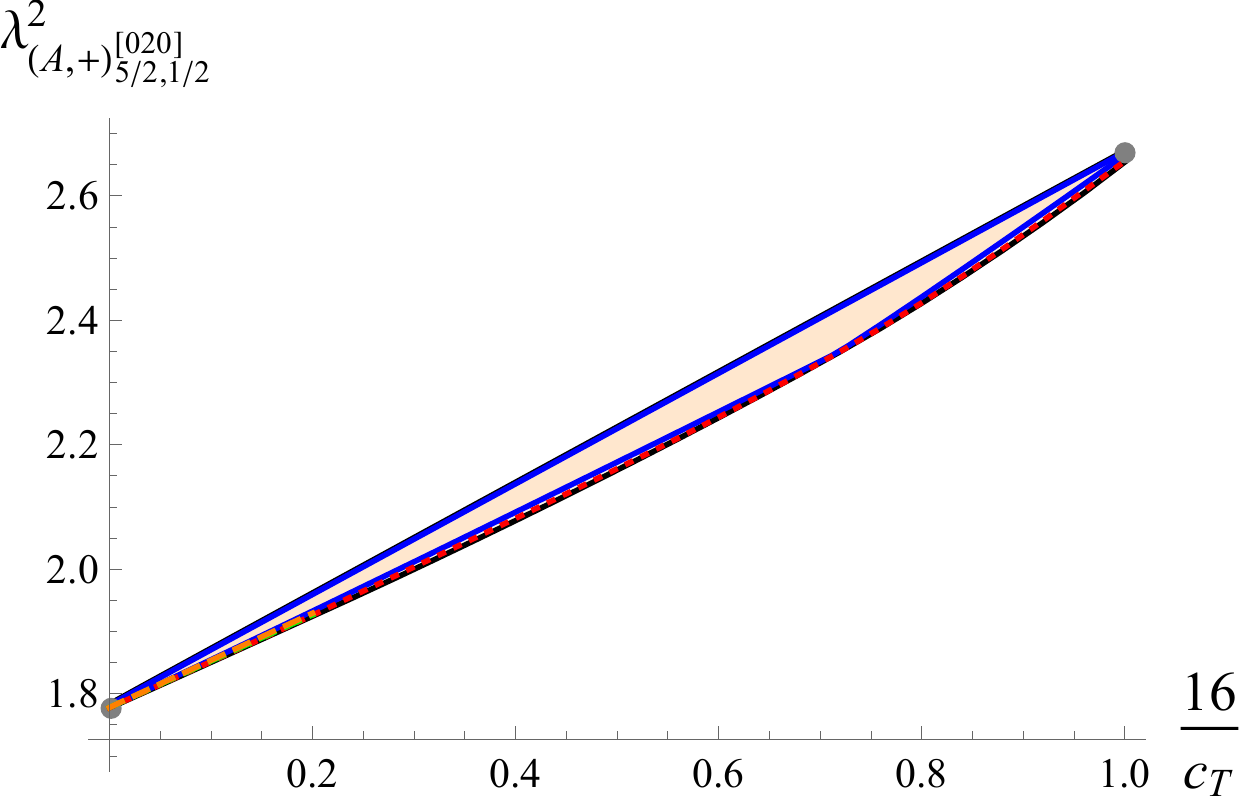} \includegraphics[width=.49\textwidth]{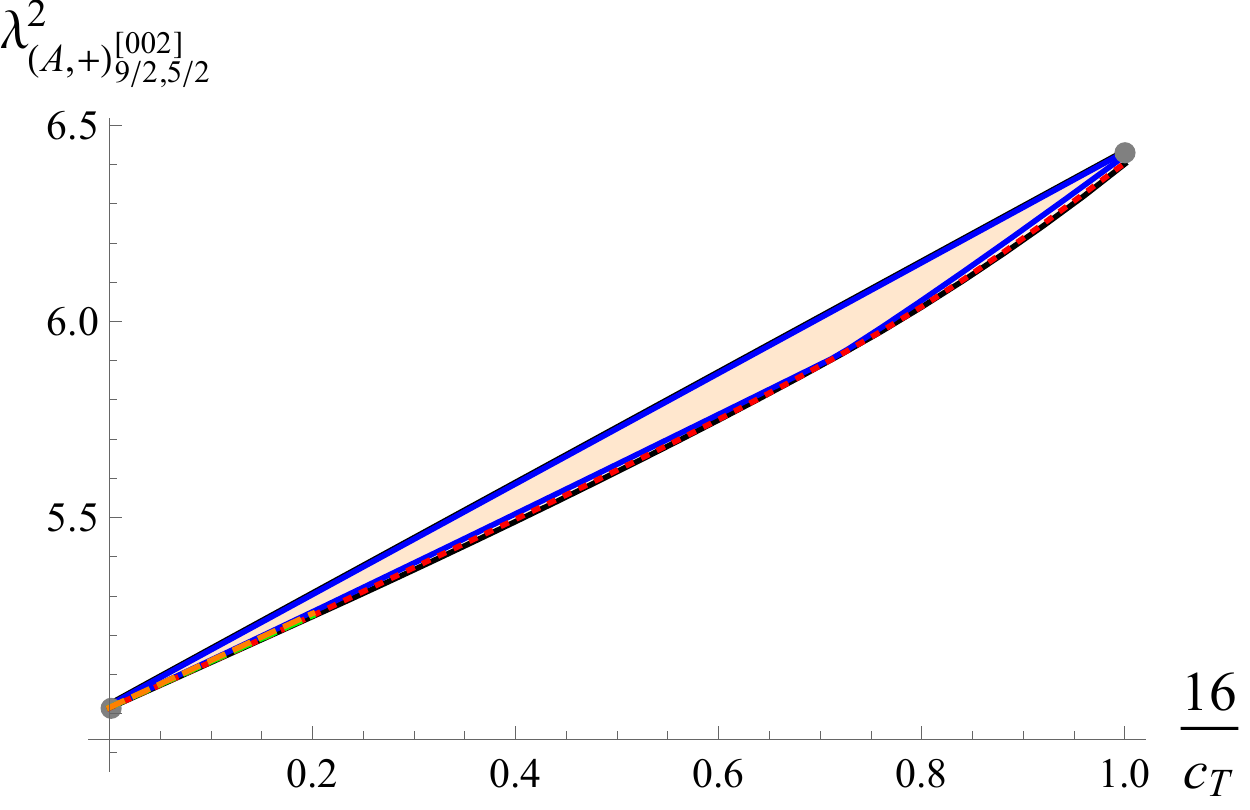}
\caption{Upper and lower bounds on various semishort OPE coefficients squared in terms of $c_T$, where the orange shaded regions are allowed, and the plots ranges from the GFFT limit $c_T\to\infty$ to the free theory $c_T=16$. The {\bf black} lines denote the $\mathcal{N}=6$ upper/lower bounds computed in \cite{Binder:2020ckj} with $\Lambda=39$, the {\bf \Blue blue} lines denote the $\mathcal{N}=8$ upper/lower bounds computed in \cite{Agmon:2017xes} with $\Lambda=43$. The {\bf\Red red} dotted lines denotes the spectrum read off from the functional saturating the lower bound on $\lambda^2_{(B,2)_{2,0}^{[022]}}$ in \cite{Binder:2020ckj}, which was identified with the $U(1)_{2M}\times U(1+M)_{-2M}$ theory. The {\bf\Green green} dashed lines denote the $O(c_T^{-1})$ correction for the $U(1)_{2M}\times U(1+M)_{-2M}$ theory computed in this work, while the {\bf\Orange orange} dashed lines denote the $O(c_T^{-1})$ correction for the supergravity limit of ABJM theory as computed in \cite{Zhou:2017zaw,Chester:2018lbz}. The gray dots denote the GFFT and free theory values.}
\label{OPEfig}
\end{center}
\end{figure}  

In Figure \ref{OPEfig}, we show numerical bootstrap bounds for the squared OPE coefficients of semishort multiplets in $S\times S$ that are isolated from the continuum of long operators. This includes all semishort multiplets in Table \ref{SupermultipletTable} except for $(A,1)_{\ell+2,\ell}^{[100],n}$. We include both the general $\mathcal{N}=6$ numerical bounds shown in black, the general $\mathcal{N}=8$ numerical bounds from \cite{Agmon:2017xes} shown in blue, and the conjectured $U(1)_{2M}\times U(1+M)_{-2M}$ spectrum shown in red. The tree level results for $U(1)_{2M}\times U(1+M)_{-2M}$ are shown in green, while the tree level results in the supergravity limit as computed in \cite{Zhou:2017zaw,Chester:2018lbz} are shown in orange. Recall that the supergravity results apply to the leading large $c_T$ correction to both the M-theory and string theory limits. As first noted in \cite{Chester:2018lbz} and visible in these plots, they match the large $c_T$ regime of the $\mathcal{N}=8$ lower bounds.\footnote{We have converted the $\mathcal{N}=8$ results in \cite{Chester:2018lbz} to $\mathcal{N}=6$ using the superblock decomposition given in Appendix D of \cite{Binder:2020ckj}.} For $\mathcal{N}=6$, we see in all these plots that the tree level results approximately match the conjectured $U(1)_{2M}\times U(1+M)_{-2M}$ spectrum in the large $c_T$ regime. Curiously, the conjectured spectrum approximately coincides with the $\mathcal{N}=6$ lower bounds for $\lambda^2_{(A,+)_{\ell+5/2,\ell+1/2}^{[020]}}$ and $\lambda^2_{(A,2)_{\ell,\ell+2}^{[011]}}$ with odd $\ell$, but not for $\lambda^2_{(A,2)_{\ell,\ell+2}^{[011]}}$ with even $\ell$.\footnote{Recall that, as described in Table~\ref{SupermultipletTable}, the superblocks for $\lambda^2_{(A,2)_{\ell,\ell+2}^{[011]}}$ have completely different structures for even/odd values of $\ell$.} 

 \begin{table}[htp]
\begin{center}
\begin{tabular}{|c|c|c|}
\hline
 \multicolumn{1}{|c|}{CFT data}  &  Numerical bootstrap spectrum & Analytic Tree $U(1)_{2M}\times U(1+M)_{-2M}$   \\
  \hline
   $\lambda^2_{(A,+)_{5/2,1/2}^{[002]}}$  & $11.9$ & $\frac{64}{9}+\frac{1280}{27 \pi ^2}\approx11.9145$\\
  $\lambda^2_{(A,+)_{9/2,5/2}^{[002]}}$   & $18.7$ & $\frac{13824}{1225}+\frac{90112}{1225 \pi ^2}\approx18.7382$\\
 $\lambda^2_{(A,2)_{2,0}^{[011]}}$   & $-45$ & $-\frac{64}{3}-\frac{256}{\pi ^2}\approx-47.2716$\\
  $\lambda^2_{(A,2)_{4,2}^{[011]}}$    & $-18$& $-\frac{1024}{175}-\frac{16384}{175 \pi ^2}\approx-15.3374$\\
  $\lambda^2_{(A,2)_{3,1}^{[011]}}$  & $-6$& $\frac{256}{45}-\frac{16384}{135 \pi ^2}\approx-6.60775$\\
  $\lambda^2_{(A,2)_{5,3}^{[011]}}$ & $-10$& $\frac{4096}{1225}-\frac{524288}{3675 \pi ^2}\approx-11.1112$\\
  $\Delta_{(0,1)}$ &  $-16$& $-\frac{160}{\pi ^2}\approx-16.2114$\\
  $\Delta_{(0,2)}$ &  $16$& $\frac{128}{\pi ^2}\approx12.9691$\\
  \hline
\end{tabular}
\end{center}
\caption{The $1/c_T$ correction to the scaling dimensions $\Delta_{0,1}$ and $\Delta_{0,2}$ for the lowest dimension $\text{Long}_{\Delta, 0}^{[000], 1}$ and $\text{Long}_{\Delta, 0}^{[000], 2}$ operators, respectively, as well as the OPE coefficients squared of various semishort operators. The numerical bootstrap results come from a large $c_T$ fit to the numerical bootstrap spectrum of \cite{Binder:2020ckj} that was conjectured to apply to the $U(1)_{2M}\times U(1+M)_{-2M}$, and corresponds to the dashed red lines in Figure \ref{OPEfig} and \ref{scalfig}. The analytic tree level results were computed in this work in the previous section. }
\label{comparisonTable}
\end{table}

The $\mathcal{N}=6$ numerics are not completely converged yet, which can be seen from the fact that at $c_T\to\infty$ the numerics do not exactly match the GFFT value shown as a grey dot. On the other hand, it has been observed in many previous numerical bootstrap studies \cite{Beem:2013qxa,Beem:2016wfs,Beem:2015aoa,Chester:2014fya,Chester:2014mea,Agmon:2017xes} that the bounds change uniformly as precision is increased, so that the large $c_T$ slope is still expected to be accurate, even if the intercept is slightly off. In Table \ref{comparisonTable}, we compare the coefficient of the $1/c_T$ term as read off from the numerics at large $c_T$ to the tree level results, and find a good match for all data. The match is especially good for the most protected quantities, which are the $1/4$-BPS $\lambda^2_{(A,+)_{\ell+5/2,\ell+1/2}^{[020]}}$. In fact, this quantity is so constrained that it is difficult to distinguish by eye between the $\mathcal{N}=8$ and $\mathcal{N}=6$ numerical and analytical results in Figure \ref{OPEfig}. Nevertheless, the exact tree correction for supergravity and tree level $U(1)_{2M}\times U(1+M)_{-2M}$ are different. For instance, compare the supergravity value $\frac{17920}{9 \pi ^2}-\frac{5120}{27}\approx12.1121$ from \cite{Chester:2018lbz} for $\lambda^2_{(A,+)_{5/2,1/2}^{[020]}}$ to the corresponding $U(1)_{2M}\times U(1+M)_{-2M}$ value $\frac{64}{9}+\frac{1280}{27 \pi ^2}\approx11.9145$ shown in Table \ref{comparisonTable}.

\begin{figure}[]
\begin{center}
          \includegraphics[width=.49\textwidth]{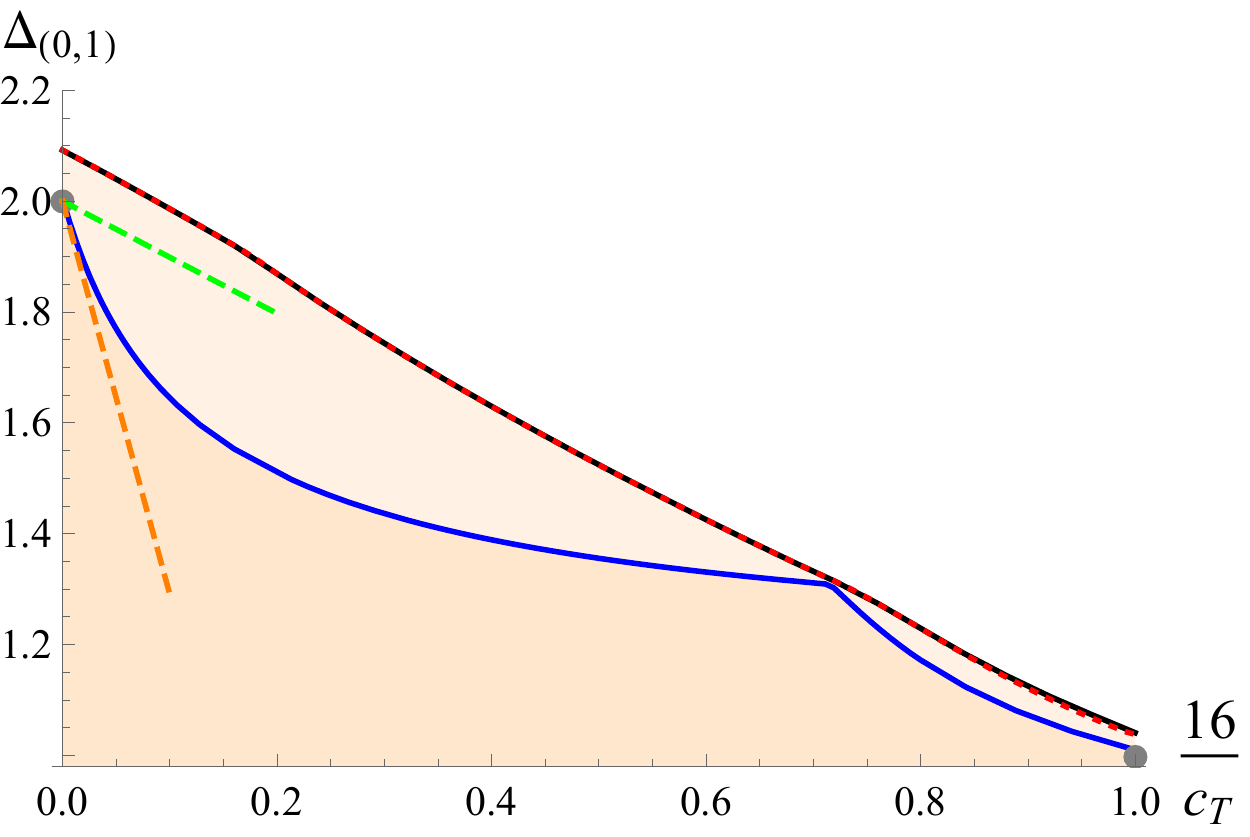} \includegraphics[width=.49\textwidth]{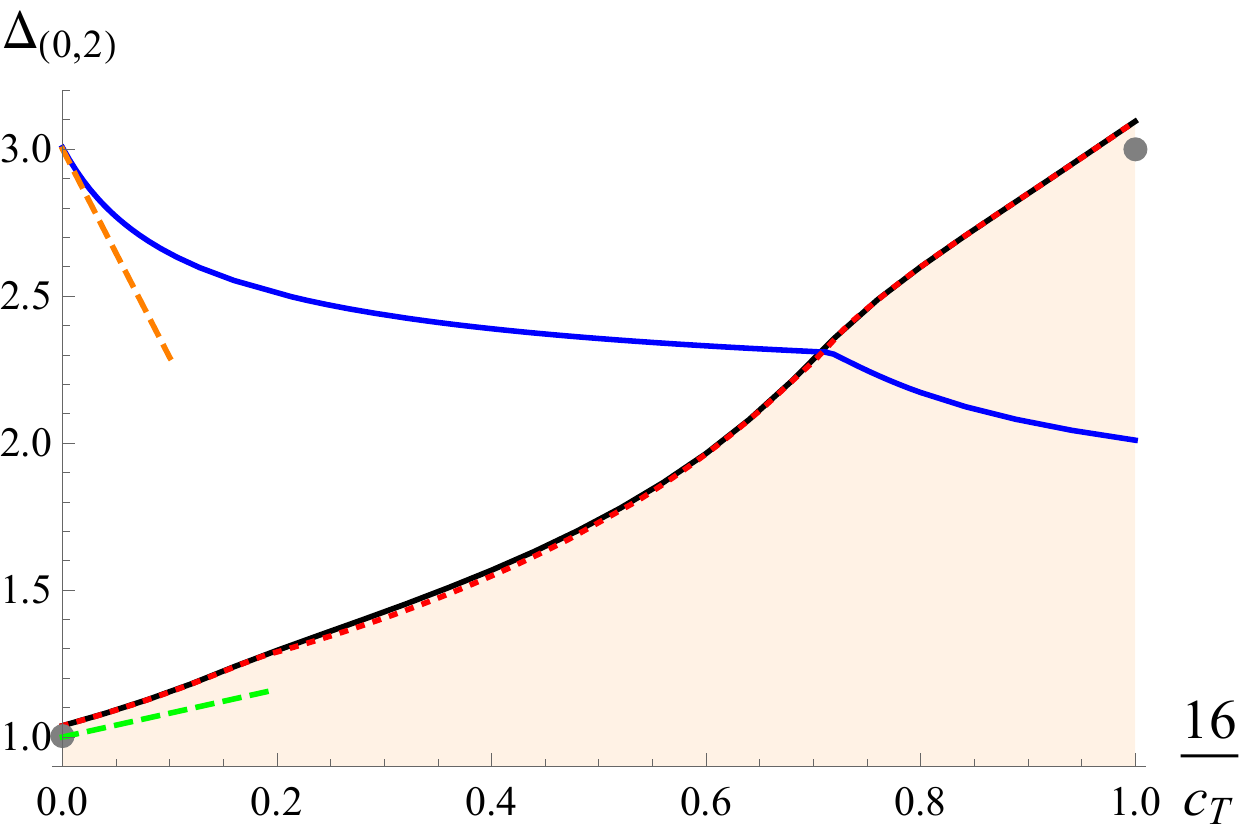}
\caption{Upper bounds on the scaling dimension of the lowest dimension $\ell=0$ long multiplet in terms of $c_T$ for the $\text{Long}_{\Delta, 0}^{[000], 1}$ ({\bf left}) and $\text{Long}_{\Delta, 0}^{[000], 2}$ ({\bf right}) superconformal structures, which for parity preserving theories has the same and opposite parity as the superprimary, respectively. The orange shaded region is allowed, and the plot ranges from the GFFT limit $c_T\to\infty$ to the free theory $c_T=16$. The {\bf black} lines denote the $\mathcal{N}=6$ upper/lower bounds computed in \cite{Binder:2020ckj} with $\Lambda=39$, the {\bf \Blue blue} lines denote the $\mathcal{N}=8$ upper/lower bounds computed in \cite{Agmon:2017xes} with $\Lambda=43$. At $c_T=16$ the $\mathcal{N}=8$ upper bound does not apply for the $\Delta_{(0,2)}$ plot, as the $\mathcal{N}=8$ superblock becomes a conserved current that does not decompose to $\text{Long}_{\Delta, 0}^{[000], 2}$. The {\bf\Red red} dotted lines denotes the spectrum read off from the functional saturating the lower bound on $\lambda^2_{(B,2)_{2,0}^{[022]}}$ in \cite{Binder:2020ckj}, which was identified with the $U(1)_{2M}\times U(1+M)_{-2M}$ theory. The {\bf\Green green} dashed lines denote the $O(c_T^{-1})$ correction for the $U(1)_{2M}\times U(1+M)_{-2M}$ theory computed in this work, while the {\bf\Orange orange} dashed lines denote the $O(c_T^{-1})$ correction for the supergravity limit of ABJM theory as computed in \cite{Zhou:2017zaw,Chester:2018lbz}. The gray dots denote the GFFT and free theory values.}
\label{scalfig}
\end{center}
\end{figure}

Finally, in Figure \ref{scalfig} we compare the conjectured $U(1)_{2M}\times U(1+M)_{-2M}$ numerical spectrum to the analytic tree level results for the scaling dimensions of the lowest dimension operators for the $\text{Long}_{\Delta, 0}^{[000], 1}$ and $\text{Long}_{\Delta, 0}^{[000], 2}$ structures, which are parity even and odd, respectively, for the parity preserving theory we are considering. Recall that, as per \eqref{LimitSpin0}, the unitarity limit of the $\text{Long}_{\Delta, 0}^{[000], 2}$ superconformal block is precisely given by the $(B,1)_{2,0}^{[200]}$ superconformal block, so the bound on $\Delta_{(0, 2)}$ that we find depends on the assumptions we make about the possibility of having a $(B,1)_{2,0}^{[200]}$ multiplet appearing in the $S \times S$ OPE. For the $\Delta_{(0,2)}$ plot, we assumed that no $(B,1)_{2,0}^{[200]}$ appear, which as shown in the plot excludes all $\mathcal{N}=8$ theories with $\frac{16}{c_T}<.71$ that were shown in \cite{Chester:2014mea} to contain an $\mathcal{N}=8$ operator that decomposes to $(B,1)_{2,0}^{[200]}$.\footnote{The only known $\mathcal{N}=8$ theories that do not contain any $(B,1)_{2,0}^{[200]}$ operators are the free theory with $\frac{16}{c_T}=1$ and the $U(1)_2\times U(2)_{-2}$ theory with $\frac{16}{c_T}=.75$.} As with the OPE coefficient plots, we again find that large $c_T$ slope of the numerics approximately matches the tree level result, as shown in Table \ref{comparisonTable}. Note that the $\Delta_{(0, 2)}$ plot describes the scalar approximately conserved current, which is parity odd. 

\section{Discussion}
\label{disc}


The main result of this paper is the expression for tree level, i.e. leading large $c_T$, $\<SSSS\>$ for any higher spin $\cN=6$ theory in terms of just two free parameters. For the $U(N)_k\times U(N+M)_{-k}$ and $SO(2)_{2k}\times USp(2+2M)_{-k}$ ABJ theories, we used localization to fix these parameters in terms of $\lk$, which is $M/k$ for the former theory and $(M+1/2)/k$ for the latter theory.\footnote{We find that the tree level result is independent of $N$.} We then successfully compared the CFT data extracted from this tree level correlator at $\lambda=1/2$ to the large $c_T\sim M$ regime of the conjectured non-perturbative numerical bootstrap solution to $\langle SSSS\rangle$ for the $U(1)_{2M}\times U(1+M)_{-2M}$ theory \cite{Binder:2020ckj}.  On the way to deriving these results, we derived superconformal Ward identities for $\<SSSP\>$ as well as an integrated relation between this correlator and $\partial_{m_\pm}^3\partial_{m_-}F\big\vert_{m_\pm=0}$ which can be computed using supersymmetric localization. 


It is instructive to compare our $\mathcal{N}=6$ correlators $\langle SSSS\rangle$ in \eqref{finalAnswer} and $\langle PPPP\rangle$ in  \eqref{finalAnswerP} to the tree level correlator of the scalar single trace quasibosonic $\cO_{qb}$ and quasifermionic $\cO_{qf}$ operators for non-supersymmetric vector models in\cite{Turiaci:2018nua}: 
\es{JT}{
\langle \cO_{qb}&(\vec x_1)\cO_{qb}(\vec x_2)\cO_{qb}(\vec x_3)\cO_{qb}(\vec x_4)\rangle=\frac{1}{x_{12}^2x_{34}^2}\frac{8}{c_{T}}\Bigg[\sqrt{U}+\sqrt{\frac{U}{V}}+\frac{U}{\sqrt{V}}\\
&-\frac{2}{\pi^{\frac52}}\sin^2\left(\frac{\pi\lambda_{qb}}{2}\right)\left(U\bar D_{1,1,\frac12,\frac12}(U,V)+U\bar D_{1,1,\frac12,\frac12}(V,U)+\bar D_{1,1,\frac12,\frac12}\big(\frac1U,\frac VU\big)\right)\Bigg]\,,
}
and
\es{JT}{
\langle \cO_{qf}(\vec x_1)&\cO_{qf}(\vec x_2)\cO_{qf}(\vec x_3)\cO_{qf}(\vec x_4)\rangle \\&=\frac{1}{x_{12}^4x_{34}^4}\frac{2}{c_{T}}\Bigg[\frac{U^2 (U-V-1)}{V^{3/2}}-\sqrt{U} (U-V+1)-\frac{\sqrt{{U}}
   (U+V-1)}{V^{\frac32}}\Bigg]\,.
}
For concreteness, we set $\frac{1}{\tilde N}=\frac{2}{c_{T}}$ and $\tilde\lambda_{qb}=\tan(\frac{\pi\lambda_{qb}}{2})$ in \cite{Turiaci:2018nua} as in \cite{Aharony:2012nh} for a $U(N_{qb})_{k_{qb}}$ Chern-Simons matter theory with $\lambda_{qb}\equiv N_{qb}/k_{qb}$ and one complex scalar for the quasibosonic case, or one complex fermion in the quasifermionic case. The quasibosonic case should be naturally compared to $\langle SSSS\rangle$, as both $S$ and $\cO_{qb}$ are scalars with $\Delta=1$ at tree level, while the quasifermionic case should be compared to $\langle PPPP\rangle$, as both $P$ and $\cO_{qf}$ are pseudoscalars with $\Delta=2$ at tree level. For all cases, the contact terms allowed by the Lorentzian inversion formula vanish. For the quasiboson and $\langle SSSS\rangle$, the tree level correlator consists of a connected free theory term and a scalar exchange term, while for the quasifermion and $\langle PPPP\rangle$, only a connected free term appears. In our case both $\langle SSSS\rangle$ and $\langle PPPP\rangle$ depend on $\lambda$ as $\sin^2(\pi\lambda)$, while in the nonsupersymetric case only the quasiboson depends on $\lambda_{qb}$, and has the slightly different periodicity $\sin^2(\frac{\pi\lambda_{qb}}{2})$.\footnote{The factor of two discrepancy in the periodicity between the ABJ case and the non-supersymmetric case is discussed in Section 6.2 of \cite{Chang:2012kt}.} For both the quasiboson and $\langle SSSS\rangle$, the exchange terms are given simply by scalar exchange Witten diagrams, though the physical origin is quite different in each case. In the quasibosonic case, \cite{Maldacena:2012sf} showed that for spin $\ell$ single trace operators $J_\ell$, all tree level $\langle\cO_{qb} \cO_{qb} J_\ell\rangle$ were the same as the free theory except for $J_0\equiv \cO_{qb}$, which depends on $\lambda_{qb}$. The scalar exchange then appears so as to compensate for the fact that tree level $\langle \cO_{qb} \cO_{qb} \cO_{qb}\rangle$ is not given by the free theory result. In our $\mathcal{N}=6$ case, we found that the tree level three-point functions between two $S$'s and a higher spin multiplet were given by the free theory result only for odd $\ell$, while for even $\ell$ they are all proportional to the same $\lambda$ dependent coefficient.  The contribution of the exchange diagrams for the even and odd spin single trace long multiplets, which at tree level coincide with conserved supermultiplets, exactly canceled so that only the scalar exchange diagrams remained.

We showed that the contact terms allowed by the Lorentzian inversion formula for $\langle SSSS\rangle$ vanished by combining localization with the $\<SSSP\>$ four-point function computed using the weakly broken Ward identity. There is in fact a possible alternative argument that only uses $\mathcal{N}=6$ superconformal symmetry, and so would apply to any $\mathcal{N}=6$ higher spin theory. Note that $\mathcal{N}=6$ superconformal symmetry only allows a single contact term with four or less derivatives, which thus contributes to spin two or less as allowed by the large $M$ Lorentzian inversion formula \cite{Caron-Huot:2017vep}. In \cite{Binder:2019mpb}, we used flat space amplitude arguments to show that this four derivative contact term for $\langle SSSS\rangle$ actually becomes a six derivative contact term in other stress tensor multiplet correlators like $\langle SSJJ\rangle$, where $J$ is the R-symmetry current, that are related to $\langle SSSS\rangle$ by supersymmetry. Since six derivative contact terms generically contribute to spin three CFT data in correlators of non-identical operators \cite{Heemskerk:2010ty}, they would be disallowed by the Lorentzian inversion formula for correlators with spin \cite{Caron-Huot:2021kjy}, which would then disallow the putative four derivative $\langle SSSS\rangle$ contact term. In fact, the $\langle SSSS\rangle$ contact term contributes to a scalar long multiplet that contains a spin three descendant. This happens to not contribute to the $\langle SSSS\rangle$ superblock \cite{Binder:2020ckj}, but could well appear in the $\langle SSJJ\rangle$ superblock. It would be interesting to derive the superconformal Ward identity that explicitly relates $\langle SSJJ\rangle$ to $\langle SSSS\rangle$, so that we could verify this alternative argument for the vanishing of the contact term. Our tree level result would then just be fixed in terms of a single free parameter, as in the non-supersymmetric case of \cite{Turiaci:2018nua,Li:2019twz}.

There are several ways we could improve our comparison between tree level higher spin $\langle SSSS\rangle$ and the numerical bootstrap results in \cite{Binder:2020ckj} for the $U(1)_{2M}\times U(1+M)_{-2M}$ theory. Numerically, it would be good to improve the precision of the numerics and compute predictions for more CFT data. In particular, we will likely need much higher precision to probe the single trace multiplets with superprimaries of odd spin, since their OPE coefficients squared scale as $O(c_T^{-1})$ and so are hard to see numerically, unlike the even spin case that scales as $O(c_T^{0})$. Analytically, we will need to generalize the large $c_T$ analytic calculation of $\langle SSSS\rangle$ to order $1/c_T^2$ if we want to extract anomalous dimensions of the odd single trace higher spin multiplets even at tree level, due to the $O(c_T^{-1})$ scaling of their OPE coefficients squared. This 1-loop calculation can in principle be computed from tree level CFT data \cite{Aharony:2016dwx}, but would require one to unmix the double trace tree level CFT data, which is difficult even in the non-supersymmetric case \cite{Aharony:2018npf}. This unmixing would similarly be required if we want to compare to numerical results for unprotected double trace operators with non-lowest twist, which are degenerate.

We could also generalize our tree level higher spin correlator calculation to a wider class of theories. For instance, the ABJ quadrality of \cite{Honda:2017nku} considered not just the 3d $\mathcal{N}=6$  $SO(2)_{2k}\times USp(2+2M)_{-k}$ theory considered in this work, but the wider class of $\mathcal{N}=5$ $O(N_1)_{2k}\times USp(N_2)_{-k}$ ABJ theories, which also have approximately broken higher spin symmetry when $N_{1},k$ are large and $N_2,\lambda\equiv N_1/k$ are finite (similarly for $N_1\leftrightarrow N_2$), and so are conjecturally related to $\mathcal{N}=5$ higher spin gravity on $AdS_4$. From the string theory perspective, these theories are obtained by orientifolding the brane construction of the $U(N)_k\times U(N+M)_{-k}$ theory, so that the $O(N_1)_{2k}\times USp(N_2)_{-k}$ theories are dual to type IIA string theory on $AdS_4\times \mathbb{CP}^3/\mathbb{Z}_2$. In the string or M-theory limit, orientifolding changes the single trace spectrum, such that certain tree level correlators vanish, and the 1-loop corrections are suitably modified \cite{Alday:2020tgi}. In the higher spin limit, however, the orientifold does not affect the single trace spectrum aside from reducing the supersymmetry when $N_1\neq 2$ from $\mathcal{N}=6$ to $\mathcal{N}=5$, so we expect that the general structure of the $\mathcal{N}=5$ tree level correlator should be very similar to our $\mathcal{N}=6$ result. The precise dependence on $\lambda$ could still be different, as that depends on the Lagrangian of the specific theory, as well as the the specific form of the $\mathcal{N}=5$ version of the $\mathcal{N}=6$ integrated constraints discussed in this work. It is possible one might also need to consider integrated constraints involving the squashed sphere, which can also be computed using supersymmetric localization as in \cite{Hama:2010av,Hama:2011ea,Chester:2020jay,Chester:2021gdw}.

The numerical bootstrap has been used so far to find conjectured numerical solutions to the $U(1)_{2M}\times U(1+M)_{-2M}$ and $U(N)_{2}\times U(N+1)_{-2}$ theories, which at large $c_T$ are dual to higher spin gravity and M-theory, respectively. These $\mathcal{N}=6$ and $\mathcal{N}=8$ solutions were found because they conjecturally saturate the lower bounds of bootstrap studies with the respective amount of supersymmetry. Ideally, we would like to numerically solve the $\mathcal{N}=6$ $U(N)_{k}\times U(N+M)_{-k}$ ABJ theory for any $N,M,k$, which would require us to input three exactly known CFT data into the bootstrap, and then see if the new bounds are saturated by known theories by comparing to a fourth known quantity. So far, we know how to input two quantities: $c_T$ and $\lambda^2_{(B,2)_{2,0}^{022}}$, which can be computed exactly from $\partial_{m_\pm}^2F\big\vert_{\pm=0}$ and $\partial_{m_\pm}^4F\big\vert_{\pm=0}$, respectively. A third quantity could be $\partial_{m_+}^2\partial_{m_-}^2F\big\vert_{\mp=0}$, which is related to a certain integral of $\langle SSSS\rangle$ as shown in \eqref{IDefs}. A fourth quantity could be a similar integrated constraint from the squashed sphere, or the tree level correlators computed in this work and \cite{Zhou:2017zaw,Binder:2018yvd,Binder:2019mpb}. Once we can study ABJ theory for any $N,M,k$, we will be able to non-perturbatively understand the relation between the higher spin and supergravity regimes. In particular, it will be interesting to see how the approximately conserved currents at finite $N$ and large $M,k$ disappear as $N$ increases.


All the discussion so far has concerned the CFT side of the higher spin AdS/CFT duality. This is mostly because supersymmetric higher spin gravity is still poorly understood. The only known formulation so far is in terms of Vasiliev theory \cite{Vasiliev:1990en,Vasiliev:1992av,Vasiliev:1995dn,Vasiliev:1999ba,Chang:2012kt,Konstein:1989ij,Vasiliev:1990bu,Vasiliev:1990vu}, which is just a classical equation of motion with no known action, and so cannot be used to compute loops. Even on the classical level, it has been difficult to regularize the calculation of various correlation functions \cite{Boulanger:2015ova,Sleight:2017pcz}.  Recently, a higher spin action has been derived in \cite{Aharony:2020omh} for the $O(N)$ free and critical vector models, which manifestly reproduces the correct CFT results to all orders in $1/N$. If this construction could be extended to $\mathcal{N}=6$, then it is possible that the bulk dual of $\langle SSSS\rangle$ could be computed and the absence of contact terms understood from the bulk perspective.

\section*{Acknowledgments} 

We thank Ofer Aharony, Rohit Kalloor, Adar Sharon, Alexander Zhiboedov, and Silviu Pufu for useful discussions and correspondence, and Silviu Pufu for collaboration at an early stage of this project. We also thank Silviu Pufu and Ofer Aharony for reading through the manuscript. DJB and MJ are supported in part by the Simons Foundation Grant No.~488653, and by the US NSF under Grant No.~1820651\@.  DJB is also supported in part by the General Sir John Monash Foundation.  SMC is supported by the Zuckerman STEM Leadership Fellowship. DJB is grateful to the ANU Research School of Physics for their hospitality during the completion of this project.

\appendix

\section{Pseudocharge Action on $S$ and $P$}

Our task in this section is to derive the action of $\tilde\dk(X)$ on the scalars $S$ and $P$. In Appendix~\ref{DELTASP} we derive \eqref{chiEq}, which gives expressions for $\tilde\dk(X)S(y,Y)$ and $\tilde\dk(X)P(y,Y)$ in terms of $\lk_{SSB_0}$ and two unknown coefficients, $\kk_1$ and $\kk_2$. We then compute these coefficients in Appendix~\ref{OPMIX}.

\subsection{Constraining the Pseudocharge Action}
\label{DELTASP}

Our task is to derive \eqref{chiEq}. We begin with $\tilde\dk(X)S(\vec y,Y)$, which we can compute by evaluating
\begin{equation}\label{Pseudo3pt}
\<\tilde\dk(X)S(0,Y)\fcy O(\hat e_3)\> = \frac1{4\pi}\left.\int_{|x|=r} dS\cdot \left\< H_1(\vec x,X)S(0,Y)\fcy O(\hat e_3)\right\>\right|_{\text{finite as }r\rightarrow\infty}
\end{equation}
for general operators $\fcy O(\hat e_3)$ located at $\hat e_3 = (0,0,1)$. We first note that the right-hand side of \eqref{Pseudo3pt} is only non-zero if $\fcy O$ is a scalar with conformal dimension 1. For this special case, conformal invariance implies that
\begin{equation}\begin{split}
\left\< H_1(\vec x,X)S(0,Y)\fcy O(\hat e_3)\right\> &= f_{S\fcy OH_1}(X,Y)\fcy C_{1,1,1}^\mu(0,\hat e_3,\vec x) \\
&= \frac{f_{S\fcy OH_1}(X,Y)}4\left(\frac{(\hat e_3-x)^\mu}{|\hat e_3 - x|^2}-\frac{x^\mu}{|x|^2}\right)\frac1{|x||x-\hat e_3|}\,,
\end{split}\end{equation}
where $f_{S\fcy OH_1}(X,Y)$ is a function of $X$ and $Y$ whose exact form depends on the $\mathfrak{so}(6)_R$ properties of $\fcy O$. Substituting this into \eqref{Pseudo3pt}, we find that
\begin{equation}\label{deltaS3pt}
\<\tilde\dk(X)S(0,Y)\fcy O(\hat e_3)\>  = -\frac14 f_{S\fcy OH_1}(X,Y)\,.
\end{equation}

The only two dimension $1$ scalars in higher spin $\mathcal{N}=6$ theories are $S(\vec y,Y)$ itself and $B_0(\vec x)$. For $S$, we apply \eqref{deltaS3pt} with $\cO(\hat e_3) = S(\hat e_3,Z)$ to find that
\begin{equation}\begin{split}
\<\tilde\dk(X)S(0,Y)S(\hat e_3,Z)\> &= -\frac{\lk_{SSH_1}\text{tr}([Y,Z]X)}{4} = -\frac{\lk_{SSH_1}}{4}\left\<S(0,[X,Y])S(\hat e_3,Z)\right\>\,,
\end{split}\end{equation}
while for $B_0$ we find that
\begin{equation}\begin{split}
\<\tilde\dk(X)S(0,Y)B_0(\hat e_3)\> &= -\frac{\lk_{SB_0H_1}\text{tr}(XY)}{4} = -\frac{\lk_{SB_0H_1}\text{tr}(XY)}{4}\<B_0(0)B_0(\hat e_3)\>\,.
\end{split}\end{equation}
However, as we will now show, $\lk_{SB_0H_1} = 0$. To see this, we compute:
\begin{equation}
\tilde\dk(X)\<S(0,Y)SH_1\> = \<\tilde\dk(X)S(0,Y)SH_1\> + \dots = -\frac{\lk_{SB_0H_1}\text{tr}(XY)}{4}\<B_0(0)S_0H_1\> + \dots\,,
\end{equation}
where the additional terms come from the variations of the second $S$ and $H_1$, and from the multiplet recombination. Note that $\tilde\dk(X)\<S(0,Y)SH_1\>$ contains a term proportional to $\lk_{SB_0H_1}^2\text{tr}(XY)$. But it is straightforward to check that no additional term appears in either $\left\<S(0,Y)\tilde\dk(X)\left(SH_1\right)\right\>$ or in $\<S\tilde PH_1\>$ with the right $R$-symmetry structure needed to cancel such a contribution, and so conclude that $\lk_{SB_0H_1} = 0$. Having exhausted the possible operators that could appear in $\tilde\delta(X)S(y,Y)$, we conclude that
\begin{equation}
\tilde\delta(X)S(\vec y,Y) = -\frac{\lk_{SSH_1}}{4}S(y,[X,Y]) = -\frac{\lk_{SSB_0}}{4}S(y,[X,Y])\,.
\end{equation}

We can constrain $\tilde\dk(X)P(\vec y,Y)$ in much the same way, except now we have to consider not only the single trace operators $P$ and $C_0$, but also double-trace operators built from $S$ and $B_0$. The most general expression we can write is
\begin{equation}\begin{split}
\tilde\dk(X)P(\vec y,Y) &= \kk_0 P(\vec y,[X,Y]) + \kk_1 S^2(\vec y,[X,Y]) + \kk_2 SB_0(\vec y,[X,Y])\\
&+ \mu_1 \text{tr}(XY)\fcy O_1(\vec y) + \mu_2\fcy O_2(\vec y,\{X,Y\}) + \mu_3 S(\vec y,X)S(y,Y)\,,
\end{split}\end{equation}
where $\fcy O_1(y)$ is some linear combination of $S^a{}_bS^b{}_a$ and $B_0^2$, and $\fcy O_2$ is some linear combination of $P$, $S^2$ and $SB_0$. By computing $\<\tilde\dk PP\>$ we find that
\begin{equation}
\kk_0 = -\frac14\lk_{PPH_1} = \frac14\lk_{SSH_1}\,.
\end{equation}
If we instead consider $\<\tilde\dk P\cO_i\>$, we find that $\mu_i$ are proportional to OPE coefficients $\lk_{PP\mu_i}$, but can then check that the $\tilde\dk\<PPH_1\>$ Ward identity is satisfied if and only if $\mu_i = 0$. This leads us to \eqref{chiEq}.

\subsection{Computing $\kk_1$ and $\kk_2$}
\label{OPMIX}

We now compute $\kk_1$ and $\kk_2$ using the variations $\tilde\dk\<SSP\>$ and $\tilde\dk\<SPB_0\>$. Let us begin with $\tilde\dk(X)\<SSP\>$. As listed in equation \eqref{scalarBlock}, supersymmetry forces both $\<SSP\>$ and $\<SPP\>$ to vanish. Expanding the left-hand side of the higher spin Ward identity, we thus find that
\begin{equation}\label{SSPL}
 \tilde\dk(X)\<SSP(y_3,Y_3)\> =  \kk_1\<SSS^2(\vec y_3,[X,Y_3])\>\,,
\end{equation}
while expanding the right-hand side we instead find that
\begin{equation}\label{eqSSPR}
 \tilde\dk(X)\<SSP(y_3,Y_3)\> = -\frac{\ak}{\sqrt {c_T}}\<SS\tilde S(\vec y_3,[X,Y_3])\>\,.
\end{equation}
Equating these two expressions, we conclude that
\begin{equation}
\kk_1 = \frac{\ak \lk_{SSS}}{4\sqrt {c_T}}\,.
\end{equation}

The variation $\tilde\dk(X)\<SB_0P\>$ is a little trickier, as $\<SPB_0\>$ does not vanish at $O(c_T^{-1/2})$. It is instead related by supersymmetry to the three-point function $\<SPH_1\>$, so that
\begin{equation}
\frac{\lk_{SPH_1}}{\lk_{SPB_0}} = \sqrt{\frac{2(\Dk+1)}{\Dk}} \underset{\Dk\rightarrow1}\longrightarrow2\,,
\end{equation}
where $\Dk$ is the conformal dimension of $B_0$. We can then in turn relate $\lk_{SPH_1}$ to $\ak$ using the multiplet recombination formula \eqref{H1Recom}, and so find that
\begin{equation}
\lk_{SPB_0} = -\frac12\lk_{SPH_1} = \frac{2\ak}{\sqrt {c_T}}\,.
\end{equation}

Now that we have computed $\lk_{SPB_0}$, let us turn to $\tilde\dk\<SPB_0\>$. Expanding this using \eqref{chiEq}, we find that
\begin{equation}\begin{split}
\tilde\dk(X)\<&S(\vec y_1,Y_1)P(\vec y_2,Y_2)B_0(\vec y_3)\> \\
&= \frac12\lk_{SSB_0}\<S(\vec y_1,Y_1)P(\vec y_2,[X,Y_2])B_0(\vec y_3)\> + \kk_2 \<S(\vec y_1,Y_1)SB(\vec y_2,[X,Y_2])B_0(\vec y_3)\>\,.
\end{split}\end{equation}
But if we instead use the multiplet recombination rule, we find that 
\begin{equation}
\tilde\dk(X)\<S(\vec y_1,Y_1)P(\vec y_2,Y_2)B_0(\vec y_3)\> = \frac{\ak}{\sqrt{c_T}}\left\<S(\vec y_1,Y_1)\tilde S(\vec y_2,[X,Y_2])B_0(\vec y_3)\right\>\,,
\end{equation}
where we dropped $\<\tilde PPB_0\>$ as it vanishes due to supersymmetry. Equating the two expressions and solving for $\kk_2$, we conclude that
\begin{equation}
\kk_2 = -\frac{\ak\lk_{SSB_0}}{4\sqrt{c_T}}\,.
\end{equation}

\section{Scalar Four-Point Functions}

\subsection{$\<SSPP\>$ and $\<PPPP\>$}
\label{SSPPApp}

Conformal and R-symmetry invariance imply that the four-point functions $\<SSPP\>$ and $\<PPPP\>$ take the form \cite{Binder:2019mpb}:
\begin{equation}\begin{split}\label{SSSScor}
\langle S(\vec x_1,X_1)  S(\vec x_2,X_2) P(\vec x_3,X_3) P(\vec x_4,X_4)\rangle &= \frac{1}{x_{12}^2 x_{34}^4} \sum_{i=1}^6 {\cal R}^i(U, V) {\cal B}_i \,,\\
\langle P(\vec x_1,X_1)  P(\vec x_2,X_2) P(\vec x_3,X_3) P(\vec x_4,X_4)\rangle &= \frac1{x_{12}^4 x_{34}^4} \sum_{i=1}^6 {\cal P}^i(U, V) {\cal B}_i \,,\\
\end{split}\end{equation}
where the $R$-symmetry structures are defined as in \eqref{BasisElems}, and where $\cR^i$ and $\cP^i$ are functions of the cross-ratios \eqref{crossRatio}. Crossing symmetry implies that
\begin{equation}\begin{split}\label{CrossingP}
\cR^3(U,V) &= \cR^2\left(\frac{U}{V},\frac{1}{V}\right)\,, \quad \cR^6(U,V) = \cR^5\left(\frac UV,\frac1V\right)  \,, \\
 \cP^2(U,V) &= U^2\cP^1\left(\frac{1}{U},\frac{V}{U}\right)\,, \quad \cP^3(U,V) = \frac{U^2}{V^2}\cP^1(V,U)  \,, \\
 \cP^5(U,V) &= U^2\cP^4\left(\frac{1}{U},\frac{V}{U}\right)\,, \quad \cP^6(U,V) = \frac{U^2}{V^2}\cP^4(V,U) \,, \\
\end{split}\end{equation}
so that $\langle PPPP\rangle$ can be uniquely specified by $\cP^1(U,V) $ and $\cP^4(U,V)$, while $\langle SSPP\rangle$ is uniquely specified by $\cR^1(U,V) $, $\cR^2(U,V) $, $\cR^4(U,V) $, and $\cR^5(U,V)$.

As shown in \cite{Binder:2019mpb}, the $\cN=6$ superconformal Ward identities fully fix $\<SSPP\>$ and $\<PPPP\>$ in terms of $\<SSSS\>$. Applying these to the various terms in our ansatz \eqref{SSSSansatz}, we find that for the generalized free field term:
\begin{equation}\begin{split}
\cR^1_{\text{GFFF}}(U,V) &= 1\,,\qquad \cR^i_{\text{GFFF}}(U,V) = 0 \text{ for } i=2\,,\dots\,,6 \\
\cP^1_{\text{GFFF}}(U,V) &= 1\,,\qquad \cP^4_{\text{GFFF}}(U,V) = 0\,, \\
\end{split}\end{equation}
for the free connected term:
\begin{equation}\begin{split}
\cR^i_{\text{free}}(U,V)   &= 0 \,,\qquad \cP^1_{\text{free}}(U,V) = 0 \,,\qquad \cP^4_{\text{free}}(U,V) = \frac{U^2(U-V-1)}{4V^{3/2}}\,, \\
\end{split}\end{equation}
for the scalar exchange term:
\begin{equation}\begin{split}
\cR^1_{\text{scal}}(U,V) &= 0\,,\qquad \cR^2_{\text{scal}}(U,V) = \frac{4U}{\pi^2}\,,\qquad \cR^4_{\text{scal}}(U,V) = \frac{2U(U-V-1)}{\pi^2 V}\,,\\
\cR^5_{\text{scal}}(U,V) &= -\frac{2U}{\pi^2 V} + \frac{U^2 \bar D_{1,2,\frac12,-\frac12}(U,V)}{2\pi^{5/2}}\,, \qquad \cP^i_{\text{scal}}(U,V) = 0\,.
\end{split}\end{equation}
Finally, for the degree 2 contact term $\<SSPP\>$ is given by
\begin{equation}\begin{split}
\cR^1_{\text{cont}}(U,V) &= -\frac{4U^2}3\left(4\bar D_{2,2,1,1}-6\bar D_{2,2,2,2}+15\bar D_{4,2,1,3}+15\bar D_{3,2,2,3}-30\bar D_{3,2,1,2}\right) \\
\cR^2_{\text{cont}}(U,V) &=  4U^2\left(\bar D_{3,2,2,3}-2\bar D_{2,2,2,2}\right)\,,\qquad \cR^4_{\text{cont}}(U,V) = 4U^2\left(5\bar D_{3,2,2,3}-2\bar D_{2,2,2,2}\right)\,,\\
\cR^5_{\text{cont}}(U,V) &= \frac{4U^2}3\left(8\bar D_{2,2,1,1}-36\bar D_{2,2,2,2}+15\bar D_{2,2,3,3}-36\bar D_{3,2,1,2}+30\bar D_{3,2,2,3}+15\bar D_{4,2,1,}\right)\,.
\end{split}\end{equation}

\subsection{Shadow Transforms of 4pt Correlators}
\label{SHADOW4PT}

In this appendix, we explain how to compute the shadow transforms of $\<SSSS\>$ and $\<SSPP\>$, which, using \eqref{shadowSSSS} we can express in terms of functions
\begin{equation}\label{tildeDef}\begin{split}
\tilde \cS^i(U,V) &= \frac{x_{34}^3x_{14}}{x_{13}}\int \frac{d^3z}{4\pi|\vec z-\vec x_4|^4} \frac{\cS^i\left(\frac{x_{12}^2|\vec x_3-z|^2}{x_{13}^2|\vec x_2-\vec z|^2},\frac{|\vec x_1-\vec z|^2x_{23}^2}{x_{13}^2|\vec x_2-\vec z|^2}\right)}{|\vec x_3-\vec z|^2}\,, \\
\tilde \cR^i(U,V) &= \frac{x_{34}^3x_{14}}{x_{13}}\int \frac{d^3z}{4\pi|\vec z-\vec x_3|^2} \frac{\cP^i\left(\frac{x_{12}^2|x_4-\vec z|^2}{|\vec x_1-\vec z|^2x_{24}^2},\frac{x_{14}^2|\vec x_2-\vec z|^2}{|\vec x_1-\vec z|^2x_{24}^2}\right)}{|\vec x_4-\vec z|^4}\,. \\
\end{split}\end{equation}

Let us begin with free connected term, for which
\begin{equation}
\cS^1_{\text{free}}(U,V) = \cR^i_{\text{free}}(U,V) = 0\,,
\end{equation}
so that the only non-trivial computation is 
\begin{equation}
\tilde\cS^4_{\text{free}}(U,V)= \frac{x_{12}^2x_{34}^3x_{14}}{x_{13}^2x_{23}}\int\frac{d^3z}{4\pi|\vec z-\vec x_4|^4} \frac1{|\vec x_1-\vec z||\vec x_2-\vec z|}\,.
\end{equation}
We can evaluate this integral using the star-triangle relation
\begin{equation}\label{starTriangle}
\int\frac{d^3z}{|\vec x_1-\vec z|^{2\Dk_1}|\vec x_2-\vec z|^{2\Dk_2}|\vec x_3-\vec z|^{2\Dk_3}} = \left(\prod_{i=1}^3\frac{\Gk\left(\frac32-\Dk_i\right)}{\Gk(\Dk_i)}\right) \frac {\pi^{3/2}}{x_{12}^{d-2\Dk_3}x_{13}^{d-2\Dk_2}x_{23}^{d-2\Dk_1}}\,,
\end{equation}
and so find that
\begin{equation}
\tilde\cS^4_{\text{free}}(U,V)  = -\frac12\sqrt{\frac{U^3}V}\,.
\end{equation}

Next we turn to the contact term. By definition, the $\bar D$ functions are related to the quartic contact Witten diagram
\begin{equation}
D_{r_1,r_2,r_3,r_4}(x_i)=\int_{AdS_{4}} dz \prod_{i=1}^4G^{r_i}_{B\partial}(z,\vec x_i)\,,\qquad G^{r}_{B\partial}(z,\vec x)=\left(\frac{z_0}{z_0^2+(\vec z-\vec x)^2}\right)^{r}
\end{equation}
via the equation
\begin{equation}
\bar D_{r_1,r_2,r_3,r_4}(U,V)=\frac{x_{13}^{\frac12\sum_{i=1}^4r_i-r_4}x_{24}^{r_2}}{x_{14}^{\frac12\sum_{i=1}^4r_i-r_1-r_4}x_{34}^{\frac12\sum_{i=1}^4r_i-r_3-r_4}}\frac{2\prod_{i=1}^4\Gamma(r_i)}{\pi^{\frac 32}\Gamma\left(\frac{-3+\sum_{i=1}^4r_i}{2}\right)}D_{r_1,r_2,r_3,r_4}(x_i)\,.
\end{equation}
Because the shadow transform of a bulk-boundary propagator is another bulk-boundary propagator \cite{SimmonsDuffin:2012uy}:
\begin{equation}
\int\frac{d^3y}{|\vec x-\vec y|^{6-2r}} G^r_{B\nb}(z,\vec y) = \frac{\pi^{3/2}\Gamma\left(r-\frac32\right)}{\Gamma(r)} G^{3-r}_{B\nb}(z,\vec x)\,,
\end{equation}
we see that the shadow transform of a $D$-function is another $D$-function:
\begin{equation}\label{shadD}
\int\frac{d^3y}{|\vec x_4-\vec y|^{6-2r_4}}  D_{r_1,r_2,r_3,r_4}(\vec x_1,\vec x_2,\vec x_3,\vec y) = \frac{\pi^{3/2}\Gamma\left(r_4-\frac32\right)}{\Gamma(r_4)}D_{r_1,r_2,r_3,3-r_4}(\vec x_1,\vec x_2,\vec x_3,\vec x_4)\,.
\end{equation}
When we write $$\frac{\cS^i_{\text{cont}}(U,V)}{x_{12}^2x_{34}^2} \text{ and } \frac{\cR^i(U,V)}{x_{12}^2x_{34}^4}$$ in terms of $D$-functions, the result is a sum of $D$-functions multiplied by rational functions of $x_{ij}^2$. Using the identity \cite{Dolan:2003hv}
\begin{equation}\label{Dr1r2ex}\begin{split}
4r_1r_2 x_{12}^2&D_{r_1+1,r_2+1,r_3,r_4} -  4r_3r_4x_{34}^2 D_{r_1,r_2,r_3+1,r_4+1}\\ &= (r_1+r_2-r_3-r_4)(3-r_1-r_2-r_3-r_4)D_{r_1,r_2,r_3,r_4}\,. \\
\end{split}\end{equation}
along with its crossings, we can always rearrange the integrands in \eqref{tildeDef} into a form such that we can apply \eqref{shadD} term by term.

Finally, we turn to the exchange term. To compute the shadow transform for this term, we first note that when $\Dk_1+\Dk_2+\Dk_3+\Dk_4=3$,
\begin{equation}\begin{split}
\int d^3x &\prod_{i=1}^4\frac1{|\vec x-\vec x_i|^{2\Dk_i}} \\
&= \frac{\pi^{3/2}}{\Gk(\Dk_1)\Gk(\Dk_2)\Gk(\Dk_3)\Gk(\Dk_4)}\frac{x_{14}^{3-2\Dk_1-2\Dk_4}x_{34}^{3-2\Dk_3-2\Dk_4}}{x_{13}^{3-2\Dk_4}x_{24}^{2\Dk_2}}\bar D_{\Dk_1,\Dk_2,\Dk_3,\Dk_4}(U,V)\,.
\end{split}\end{equation}
We thus find that
\begin{equation}\begin{split}
\tilde \cS^1_{\text{scal}}(U,V) &= \frac{-2}{\pi^{5/2}}\frac{x_{34}^3x_{14}x_{12}^2}{x_{13}^3}\int \frac{d^3z}{4\pi|\vec z-\vec x_4|^4}\frac{\bar D_{1,1,\frac12,\frac12}\left(\frac{x_{12}^2|\vec x_3-\vec z|^2}{x_{13}^2|\vec x_2-\vec z|^2},\frac{|\vec x_1-\vec z|^2x_{23}^2}{x_{13}^2|\vec x_2-\vec z|^2}\right)}{|\vec x_2-\vec z|^2} \\
&= -\frac1{2\pi^4}\frac{x_{12}^2x_{14}x_{34}^3}{x_{13}} \int \frac{d^3z\,d^3w\,}{|\vec z-\vec x_4|^4|\vec w-\vec x_1|^2|\vec w-\vec x_2|^2|\vec w-\vec x_3||\vec w-\vec z||\vec z-\vec x_3|}\,.
\end{split}\end{equation}
Performing the integral over $z$ using the star-triangle relation \eqref{starTriangle}, we then find that the integral over $w$ can also be performed using the star-triangle relation, and so 
\begin{equation}
\tilde \cS^1_{\text{scal}}(U,V) = \sqrt U\,.
\end{equation}
We can evaluate $\tilde \cS^4_{\text{scal}}$ in a similar fashion, finding that
\begin{equation}
\tilde \cS^4_{\text{scal}}(U,V) = -\frac12 U\left(1+\frac1{\sqrt V}\right)\,.
\end{equation}

Now we turn to computing $\tilde\cR^i(U,V)$. Because ultimately our goal is to compute $\<SSSP\>$, we only need $\tilde\cR^1_{\text{scal}}(U,V)$ and $\tilde\cR^4_{\text{scal}}(U,V)$, as these suffice to compute $\tilde\cT^1(U,V)$ and $\tilde\cT^4(U,V)$. But $\cR^1_{\text{scal}}(U,V) = 0$, and $\tilde\cR^4_{\text{scal}}(U,V)$ can be computed by using the star-triangle relation on each term by term, so that
\begin{equation}
\tilde\cR^1(U,V) = 0\,\qquad \tilde\cR^4 = \frac{(\sqrt U-\sqrt V - 1)}{2\sqrt V}\,.
\end{equation}

\section{Computing Integrated Correlators}
\label{INTAPP}

In this appendix, we compute the integrated correlators needed in Section~\ref{INTCOR}. Many of these calculations are performed most conveniently in Mellin space, and so we begin by first reviewing this topic.

\subsection{Mellin Space Review}
\label{MelReview}

Holographic correlators take simple forms in Mellin space. For $\<SSSS\>$ we define
the Mellin amplitudes $M^i(s,t)$ through
\begin{equation}\begin{split}\label{melDef}
\cS^i(U,V)=&\int_{-i\infty}^{i\infty}\frac{ds\, dt}{(4\pi i)^2}\ U^{\frac s2}V^{\frac u2-1} \Gamma^2\left[1-\frac s2\right]\Gamma^{2}\left[1-\frac t2\right]\Gamma^{2}\left[1-\frac u2\right] M^i(s,t)\,,
\end{split}\end{equation}
where $u = 4 - s - t$, while for $\<SSSP\>$ we define the Mellin amplitudes $N^i(s,t)$ through
\begin{equation}\begin{split}\label{melDef2}
\fcy T^i(U,V) &= \int \frac{ds\,dt}{(4\pi i)^2} N^i(s,t) U^{s/2}V^{u/2-1}\\
&\times\Gk\left(1-\frac s2\right)\Gk\left(\frac{3-s}2\right)\Gk\left(1-\frac t2\right)\Gk\left(\frac{3-t}2\right)\Gk\left(1-\frac u2\right)\Gk\left(\frac{3-u}2\right)\,,
\end{split}\end{equation}
where $u = 5-s-t$. In either case the two integration contours are chosen such that\footnote{This is the correct choice of contour provided that $M^i(s, t)$ does not have any poles with $\Re(s) <2$ or $\Re(t) < 2$ or $\Re(u) < 2$.  If this is not the case (such as for the exchange diagram), the integration contour will have to be modified in such a way that the extra poles are on the same side of the contour as the other poles in $s$, $t$, $u$, respectively.}
\begin{equation}\label{contour}
\text{Re}(s) < 2\,,\quad \text{Re}(t) < 2\,,\quad \text{Re}(u) < 2\,,
\end{equation}
which include all poles of the Gamma functions on one side or the other of the contour \cite{Mack:2009mi}. Crossing symmetry implies that
\es{CrossingM}{
M^1(s,t) &= M^1(s,u)\,, \qquad M^2(s,t) = M^1(t,s)\,, \qquad M^3(s,t) = M^1(u,t)  \,, \\
M^4(s,t) &= M^4(s,u)\,, \qquad M^5(s,t) = M^4(t,s)\,, \qquad M^6(s,t) = M^4(u,t)\,, 
}
with identical formulas for $N^i(s,t)$, as can be derived from the crossing equations \eqref{CrossSSSS} and \eqref{CrossSSSP}.

The Mellin transform of the disconnected term $\cS_{\text{GFFT}}^i(U,V)$ and free connected term $\cS_{\text{free}}^i(U,V)$ are singular, while for $\cS_{\text{scal}}^i(U,V)$ and $\cS_{\text{cont}}^i(U,V)$ we find that
\begin{equation}\begin{aligned}\label{SSSSMel}
M^1_{\text{scal}}(s,t) &= -\frac{2\Gk\left(\frac{1-s}2\right)}{\pi^{5/2}\Gk\left(\frac{2-s}2\right)}\,,\qquad &M^4_{\text{scal}}(s,t) &= \frac{\Gk\left(\frac{1-t}2\right)}{\pi^{5/2}\Gk\left(\frac{2-t}2\right)}+\frac{\Gk\left(\frac{1-u}2\right)}{\pi^{5/2}\Gk\left(\frac{2-u}2\right)}\,, \\
M^1_{\text{cont}}(s,t) &= (t-2)(u-2)\,,\qquad &M^4_{\text{cont}}(s,t) &= (s-2)\left(s-\frac43\right)\,.
\end{aligned}\end{equation}
For $\<SSSP\>$ the Mellin transforms of $\fcy T_{\text{free}}$ and $\fcy T_{\text{scal}}$ are singular, but for the contact term we find that
\begin{equation}\label{SSSPMel}
N^1_{\text{cont}}(s,t) = -\frac{2\pi^{1/2}}{3}(t-2)(u-2)\,,\qquad N^4_{\text{cont}}(s,t) = -\frac{2\pi^{1/2}}3(s-2)^2\,.
\end{equation}

\subsection{Computing $I_{+-}[\cS^i_{\text{scal}}]$ and $I_{++}[\cS^i_{\text{scal}}]$}
\label{SDevenLoc}

We begin with $I_{++}[\cS^i_{\text{scal}}]$, which as noted in \eqref{IDefs} computes the coefficient of $\lk_{(B,2)^{[022]}_{2,0}}^2$ OPE coefficient in $\<SSSS\>$. The ${(B,2)^{[022]}_{2,0}}$ superconformal primary is an $1/3$-BPS operator transforming in the $\bf{84}$. But by inverting \eqref{toSi}, we find that $\cS_{\text{scal},\bf{84}_s}$ vanishes for the scalar exchange diagram, and so conclude that
\begin{equation}
I_{++}[\cS^i_{\text{scal}}] = 0\,.
\end{equation}

To compute $I_{+-}[\cS^i_{\text{scal}}]$, we first use the Mellin space form of the integral
\begin{equation}\label{stInt}
\begin{split}
I_{+-}[{\cal S}^i] & = \int \frac{ds\ dt}{(4\pi i)^2} \frac{2\sqrt{\pi}}{(2-t)(s+t-2)}M^1(s,t) \\
&\times \Gamma \left[1-\frac{s}{2}\right] \Gamma \left[\frac{s+1}{2}\right] \Gamma \left[1-\frac{t}{2}\right] \Gamma \left[\frac{t-1}{2}\right] \Gamma \left[\frac {s+t-2}{2}\right] \Gamma \left[\frac{3-s-t}2\right]\,.
\end{split}\end{equation}
derived in \cite{Binder:2019mpb}. Using \eqref{SSSSMel} and then performing the change of variables $s\rightarrow4-t-u$, we find that
\begin{equation}
I_{+-}[{\cal S}^i_{\text{scal}}] = -\frac4\pi\int\frac{dt\,du}{(4\pi i)^2}\frac{\sec \frac{\pi(t+u)}2}{(t-2)(u-2)}\Gk\left(\frac{2-t}2\right)\Gk\left(\frac{t-1}2\right)\Gk\left(\frac{2-u}2\right)\Gk\left(\frac{u-1}2\right)\,.
\end{equation}
Using \verb|Mathematica|, we can evaluate this integral numerically to arbitrary high precision, and find that, to within the numerical error,
\begin{equation}
I_{+-}[{\cal S}^i_{\text{scal}}] = -\pi^2\,.
\end{equation}

\subsection{Parity Odd Mixed Mass Derivatives}
\label{MIXEDODD}
In this appendix we derive \eqref{oddIRes} from \eqref{parOddDer}. To do so, we first note that, as explained in \cite{Binder:2019mpb}, we can write
\begin{equation}
4\pi\int\frac{dx}{1+\frac{x^2}4} i \tilde J(x) =\int d^3x\sqrt{g}(iJ_+(\vec x)+K_+(\vec x))+ (Q\text{-exact terms})\,,
\end{equation}
where $\tilde J(x)$ is defined by
\begin{equation}\begin{split}
\tilde J(x) &= \frac{\sqrt{c_T}}{64\pi}\\
&\times\left( \frac{\left( 1 + \frac{i x}{2} \right)^2}{1 + \frac{x^2}{4}} S_1{}^2(0, 0, x) - \frac{\left( 1 - \frac{i x}{2} \right)^2}{1 + \frac{x^2}{4}} S_2{}^1(0, 0, x) +  S_1{}^1(0, 0, x) - S_2{}^2(0, 0, x) \right)\,.
\end{split}\end{equation}
We can hence replace the $iJ_++K_+$ terms in \eqref{parOddDer} with $\tilde J$, and so find that
\begin{equation}\label{parOddDer2}
\frac{\nb^4\log Z}{\nb^3m_+\nb m_-} = (4\pi)^3\left\<\left(\int \frac{dx}{1+\frac{x^2}4}\tilde J(x)\right)^3\left(\int d^3x\sqrt{g}(iJ_-(\vec x)+K_-(\vec x))\right)\right\>\,.
\end{equation}
Next note, again as explained in \cite{Binder:2019mpb}, that correlators of $\tilde J(x)$ are topological. We can thus place the three $\tilde J$ operators at $0$, $1$, and infinity, so that 
\begin{equation}\label{parOddDer3}
\frac{\nb^4\log Z}{\nb^3m_+\nb m_-} = 2^9\pi^6\left\<\tilde J(0)\tilde J(1)\tilde J(\infty)\left(\int d^3x\sqrt{g}(iJ_-(\vec x)+K_-(\vec x))\right)\right\>\,.
\end{equation}
Next we expand the right-hand correlator using \eqref{JKDef}, \eqref{SSSSSt} and \eqref{SSSPSt}. The $\<\tilde J\tilde J\tilde J J_-\>$ automatically vanish, and so
\begin{equation}\begin{split}
&\frac{\nb^4 \log Z}{\nb^3 m_+\nb m_-} = -\frac{ic_T^2\pi^2}{2^{12}\sqrt 2}\int\frac{d^3\vec x}{(4+|\vec x|^2)|\vec x|}\Bigg(\cT^1\left(\frac1{|\vec x-\hat e_3|^2},\frac{|\vec x|^2}{|\vec x-\hat e_3|^2}\right) + 5\cT^2\left(\frac1{|\vec x-\hat e_3|^2},\frac{|\vec x|^2}{|\vec x-\hat e_3|^2}\right) \\
&\quad +\cT^3\left(\frac1{|\vec x-\hat e_3|^2},\frac{|\vec x|^2}{|\vec x-\hat e_3|^2}\right) + 8\cT^4\left(\frac1{|\vec x-\hat e_3|^2},\frac{|\vec x|^2}{|\vec x-\hat e_3|^2}\right) + 2\cT^6\left(\frac1{|\vec x-\hat e_3|^2},\frac{|\vec x|^2}{|\vec x-\hat e_3|^2}\right)  \Bigg)\,,
\end{split}\end{equation}
where $\hat e_3 = (0,0,1)$. We can then use the superconformal Ward identity \eqref{SSSPWard} to eliminate $\cT^6$, so that the integral then simplifies to:
\begin{equation}\begin{split}\label{oddIResv2}
\frac{\nb^4 \log Z}{\nb^3 m_+\nb m_-} = &-\frac{ic_T^2\pi^2}{2^{13}\sqrt 2}\int d^3\vec x\,\Bigg(2 \fcy T^2\left(\frac1{|\vec x-\hat e_3|^2},\frac{|\vec x|^2}{|\vec x-\hat e_3|^2}\right)\\
&+2 \fcy T^3\left(\frac1{|\vec x-\hat e_3|^2},\frac{|\vec x|^2}{|\vec x-\hat e_3|^2}\right)+4 \fcy T^4\left(\frac1{|\vec x-\hat e_3|^2},\frac{|\vec x|^2}{|\vec x-\hat e_3|^2}\right) \Bigg)\,.
\end{split}\end{equation}
Switching to spherical coordinates $\vec x = r\left(\sin(\qk)\sin(\fk),\sin(\qk)\cos(\fk),\cos(\qk)\right)$ and then integrating over $\phi$, we arrive at \eqref{oddIRes}.

\subsection{Computing $I_{\text{odd}}[\cT^i]$}
\label{IODD}
We will now compute $I_{\text{odd}}[\cT^i]$ for the various terms contributing to $\<SSSP\>$ at $O(c_T^{-1})$. Let us begin with the free connected and scalar exchange terms. Using \eqref{freeSSSP} and \eqref{oddIDef}, we can directly compute
\begin{equation}
I_{\text{odd}}\left[\fcy T^i_{\text{free}}\right] = -I_{\text{odd}}\left[\fcy T^i_{\text{scal}}\right]  = -4\pi\int dr\,d\qk\,\frac{\sin\qk}{r^2-2r\cos\qk+1} = -2\pi^3\,.
\end{equation}

To evaluate $I_{\text{odd}}[\cT^i]$, we find it most convenient to work in Mellin space. By using \eqref{melDef2} to rewrite $\cT^i$ in terms of its Mellin transform $N^i(s,t)$ and then integrating over $r$ and $\theta$, we find that
\begin{equation}
\tilde I_{\text{odd}}[\fcy T^i] = -8\pi^{9/2}\int\frac{ds\,dt}{(4\pi i)^2}\frac{N^i(s,t)\csc(\pi s)\csc(\pi t)\csc(\pi u)\left(\sin(\pi s) + \sin(\pi t) + \sin(\pi u)\right)}{(s-2)(s-3)} \,.
\end{equation}
We can now use \eqref{SSSPMel} to compute
\begin{equation}\begin{split}
I_{\text{odd}}[\fcy T^i_{\text{cont}}] &= -\frac{64\pi^5}{3}\int\frac{ds\,dt}{(4\pi i)^2} \csc(\pi s)\csc(\pi t)\csc(\pi u)\left(\sin(\pi s) + \sin(\pi t) + \sin(\pi u)\right) \\
&= -64\pi^5\int\frac{ds\,dt}{(4\pi i)^2} \csc(\pi s)\csc(\pi t) \\
&= -4\pi^3 \,.
\end{split}\end{equation}

\section{Localization in the Higher-Spin Limit}
\label{locHS}

In this appendix we evaluate the ${U(N)_k\times U(N+M)_{-k}}$ and ${SO(2)_{2k}\times USp(2+2M)_{-k}}$ sphere partition functions at large $M$. We begin with the former case, evaluating the integral 
\begin{equation}\begin{split}
Z_{M,N,k}(m_+,m_-)\qquad\qquad& \\
= \frac{e^{-\frac \pi2MN m_-}Z_0}{\cosh^N \frac{\pi m_+}2}&\int d^Ny\prod_{a<b} \frac{\sinh^2\frac{\pi(y_a-y_b)}{k}}{\cosh\left[\frac{\pi (y_a-y_b)}k +\frac{\pi m_+}{2}\right]\cosh\left[\frac{\pi (y_a-y_b)}k -\frac{\pi m_+}{2}\right]}\\
&\times\prod_{a=1}^N\left(\frac{e^{i\pi y_am_-}}{2\cosh\left(\pi y_a\right)}\prod_{l=0}^{M-1}\frac{\sinh\left[\frac{\pi\big(y_a+i(l+1/2)\big)}{k}\right]}{\cosh\left[\frac{\pi\big(y_a+i(l+1/2)\big)}k-\frac{\pi m_+}2\right]}\right)\,,
\end{split}\end{equation}
at large $M$, holding $\lk = M/k$ fixed. To begin, let us define
\begin{equation}\begin{split}
F_1(x) &= \sum_{l=-\frac{M-1}2}^{\frac{M-1}2}\log\tanh\left[\frac{\pi\big(x+il\big)}{k}\right] - \log R(x)\,, \\
F_2(x) &= \sum_{l=-\frac{M-1}2}^{\frac{M-1}2}\log\cosh\left[\frac{\pi\big(x+il\big)}{k}\right]\
\end{split}\end{equation}
where $R(x) = \cosh\left(\pi x\right)$ if $M$ is even and $R(x) = \sinh\left(\pi x\right)$ if $M$ is odd, and
\begin{equation}\begin{split}
G(x,\hat m_+) &= \log\left(\frac{k\sinh^2\frac{\pi x}{\sqrt k}}{\pi^2x^2}\sech\left[\frac{2\pi x+\pi \hat m_+}{2\sqrt k}\right]\sech\left[\frac{2\pi x-\pi \hat m_+}{2\sqrt k}\right]\right)\,,\\
\end{split}\end{equation}
where $\hat m_\pm = k^{-1/2} m_\pm$. After a change of variables $y_a\rightarrow \sqrt k\left(x_a-\frac{i M}2\right)$, we find that
\begin{equation}\begin{split}
Z_{M,N,k}&(\hat m_+,\hat m_-) \propto \frac1{\cosh^N \frac{\pi\hat m_+}{2\sqrt k}}\int d^Nx\prod_{a<b}(x_a-x_b)^2 \exp\left(G(x_a-x_b,\hat m_+)\right) \\
&\times\exp\left(\sum_a i\pi x_a\hat m_- + F_1\left(x_a\sqrt k\right)+F_2\left(x_a\sqrt k\right)-F_2\left(\frac{\sqrt k}{2}(2x-\hat m_+)\right)\right)\,.
\end{split}\end{equation}
We now expand $F_1(x)$, $F_2(x)$ and $G(x)$ at large $M$ and $k$, holding $x$, $\hat m_\pm$ and $\lk$ fixed. The large $M$ expansion of $F_1(x)$ has already been computed in \cite{Hirano:2015yha}, where it was shown that
\begin{equation}\begin{split}
F_1(x)\equiv \sum_{l = -\frac{M-1}2}^{\frac{M-1}2}&\log\tanh \frac{\pi(x+ il)}{k} - R(x) \sim \frac{\cos\frac{2x\nb_\lk}{k}}{\sinh \frac{\nb_\lk}k }\log\tan\frac{\pi \lk}2
\end{split}\end{equation}
The right-hand expression should be understood as a formal series expansion, which can be written more verbosely as
\begin{equation}\begin{split}
F_1(x) &= \sum_{n = 0}^\infty \frac{(-1)^nf_{2n}(k,\lk)}{(2n)!}\frac{x^{2n}}{k^{2n-1}}\,,\\
\text{ where }f_{2n}(k,\lk) &= \sum_{p = 0}^\infty \frac{4^n(2-4^p)B_{2p}}{(2p)!k^{2p}} \nb_\lk^{2p+2n-1}\log\tan\frac{\pi\lk}2\,,
\end{split}\end{equation}
and so we find that 
\begin{equation}
F_1\left(x\sqrt k\right) = \text{cons.} -2\pi\csc(\pi\lk)x^2 + \frac13 \pi^3(\cos(2\lk \pi)+3)\csc^3(\lk\pi)\frac{x^4}{k}+O(k^{-2})\,,
\end{equation}
Next we expand $F_2(x)$ using the Euler-MacLaurin expansion, finding that
\begin{equation}\begin{split}
F_2\left(x\right) &= \sum_{l=-\frac{M-1}2}^{\frac{M-1}2}\log\cosh\left[\frac{\pi\big(x+il\big)}{k}\right] \\
&= \frac{\pi x^2\tan\frac{\pi\lk}2}{k} - \frac{2\pi^3x^2(2x^2+1)\sin^4\frac{\pi\lk}2}{k^3\cos\pi\lk} + O(k^{-5})\,.
\end{split}\end{equation}
Finally, we can expand $G(x)$ by simply using the Taylor series expansion around $k^{-1/2} = 0$, so that
\begin{equation}
G(x,\hat m_+) = -\frac{\pi^2(8x^2+3\hat m_+^2)}{12k} + \frac{\pi^4(224x^4+360x^2\hat m_+^2+15\hat m_+^4)}{1440k^2} + O(k^{-3})\,.
\end{equation}

Putting everything together, we find that
\begin{equation}\begin{split}
Z_{M,N,k}&(\hat m_+,\hat m_-) \\
&\propto \frac1{\cosh^N \frac{\pi\hat m_+}{2\sqrt k}}\int d^Nx\prod_{a<b}(x_a-x_b)^2 \exp\left(-2\pi\csc(\pi\lk)\sum_a x_a^2 + O(k^{-1}) \right)\,.
\end{split}\end{equation}
where all higher order terms are polynomial in $x$ and $\hat m_\pm$. We thus find that to compute
\begin{equation}
\left.\frac{\nb^{n_1+n_2}Z_{M,N,k}(\hat m_+,\hat m_-)}{\nb^{n_1}\hat m_+\nb^{n_2}\hat m_-}\right|_{\hat m_\pm = 0}
\end{equation}
at each order in $k^{-1}$, all we must do is evaluate Gaussian integrals of the form
\begin{equation}
\int d^Nx\,  p(x_a)\prod_{a<b}(x_a-x_b)^2 \exp\left(-2\pi\csc(\pi\lk)\sum_a x_a^2\right)\,,
\end{equation}
where $p(x_a)$ is a polynomial in $x_a$. These are just polynomial expectation values in a Gaussian matrix model. They can be computed at finite $N$ as sums of $U(N)$ Young tableux \cite{Itzykson:1990zb}, as described in detail in Appendix B of \cite{Chester:2021gdw}. After computing these integrals, we find the explicit results given in \eqref{ctUN}, \eqref{UNs}. and \eqref{fourDivUN}.

We now turn to the large $M$ expansion of ${SO(2)_{2k}\times USp(2+2M)_{-k}}$ sphere partition function
\begin{equation}\begin{split}
&Z_{M,k}(m_+,m_-)\\
&\propto \frac 1 {\cosh \frac{\pi m_+}{2} }\int d y\  \frac{e^{i \pi m_- y}\cosh\left[\frac{\pi y}{2k}\right]\cosh\left[\frac{\pi y}{2k}+\frac{\pi m_+}2\right]}{\sinh\left[\pi y\right]\cosh\left[\frac{\pi y}{k}+\frac{\pi m_+}{2}\right]}\prod_{l = -M}^M\frac{\sinh\left[\frac{\pi( y+i l)}{2k}\right]}{\cosh\left[\frac{\pi( y+ i l)}{2k}+\frac{\pi m_+}2\right]}\,,
\end{split}\end{equation}
this time holding $\lk = \frac{2M+1}{2k}$ fixed. Defining 
\begin{equation}\begin{split}
\tilde F_1(x) &= \sum_{l=-M}^{M}\log\tanh\left[\frac{\pi\big(x+il\big)}{2k}\right] - \log \sinh\left(\pi x\right)\,, \\
\tilde F_2(x) &= \sum_{l=-M}^{M}\log\cosh\left[\frac{\pi\big(x+il\big)}{2k}\right]\,, \\
\hat G(x,\hat m_+) &= \log\left(\frac{\cosh\left[\frac{\pi x}{2\sqrt k}\right]\cosh\left[\frac{\pi(x+\hat m_+)}{2\sqrt k}\right]}{\cosh \frac{\pi \hat m_+}{2\sqrt k}\cosh\left[\frac{\pi (2x+\hat m_+)}{2\sqrt k}\right]}\right)\,,\\
\end{split}\end{equation}
and then performing a change of variables $x = k^{-1/2}y$, we find that
\begin{equation}
Z_{M,k}(m_+,m_-) \propto \int dx\ \exp\left(i\pi \hat m_- x + \tilde G(x,\hat m_+)  + \tilde F_1(x)+\tilde F_2(x) - \tilde F_2(x+\hat m_+)\right)\,.
\end{equation}
Each of $\tilde F_1(x)$, $\tilde F_2(x)$ and $\tilde G(x,\hat m_-)$ can be expanded at large $k$ with $x$ and $\hat m_\pm$ fixed in a completely analogous fashion to $F_1(x)$, $F_2(x)$ and $G(x,\hat m_-)$ respectively. We find that
\begin{equation}
Z_{M,k}(m_+,m_-) \propto \int dx\ \exp\left(-\pi\csc(\pi\lk) x^2 + \dots\right)
\end{equation}
where at each order in $k^{-1}$ and $\hat m_\pm$ the terms in the exponent are polynomial in $x$. Derivatives of $Z_{M,k}(m_+,m_-)$ at $m_\pm = 0$ reduce to a number of Gaussian integrals at each order in $k^{-1}$. After computing these integrals, we find the explicit results given in \eqref{cTLamSO} and \eqref{mixedSO}.

\section{Expansion of Conformal Blocks and $\bar{D}_{r_1,r_2,r_3,r_4}(U,V)$}
\label{Dolan}

In this appendix, we review the $U\sim0$ and $V\sim1$ expansion that we use to extract CFT data in Section \ref{numSec}. First we expand the superblocks as defined in \eqref{SuperconfBlock} at small $U$ as
 \es{SuperconfBlock}{
  \mathfrak{G}_I^{\bf r}(U, V)= \sum_{\substack{\text{conf primaries } \\ \text{${\cal O}_{\Delta, \ell, {\bf r}} \in {\cal M}_{\Delta_0, \ell_0}^{{\bf r}_0}$ }}} 
    a^I_{\Delta, \ell, {\bf r}}  \sum_{l=0}^\infty  U^{\frac{\Delta-\ell}{2}+l}g^{[l]}_{\Delta, \ell}(V)  \,,
 }
where the light-cone blocks $g^{[l]}_{\Delta, \ell}(V)$ are given explicitly in \cite{Li:2019cwm}. In our conventions, the lowest couple are
\es{lightconeBlock}{
g_{\Delta,\ell}^{[0]}(V)&=\frac{\Gamma(\ell+1/2)}{4^\Delta\sqrt{\pi}\ell!}(1-V)^\ell \,{}_2F_1\left(\frac{\Delta+\ell}{2},\frac{\Delta+\ell}{2},\Delta+\ell,1-V\right)\,,\\
g_{\Delta,\ell}^{[1]}(V)&=\frac{\Gamma(\ell+1/2)(1-V)^{\ell-2}}{2(2\ell-1)(2\Delta-1) 4^\Delta\sqrt{\pi}\ell! }\left[
2(\ell+\Delta)(\ell+\Delta-2\ell \Delta)\, {}_2F_1\left(\frac{\Delta+\ell-2}{2},\frac{\Delta+\ell}{2},\Delta+\ell,1-V\right)\right.\\
&\left.-(1+V)( \Delta^2+\ell^2(2\Delta-1)-2\ell(\Delta^2+\Delta-1) ) \,{}_2F_1\left(\frac{\Delta+\ell}{2},\frac{\Delta+\ell}{2},\Delta+\ell,1-V\right)
\right]
\,.\\
}
The expansion in $U\sim0$ organizes operators by twist, and we can then also expand in $V\sim1$ to organize operators by spin. After taking the $\Delta$ derivative in \eqref{treeExp}, we see that the anomalous dimensions come multiplied by $\log U$, while the OPE coefficients have no such coefficient. We can then compare this $U\sim0$ and $V\sim1$ expansion to the explicit tree level correlator given in \eqref{finalAnswer}. The expansion of the connected free term is trivial. For the exchange term, we use the $U\sim0$ and $V\sim1$ of $\bar{D}_{r_1,r_2,r_3,r_4}(U,V)$ given in \cite{Dolan:2000ut}, which in our case takes the form:
\es{DbarExp}{
\bar{D}_{1,1,\frac12,\frac12}(U,V)&=\sum_{m,n=0}^\infty \frac{\pi  U^m (1-V)^n \left(\frac{\Gamma \left(m+\frac{1}{2}\right)
   \Gamma \left(m+n+\frac{1}{2}\right)^2}{\sqrt{U} \Gamma (2
   m+n+1)}-\frac{\Gamma (m+1)^2 \Gamma (m+n+1)^2}{\Gamma
   \left(m+\frac{3}{2}\right) \Gamma (2 m+n+2)}\right)}{ m!
    n!}\,,\\
   \bar{D}_{\frac12,1,1,\frac12}(U,V)&=-\sum_{m,n=0}^\infty\Bigg[\frac{2 U^m(1-V)^n \Gamma \left(m+\frac{1}{2}\right)^2  
   (m+n)!^2 }{ m!^2  n! \Gamma \left(2
   m+n+\frac{3}{2}\right)}  \Big(\psi(m+n+1)-\psi\left(2
   m+n+\frac{3}{2}\right)\\
   &\qquad\qquad\qquad+\psi\left(m+\frac{1}{2}\right)-\psi
   ^{(0)}(m+1)\Big)+\frac12\log U  \Bigg]\,,\\
    \bar{D}_{1,\frac12,1,\frac12}(U,V)&=\sum_{m,n=0}^\infty \Bigg[\frac{\sqrt{\pi }  U^m(1-V)^n \Gamma \left(m+\frac{1}{2}\right)
   \Gamma (2 m+2 n+1) }{ 4^{m+n}m!  n! \Gamma \left(2
   m+n+\frac{3}{2}\right)}\Big(2 \psi\left(2
   m+n+\frac{3}{2}\right)\\
   &-2 \psi(2 m+2 n+1)-\psi\left(m+\frac{1}{2}\right)+\psi(m+1)+\log
   (4)\Big)+\frac12\log U \Bigg]\,,\\
}
where $\psi(x)$ is the Digamma function, and note the $\log U$ dependence in the last two expressions. It is then straightforward to compare these expansions to the superblock expansion around $U\sim0$ and $V\sim1$ to get the CFT data given in the main text.

\bibliographystyle{ssg}
\bibliography{N6bootdraft}

\end{document}